\documentclass[12pt]{report}
\usepackage{graphicx}
\usepackage{ODUthesis}
\usepackage{indentfirst}
\newcommand{\Pom}{I$\!$P} 		% gives pomeron symbol
\usepackage[sort&compress,numbers]{natbib}			% for bibtex
\usepackage{amsmath}			% math symbols
\usepackage{hyperref}			% for dynamic links
\usepackage{pdflscape}			% for landscape
\usepackage{multirow}			% to have a value take up multiple rows in a tables
\usepackage{bm} 			% for boldmath
\usepackage{caption}
\newcommand{\physrep}{Physics Reports}	% needed for bibliography stuff
\begin{document}

\title{Nucleon-Nucleon Spin Dependent Scattering Amplitudes to Describe Final State Interactions in Electromagnetic and Electroweak Nuclear Processes}

\author{William P. Ford}
\principaladviser{J. Wallace Van Orden}
\member{Jo Dudek}
\member{Sebastian Kuhn}
\member{Geoffrey Kraft}
\member{Ruhai Zhou}

\degrees{B.S. June 2004, Wright State University\\
         M.S. August 2006, Wright State University}
\dept{Physics}
%\copyrightfalse

\submitdate{August 2013}

\abstract{There are currently no models readily available that provide nucleon-nucleon spin dependent scattering amplitudes at high energies ($s \geq 6$   GeV$^2$). 
This work aims to provide a model for calculating these high-energy scattering amplitudes. 
The foundation of the model is Regge theory since it allows for a relativistic description and full spin dependence. 
A parameterization of the amplitudes is presented, and comparisons of the solution to the assembled data set are shown.  
In addition, an application of the model to describe final state interactions in deuteron electrodisintegration is presented.
Overall the model works as intended, and provides an adequate description of the nucleon-nucleon interaction at these energies.}

    \beforepreface
    \prefacesection{Acknowledgments}
        I would like to thank my advisor Wally Van Orden for all his patience, help, encouragement, suggestions, insight, etc... over the course of this work. The ODU nuclear theory group, Anatoly Radyushkin, Rocco Schiavilla, Jo Dudek, and Ian Balitsky for the various mentoring roles that they have provided. The Jefferson Lab theory group and ODU physics department for providing such wonderful environments to learn about physics. All my ODU classmates from whom I was able to learn so much from while working together.
    \afterpreface

\chapter{Introduction}\label{sec:intro}
The goal of this work is to develop a model for calculating elastic, spin-dependent scattering amplitudes for the nucleon-nucleon system. 
While much work has been applied to this topic over the years there is no analysis available for both proton-proton and proton-neutron 
in the mid to high energy range,  Mandelstam $s>6$ GeV$^2$. 
Thus far the most complete, highest energy, and readily available work is the Scattering Analysis Interactive Dial-in (SAID) \cite{SAIDpaper,SAIDdata} 
analysis which provides the proton-neutron amplitudes to $s \approx 6$ GeV$^2$ and the proton-proton amplitudes up to $s \approx 9.8$ GeV$^2$. 
The objective is to calculate the amplitudes at higher energies. In order to accurately describe the nucleon-nucleon system at these energies 
a fully relativistic, spin dependent model is required. Furthermore, due to the scarcity of data, particularly in the proton-neutron case, the model 
should provide confidence in extrapolating the results to higher energies. 

\begin{figure}
    \centering{\includegraphics[width=4in]{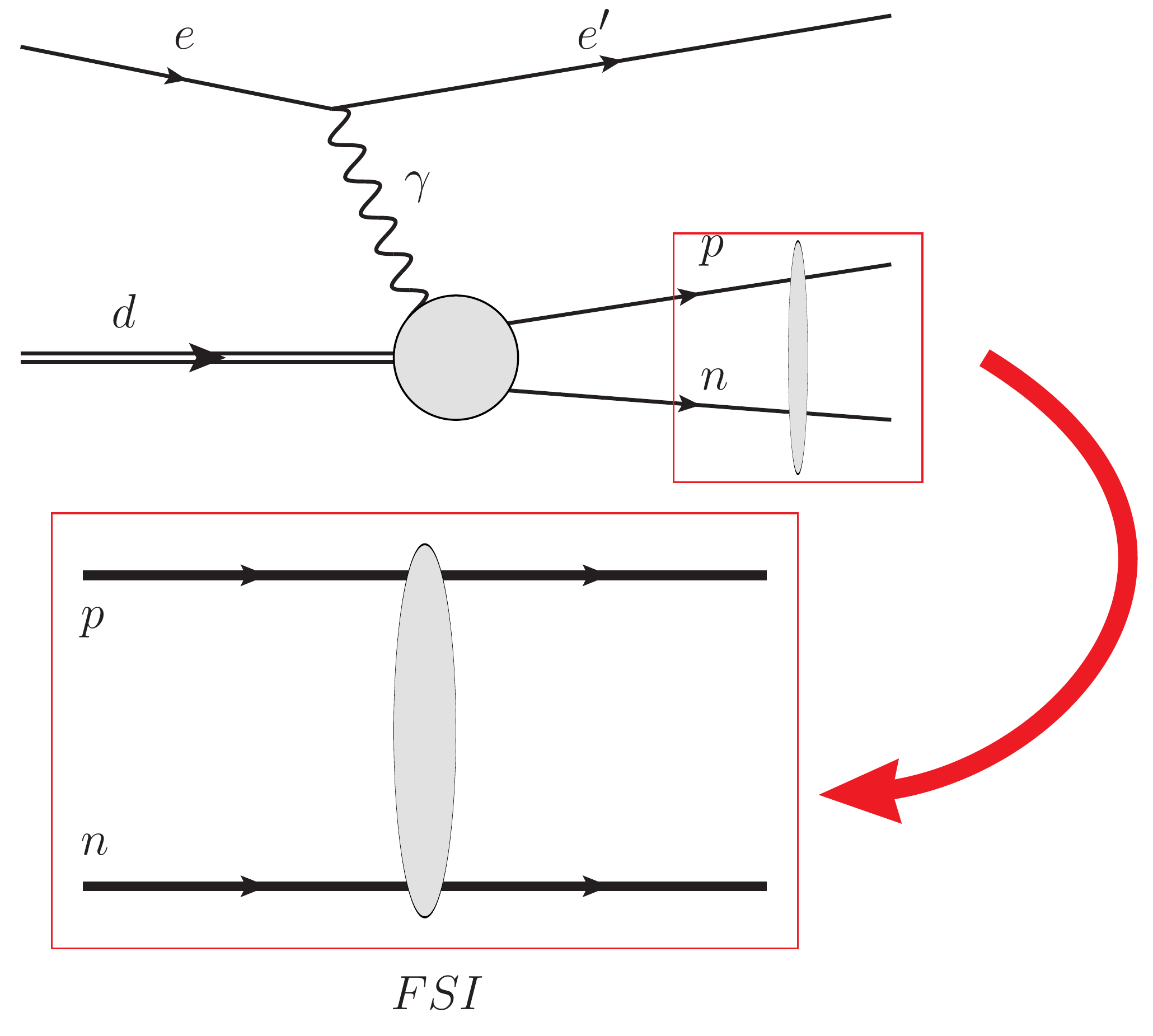}}
    \caption{Representation of deuteron electrodisintegration. The electron scatters off the deuteron via a virtual photon, breaking it into proton and neutron which interact in the final state. The box represents proton neutron scattering amplitudes which are input to the calculation.}\label{fig:deepn}
\end{figure}

The primary motivation in building this model is to utilize the amplitudes to describe final state interactions  in deuteron electrodisintegration, 
\begin{equation}
	e+d\rightarrow e+p+n. 
\end{equation}
A diagram of this reaction is shown in Fig. \ref{fig:deepn}.
Understanding this reaction at high energies is interesting in itself, and is an important stepping stone to describing heavier nuclei.
Additional motivation for investigating this process is to answer several open questions. 
At some point it is expected that a description with nucleon degrees of freedom will break down, and a description involving quark degrees of freedom will be necessary; at what energies does this occur? 
Is it possible to determine high momentum components of the deuteron wave function?
Also, since there are no free neutrons, can this process shed light on understanding properties of the neutron better, i.e. form factors? 
To have a chance at answering any of these questions, it is necessary to have an adequate theoretical description of the reaction mechanism.

At these energies a relativistic description is necessary, and there are several phenomena that can contribute. 
It is known that there will be final state interactions between the proton and neutron after the deuteron has broken apart.
There can also be isobar contributions, where a delta resonance occurs in an intermediate state.
Two body currents involving photon interactions with exchange forces between the two nucleons can also play a role in this process. 
Of these final state interactions are an important contribution, and are the focus of this work.

It has been shown that a complete description of the final state interactions is necessary in order to accurately describe this process \cite{JVO_2008_newcalc,JVO_2009_tar_pol,JVO_2009_ejec_pol}. 
This calculation, which shall be referred to as the JVO (authors Jeschonnek and Van Orden) model, requires the five spin dependent nucleon-nucleon amplitudes as input to the model.
The final state interactions are implemented via the SAID amplitudes, however, the kinematics at Jefferson Lab, where experiments have been performed, 
allow for final state nucleons with energies greater than can be described by SAID. 
More experiments are expected to be performed in the future once Jefferson Lab completes the 12 GeV upgrade, and there is anticipation of an even greater need to cover a larger kinematic regime.
The goal of this work is to construct a model for the nucleon-nucleon spin dependent amplitudes in order to extend the kinematic range of the JVO model. 

%\section{Regge Phenomenology}
The approach this work follows is to parameterize the spin dependent nucleon-nucleon amplitudes in terms of Regge poles or exchanges \cite{regge1959,martin,perl,collins,irving_Regge_phenom}.
Regge theory has had great phenomenological success as a parametrization method for many processes including nucleon-nucleon interactions. 
The theory is fully relativistic and allows for a complete spin-dependent description. 
Regge theory is also ideal for this application since it readily scales to higher energies, allowing one to have confidence in extrapolating the results. Furthermore, Regge theory has been utilized in the past to model 
proton-neutron scattering at mid-range energies with good results \cite{irvingNN}. Overall Regge theory provides us a systematic method of parameterizing 
the scattering amplitudes, while meeting all the criteria of the model.

The fundamental idea of Regge theory is to study the analytic behavior of the amplitudes, when one allows the angular momentum $J$ to be continuous and complex. 
While the analysis is rigorous\cite{regge1959} for non-relativistic scattering, the relativistic case is based on a series of assumptions. 
However, it is in the relativistic case, by exploiting crossing symmetry, which is discussed in more detail later, that Regge theory is useful as a parameterization method. 

While Regge theory excels at high energies, at lower energies it becomes more difficult to implement, as more and more Regge exchanges can contribute. 
Because of this feature, however, it naturally lends itself as a method for extrapolating to higher energies, 
since as one increases in energy the low energy exchanges are suppressed. 
In this work, fits are performed to the low energy nucleon-nucleon data, and the results are extrapolated to higher energy regions where data are unavailable.

%\section{Outline}
Chapter \ref{sec:theoretical framework} discusses the method of parameterizing the helicity amplitudes in terms of Regge poles. Then in Chapter \ref{sec:NNresults} 
the fitting procedure that was developed is discussed, and results of the fit to the data set of nucleon-nucleon observables is presented. 
In Chapter \ref{sec:deepfsi} an application of the model to describe final state interactions for deuteron electrodisintegration is shown. Observables for this process are calculated for kinematics which overlap with the SAID analysis, and comparisons between the two FSI models are presented. 
Chapter \ref{sec:conclusions} concludes this thesis with a summary of the results and an outlook of future work. 

\chapter{Regge Theory Applied to the Nucleon-Nucleon System}
\label{sec:theoretical framework}
This work assumes an understanding of scattering theory. 
For an excellent introduction to the topic \cite{martin,perl} are recommended, and were heavily utilized for this project. 

In developing this model, isospin ($I$) symmetry is assumed in order to describe both proton-proton and proton-neutron scattering. This allows one to treat protons and neutrons as identical particles, nucleons.
All observables in the nucleon-nucleon ($NN$) system can be described by five independent amplitudes \cite{Bystricky}. 
To keep track of spin the helicity basis is utilized. Helicity is the projection of spin onto the direction of the momentum, $\lambda = \vec{S}\cdot\hat{p}$.
For nucleons $\lambda = \pm \frac{1}{2}$. There are a total of sixteen amplitudes based on different helicity configurations, however, due to the symmetries of parity, time reversal, and isospin the number of independent amplitudes is reduced to five.
These five independent helicity amplitudes are given as,
\begin{align} \label{eq:amplitudes(abcde)}
a &= \phi_1  = T_{++;++} = \langle ++ |T| ++ \rangle \nonumber\\
b &= \phi_5  = T_{++;+-} = \langle ++ |T| +- \rangle  \nonumber\\
c &= \phi_3  = T_{+-;+-} = \langle +- |T| +- \rangle  \\
d &= \phi_2  = T_{++;--} = \langle ++ |T| -- \rangle  \nonumber\\
e &= \phi_4  = T_{+-;-+} = \langle +- |T| -+ \rangle , \nonumber
\end{align}
where the $+,-$ correspond to the helicities ($\pm \frac{1}{2}$) of the initial and final nucleons, 
%$\langle \mu_1,\mu_2 |T| \mu_1',\mu_2' \rangle$, where $\mu$ is used to clarify 
and for clarity the amplitudes are expressed in just some of the numerous forms that abound in the literature.

It is now useful to define the kinematics of the system. 
For elastic scattering of equal mass particles the center of momentum (cm) frame is convenient to work in. The amplitudes are functions of two variables, the cm energy ($E$) and scattering angle ($\theta$). 
It is preferable, since this is a relativistic system, to instead work with the invariant variables Mandelstam $s$, $t$, and $u$. 
The Mandelstam variables are defined as,
\begin{align}\label{eq:mandelstam_var}
s =& (p_1 + p_2)^2 = 4E^2  							\\
t =& (p_1 - p_1')^2 = -2 \lvert\vec{p\,} \rvert^2(1-\cos(\theta))		\\
u =& (p_1 - p_2')^2 = -2|\vec{p\,}|^2(1+\cos(\theta)),
\end{align}
where $p_1,p_2,p_1',p_2'$ are the incoming and outgoing four-momenta respectively and $\vec{p}$ is the three momentum of cm system. 
Note that $u$ is simply used for convenience as there are only two independent variables since,
\begin{equation}
s+t+u = 4m^2,
\end{equation}
where $m$ is the nucleon mass.

\begin{figure}
    \centerline{\includegraphics[height = 8cm]{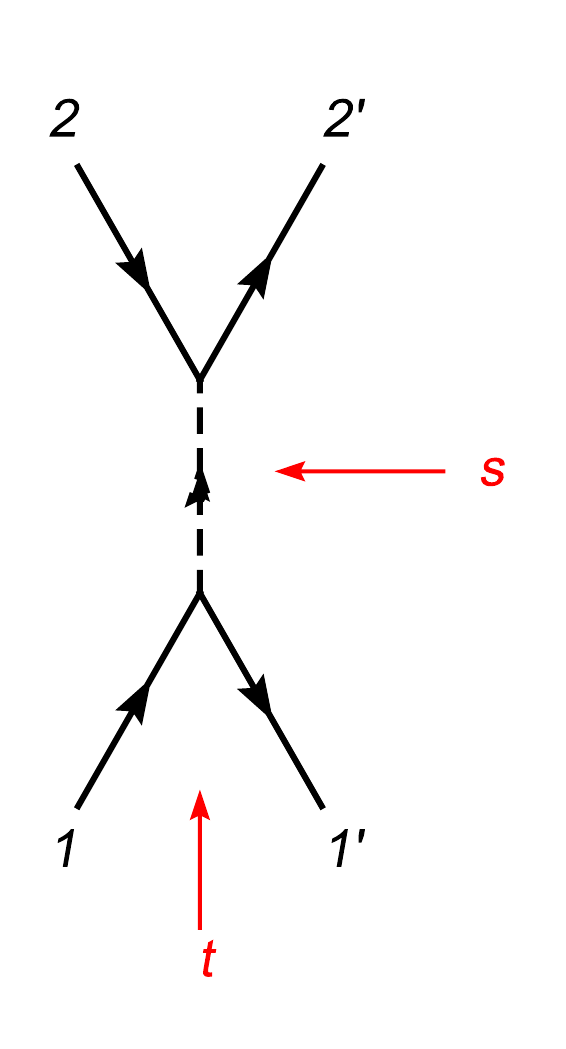}}
    \caption{Diagram depicting crossing symmetry between the $s$-channel process $N + N\rightarrow N + N$ and the $t$-channel process $N+\bar{N} \rightarrow N+\bar{N}$.}\label{fig:tcm}
\end{figure}
Now, when performing the Regge analysis it is necessary to exploit crossing symmetry. 
Crossing symmetry allows one to relate the amplitudes of the $NN$ system to the nucleon anti-nucleon system ($N\bar{N}$).
An example of crossing symmetry is shown in the diagram in Fig. \ref{fig:tcm}.  
Motivation for this will be addressed in a following discussion, but for now it is necessary to define the kinematics of the $N\bar{N}$ system. 
The $NN$ system will be referred to as the direct or $s$-channel process, and the $N\bar{N}$ system as the crossed or $t$-channel process.
In the $N\bar{N}$ system the Mandelstam variables become,
\begin{align}\label{eq:mandelstam_s_tchannel}
s =&  -2 \lvert\vec{p_t} \rvert^2(1 + \cos(\theta_t))  \\ \label{eq:mandelstam_t_tchannel}
t =&  4E_{t}^2 ,
\end{align}
where in the $t$-channel cm frame $E_t$ is the energy, $\vec{p_t}$ is the three momentum, and $\theta_t$ is the scattering angle.
Note that the $t$-channel cm angle is defined $90^{\circ}$ out of phase compared with conventional approaches. This was done in order to exploit a relation between the intitial and final $t$-channel states, (\ref{eq:ifrelation}), when symmetrizing the amplitudes. 

\begin{figure}
    \centerline{\includegraphics[width=6in]{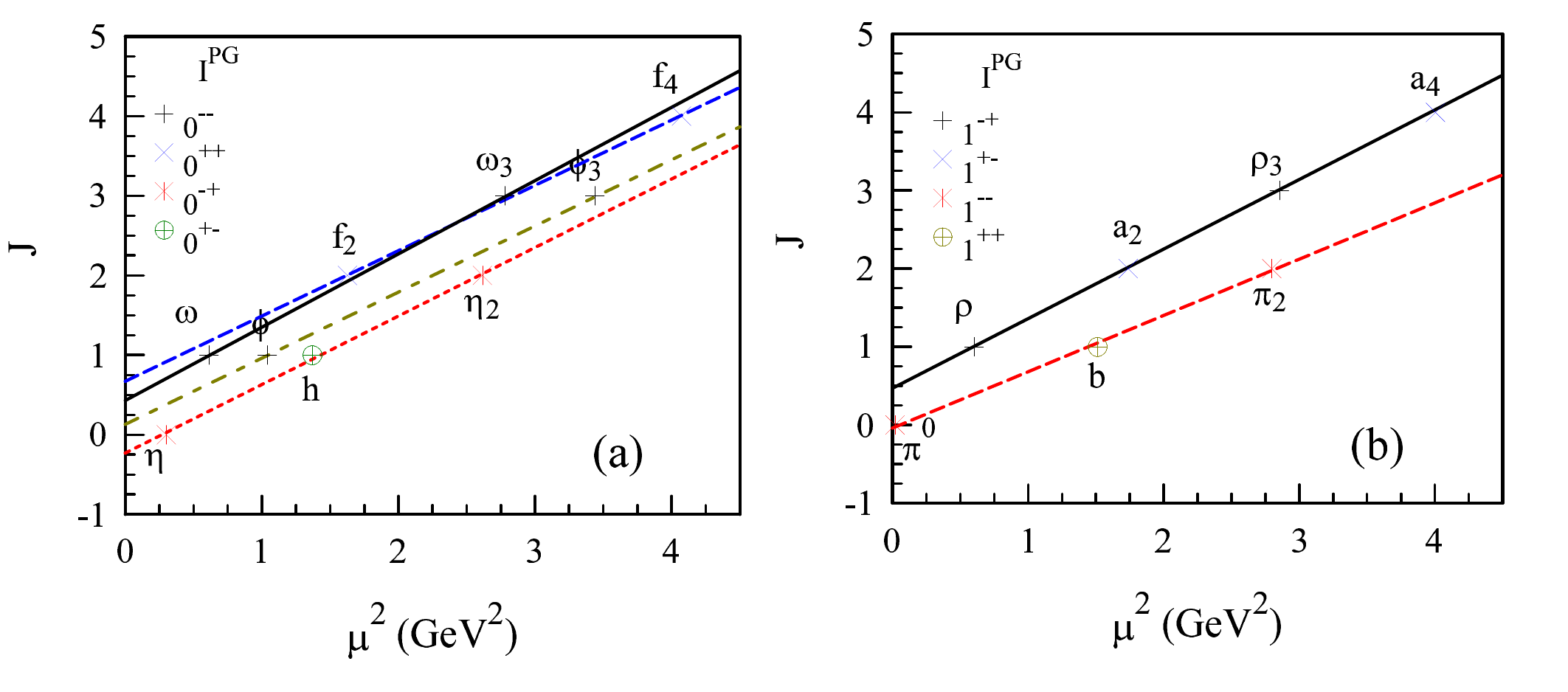}}
    \caption{Isoscalar (a) and isovector (b) mesons with $J$ plotted versus the square of the meson masses $\mu$. The various lines correspond to the Regge trajectories used in the fit to $NN$ scattering. An additional trajectory, the Pomeron, with $0^{++}$ is required to fit the large $s$ data.  It has an intercept of 1.08 and a slope of 0.25 GeV$^2$.}\label{fig:trajectories}
\end{figure}

To calculate the amplitudes Regge theory is used to provide a parameterization method. To motivate the discussion consider a meson exchange with fixed parity ($P$), $G$-parity ($G$), and isospin ($I$). Since the idea of Regge theory is to analyze the amplitudes in terms of continuous angular momentum, a plot of the spin $J$ vs. mass squared $\mu^2$ for the mesons which can contribute to the $NN$ system is shown in Fig. \ref{fig:trajectories}, and it is observed that the mesons fall on smooth curves. The curves can be represented as
\begin{equation}
J=\alpha(\mu^2)\,
\end{equation}
where $\alpha(\mu^2)$ is some function of the square of the meson mass $\mu^2$.  In the case of the well established mesons, the function is consistent with a straight line. The interpolating functions $\alpha_i$ describe Regge trajectories. Regge theory describes the $NN$ scattering amplitudes in terms of the exchange of Regge trajectories, or entire families of mesons rather than individual mesons. This discussion is to serve as motivation of the Regge analysis, as well as to provide a relationship between Regge exchanges and the physical mesons. In Section \ref{sec:ReggeAnalysis} it will be shown how the amplitudes can be parametrized in terms of Regge exchanges. 

It will also be important to perform the Regge analysis in the $t$-channel nucleon - antinucleon ($N\bar{N}$) center of momentum (cm) frame. To argue this again consider a meson exchange with a propagator proportional to $\frac{1}{\mu^2 - t}$. In the $s$-channel cm frame $t<0$ and this pole will never be reached. However, in the $t$-channel cm frame $t = 4E_{t}^{2}$, so for mesons with sufficiently large mass, the pole will be reached. This reasoning aims to provide an intuitive argument for why the Regge analysis will be performed in the $t$-channel cm frame, however, it will be seen when performing the calculation that it is also necessary for technical reasons as well.

\begin{figure} \centering
    \includegraphics[width=8.6cm]{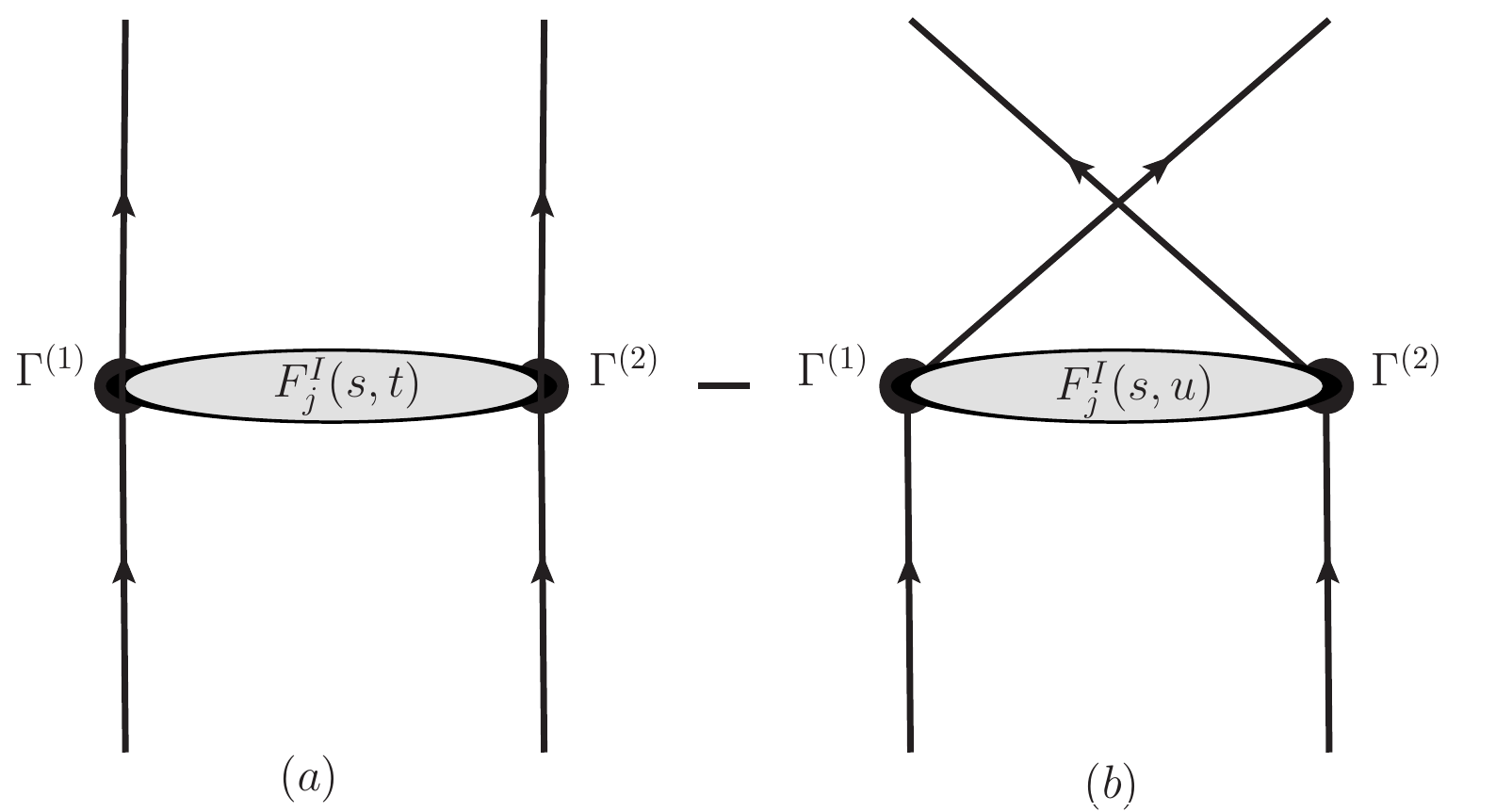} 
    \caption{Pictorial representation of the helicity amplitudes in terms of the Fermi invariants. $\Gamma$ represents the various gamma matrices which contribute to this process.}
    \label{fig:stat_wFI}
\end{figure}

From the previous discussion it can be noted that applying Regge theory to the nucleon-nucleon system presents some challenges, primarily due to the inclusion of spin. The Regge analysis should be performed in the crossed ($t$) channel, and the result analytically continued back to the $s$-channel, and because of the many helicity configurations the crossing relations are complicated. 
It has also been argued that Regge exchanges have definite quantum numbers, $P$, $G$, and $I$, which must be taken into account, and because of the symmetries of the nucleon-nucleon system, any non-strange mesonic Regge exchange, with $I = 0$ or $I = 1$, can contribute. 
Finally, since nucleons are fermions it is necessary to properly take into account Fermi statistics, that is the amplitudes must be antisymmetric. This can be accomplished by simply interchanging the labels on the final state particles as shown in Fig. \ref{fig:stat_wFI} (b).

Fortunately these complications can either be avoided, or at least simplified,  by relating the Regge exchanges to the Fermi invariants \cite{gribov,sharp},
\begin{align}\label{eq:FI}
   \hat{T} &= {F_{S}^I(s,t)}1^{(1)}  1^{(2)} - {F_P^I(s,t)} (i\gamma_5)^{(1)}  (i\gamma_5)^{(2)} \nonumber \\
           &+ {F_V^I(s,t)} \gamma^{\mu(1)} \gamma_{\mu}^{(2)} + {F_A^I(s,t)} (\gamma_5\gamma^{\mu})^{(1)} (\gamma_5\gamma_{\mu})^{(2)}  \\
           &+ {F_T^I(s,t)} \sigma^{\mu \nu (1)} \sigma_{\mu \nu}^{(2)}  \nonumber
\end{align}
where $s$ and $t$ are the Mandelstam variables, $I$ is an isospin label, $1$ and $2$ correspond to the vertices shown in Fig. \ref{fig:stat_wFI}.
This is an immediate benefit, since it gets all spin dependence ``out in the open''. The Dirac algebra can be performed straightforwardly to get the $s$-channel helicity amplitudes, 
\begin{align} \label{eq:Tpp}
T_{i}^{pp \rightarrow pp} &= \sum_j \left\{  C_{ij}^{t} \left[ F_{j}^{0}(s,t) + F_{j}^{1}(s,t) \right] 
                                           -C_{ij}^{u} \left[ F_{j}^{0}(s,u) + F_{j}^{1}(s,u) \right]  \right\} \\ \label{eq:Tpn}
T_{i}^{pn \rightarrow pn} &= \sum_j \left\{  C_{ij}^{t} \left[ F_{j}^{0}(s,t) - F_{j}^{1}(s,t) \right] 
                                           -2C_{ij}^{u} F_{j}^{1}(s,u)  \right\} ,
\end{align}
where $i$ corresponds to the helicity configurations $(++;++)$, $(++;+-)$, $(+-;+-)$, $(++;--)$, $(+-;-+)$, and $j$ to the different types of Fermi invariants $ S,V,T,P,A$. 
The matrices $C^{t}_{ij}$ and $C^u_{ij}$, containing all the spin dependence, are obtained from performing the Dirac algebra, and are given in the appendix (\ref{eq:C_t}), (\ref{eq:C_u}). 
For convenience, Mandelstam $u$ is used in the terms corresponding to the interchange of the final state particles necessary to account for Fermi statistics, Fig. \ref{fig:stat_wFI}(b).

This approach forgoes Reggeizeing the helicity amplitudes directly, and instead parameterizes the Fermi invariants in terms of Regge exchanges. Therefore the goal will be to determine which helicity amplidudes of definite $P$ and $G$ contribute to each invariant and Reggeize those combinations of amplitudes.
This is extremely beneficial since the crossing relations become trivial as the Fermi invariants can simply be analytically continued between the $s$ and $t$-channel cm frames, and since spin is taken care of explicitly, 
it will be seen that the Regge analysis will reduce to the spinless case. 

\section{Symmetric Amplitudes in terms of the Fermi Invariants}
Regge exchanges are found in the crossed $t$-channel cm frame, and have definite quantum numbers $P$, $G$, and $I$. The goal is to find a Regge approximation to the Fermi invariants, therefore it is necessary to symmetrize the amplitudes so one can ensure that a Regge exchange with definite quantum numbers contributes to the appropriate invariant. 
$I$ is easily factored out for the most part, and is taken care of in (\ref{eq:Tpp}) and (\ref{eq:Tpn}), 
so one simply needs to retain the label here, and it will be seen that it still plays a role when considering $G$-parity. Also, only exchanges related to Fig. \ref{fig:stat_wFI}(a) are necessary to perform explicitly, 
since Reggeization of Fig. \ref{fig:stat_wFI}(b) can easily be obtained by interchanging $t \leftrightarrow u$ in the final result.

\begin{figure} \centering
    \includegraphics[width=5cm]{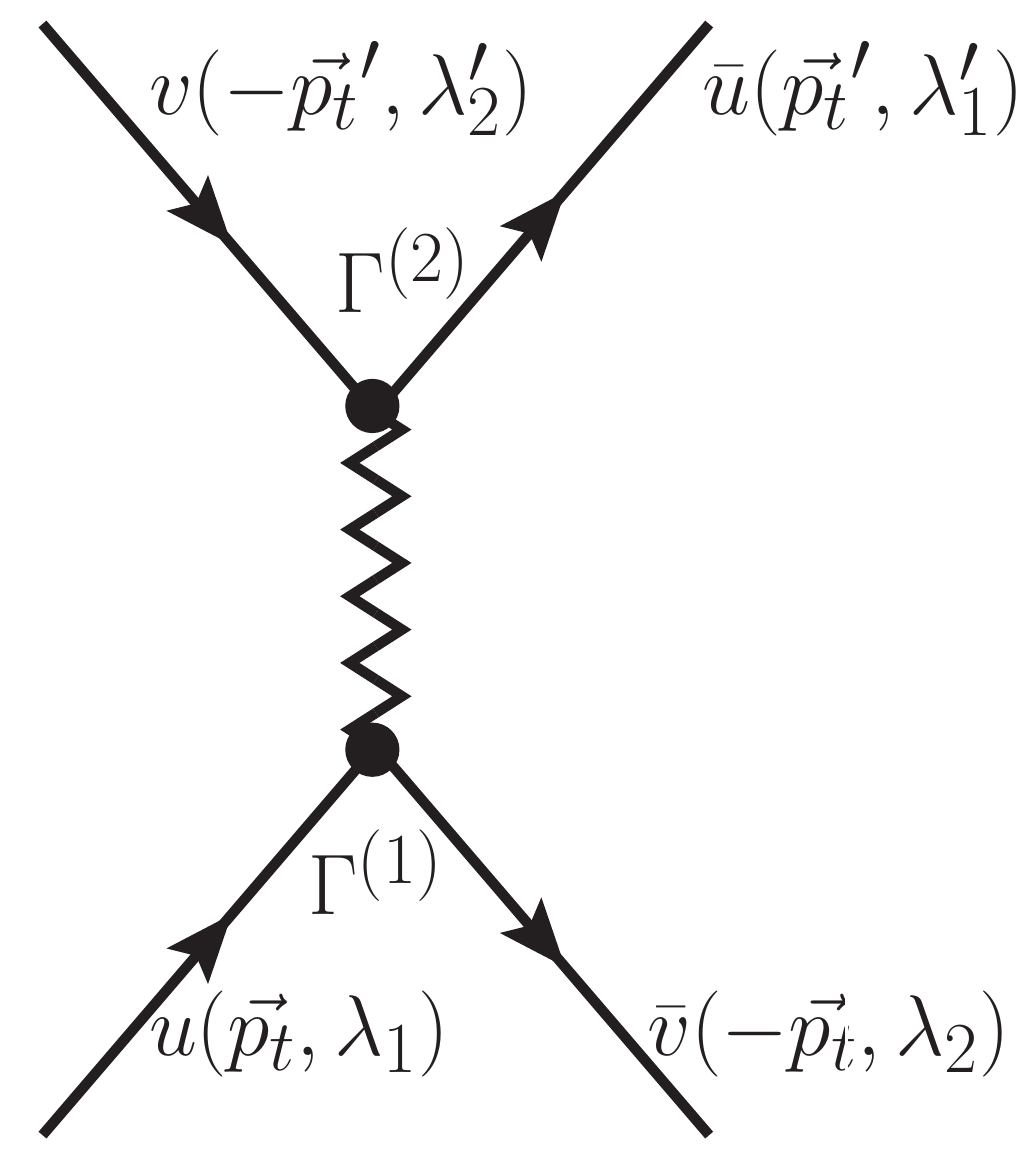} 
    \caption{Diagram depicting the $N\bar{N}$ annihilation process in the $t$-channel cm frame where the Regge analysis is performed.}
    \label{fig:tchannelexchange}
\end{figure}

In order to find the Regge contributions to the nucleon-nucleon ($NN \rightarrow NN$) system, it is necessary to analyze the $t$-channel, 
nucleon-anti-nucleon ($N\bar{N} \to N\bar{N}$) amplitudes of definite $P$ and $G$.
The $N\bar{N} \to N\bar{N}$ scattering process, in the $t$-channel cm frame where this analysis is performed is shown in Fig. \ref{fig:tchannelexchange}.

The focus will be on the initial state since there is a simple relation between the initial and final states. 
The basic two particle state that will be worked with is,
\begin{equation} \label{eq:psi_in}
 (\psi_{in})_{\alpha \beta} = \bar v_{\alpha}(-\vec{p_t},\lambda_2) u_{\beta}(\vec{p_{t}},\lambda_1) |I,M_I  \rangle,
\end{equation}
where $p_t$ is the $t$-channel center of mass momentum, $\lambda_1$ and $\lambda_2$ are the helicities of the incoming particles 1, and 2, 
$I$ is the total isospin of the state, $M_I$ is the $3^{rd}$ component of isospin, and the Dirac indices have been labeled explicitly with $\alpha$ and $\beta$. 
The goal is to symmetrize this state in terms of parity and $G$-parity. Beginning with parity, in Dirac space the one body parity operator is $\hat{P} = \gamma_0$. 
Parity acting on the two particle state yields,
\begin{equation}
  \hat{P} (\psi_{in})_{\alpha \beta} = -\bar{ v}_{\alpha}(\vec{p_t},-\lambda_2) u_{\beta}(-\vec{p_{t}},-\lambda_1) |I,M_I \rangle .
\end{equation}
A vertex with definite parity can then be constructed,
\begin{equation}\label{eq:def_P_expanded}
 (\psi^{P}_{in})_{\alpha \beta}\Gamma^{(1)}_{\alpha \beta} = \frac{1}{\sqrt{2}}\left[ \bar v(-\vec{p_t},\lambda_2)\Gamma^{(1)} u(\vec{p_{t}},\lambda_1) -P \bar{ v}(\vec{p_t},-\lambda_2) \Gamma^{(1)} u(-\vec{p_{t}},-\lambda_1) \right] |I,M_I \rangle
\end{equation}
Defining $\gamma_0 \Gamma \gamma_0 = P_{\Gamma}\Gamma$, where $P_{\Gamma}$ is $\pm1$, (\ref{eq:def_P_expanded}) can be simplified to,
\begin{align}\label{eq:def_P}
 (\psi^{P}_{in})_{\alpha \beta}\Gamma^{(1)}_{\alpha \beta} = \frac{1}{\sqrt{2}} (1+P_{\Gamma}P) \bar v(-\vec{p_t},\lambda_2)\Gamma^{(1)} u(\vec{p_{t}},\lambda_1) |I,M_I \rangle
\end{align}
Now moving on to $G$-parity. $G$-parity is defined as, $\hat{G} = \hat{\mathcal{C}}e^{i \pi \hat{I}_2} $, 
where $\hat{I}_2$ is the $y$ rotation matrix in isospin space and $\hat{\mathcal{C}}$ is the charge conjugation operator, 
given in Dirac space as, $\hat{\mathcal{C}} = \mathcal{C} \gamma_0 K$, where $K$ is the complex conjugation operator, and 
\begin{equation}
\mathcal{C} = \left( \begin{array}{cc} 0 & -i\sigma_2 \\ -i\sigma_2 & 0 \end{array} \right), \quad 
\sigma_2 = \left( \begin{array}{cc} 0 & -i \\ i & 0 \end{array} \right). 
\end{equation}
acting on a two particle state, (\ref{eq:psi_in}), yields,
\begin{equation}
 \hat{G} (\psi_{in})_{\alpha \beta} =  \bar{u}_{\alpha}(-\vec{p_t},\lambda_2) v_{\beta}(\vec{p_{t}},\lambda_1) (-1)^I\eta_{\lambda_2}\eta_{\lambda_1} \eta_C |I,M_I \rangle
\end{equation}
where $\eta_{ \lambda} = (-1)^{1/2 - \lambda}$, and $\eta_C$ is an arbitrary phase which is convenient to define as $\eta_C = (-1)^I\eta_{\lambda_2}\eta_{\lambda_1}$. 
Then a vertex of definite parity and $G$-parity can be constructed as,
 
\begin{align}
 (\psi^{PG}_{in})_{\alpha \beta}\Gamma^{(1)}_{\alpha \beta} =  \frac{1}{{2}} (1+P_{\Gamma}P)[
             \bar v(-\vec{p_t},\lambda_2)\Gamma^{(1)}u(\vec{p_{t}},\lambda_1) + G  \bar{u}(-\vec{p_t},\lambda_2)\Gamma^{(1)} v(\vec{p_{t}},\lambda_1)]|I,M_I \rangle
\end{align}
Defining $C_\Gamma \Gamma =  \mathcal{C} \gamma_0 K \Gamma  \mathcal{C} \gamma_0 K $, where $C_{\Gamma}$ is $\pm1$, this simplifies to,
\begin{equation}\label{eq:def_PG}
 (\psi^{PG}_{in})_{\alpha \beta}\Gamma^{(1)}_{\alpha \beta} = \frac{1}{2}(1+P_{\Gamma}P)(1+\eta_{\lambda_1}\eta_{\lambda_2}\mathcal{C}_{\Gamma}G)\bar v(-\vec{p_t},\lambda_2)\Gamma^{(1)} u(\vec{p_{t}},\lambda_1) |I,M_I \rangle.
\end{equation}
$P_{\Gamma}$ and $\mathcal{C}_{\Gamma}$ for the available couplings are given in Table \ref{ta:gamma_sym}. Note that the decomposition $\sigma^{\mu \nu (1)} \sigma^{(2)}_{\mu \nu} = -2 \vec{\alpha}^{(1)} \cdot \vec{\alpha}^{(2)} +2 \vec{\Sigma}^{(1)} \cdot \vec{\Sigma}^{(2)}$ is utilized. 

\begin{table}\centering
\caption{Symmetries of $\gamma$ matrices.}
 \begin{tabular}{ c c c c c c c c c} 
 \hline \hline
$\Gamma$            &$I$ & $\gamma^{5}$ & $\gamma^{0}$ & $\vec{\gamma}$ & $\gamma^{0}\gamma^{5} $ & $\vec{\gamma}\gamma^{5}$& $i\vec{\alpha}$& $\vec{\Sigma}$  \\ 
\hline
$P_\Gamma$          & +  &     -        &     +        &      -         &           -             &            +            &       -        &        +         \\
$\mathcal{C}_\Gamma$& +  &     -        &     -        &      -         &           +             &            +            &       -        &	      -         \\
	
\hline  \hline
\end{tabular}
\label{ta:gamma_sym}
\end{table}

The Dirac algebra can then be worked out for each term of the Fermi invariants, 
\begin{align}
%scaler
 (\psi^{PG}_{in})_{\alpha \beta}(I)^{(1)}_{\alpha \beta}          =& \frac{-|\vec{p}_t|}{2m}(1+P)(1+G) \eta_{\lambda_1} \delta_{\lambda_1,\lambda_2} \\
%psuedo-scaler
 (\psi^{PG}_{in})_{\alpha \beta}(i\gamma_5)^{(1)}_{\alpha \beta}  =& \frac{-iE_t}{2m}(1-P)(1-G) \delta_{\lambda_1,\lambda_2}        \\
%vector
%0_comp
(\psi^{PG}_{in})_{\alpha \beta}(\gamma^{0})^{(1)}_{\alpha \beta}  =& 0  \\
%3_comp
(\psi^{PG}_{in})_{\alpha \beta}(\vec{\gamma})^{(1)}_{\alpha \beta} =& -\frac{E_{t}+m}{4m}(1-P)(1-\eta_{\lambda_1}\eta_{\lambda_2}G) \left(1-4\lambda_1 \lambda_2 \frac{E_{t}-m}{E_{t}+m}\right) \nonumber \\ & \times \left( \chi^{\dag}_{\lambda_2}(\hat{p_{t}})\vec{\sigma} \chi_{\lambda_1}(\hat{p_{t}}) \right)   \\
%psuedo-vector
%0_comp 
(\psi^{PG}_{in})_{\alpha \beta}(\gamma^{0}\gamma_{5})^{(1)}_{\alpha \beta} =& \frac{1}{2}(1-P)(1+G) \delta_{\lambda_1,\lambda_2} \\
% 3_comp
(\psi^{PG}_{in})_{\alpha \beta}(\vec{\gamma}\gamma_{5})^{(1)}_{\alpha \beta} =& \frac{|\vec{p}_{t}|}{2m}(1+P)(1-G) \eta_{\lambda_1}\delta_{\lambda_1,-\lambda_2} \left( \chi^{\dag}_{\lambda_2}(\hat{p_{t}})\vec{\sigma} \chi_{\lambda_1}(\hat{p_{t}}) \right)  \\
%tensor
%alpha
(\psi^{PG}_{in})_{\alpha \beta}(i\sqrt{2}\vec{\alpha})^{(1)}_{\alpha \beta} =& \frac{-i\sqrt{2}(E_{t}+m)}{4m}(1-P)(1-\eta_{\lambda_1}\eta_{\lambda_2}G) \left(1+4\lambda_1 \lambda_2 \frac{E_{t}-m}{E_{t}+m}\right) \nonumber  \\ & \times \left( \chi^{\dag}_{\lambda_2}(\hat{p_{t}})\vec{\sigma} \chi_{\lambda_1}(\hat{p_{t}}) \right)  \\
% sigma
 (\psi^{PG}_{in})_{\alpha \beta}(\sqrt{2}\vec{\Sigma})^{(1)}_{\alpha \beta} =& \frac{\sqrt{2}|\vec{p}_{t}|}{2m}(1+P)(1-G)\eta_{\lambda_1} \delta_{\lambda_1,\lambda_2}\left( \chi^{\dag}_{\lambda_2}(\hat{p_{t}})\vec{\sigma} \chi_{\lambda_1}(\hat{p_{t}}) \right)
\end{align}

The result, (\ref{eq:def_PG}) can also be used to calculate outgoing states of definite parity and $G$-parity by utilizing the relations, 
\begin{align} \label{eq:ifrelation}
\left(\bar{u}(\vec{p},\lambda_1)\Gamma v(-\vec{p},\lambda_2) \right)^{\ast} = v^{\dag}(-\vec{p},\lambda_2)  \Gamma^{\dag} \bar{u}^{\dag}(\vec{p},\lambda_1) = \bar{v}(-\vec{p},\lambda_2)\Gamma u(\vec{p},\lambda_1).
\end{align}

With these results the symmetrized amplitudes can be constructed,  
\begin{align} \label{eq:FI_Dirac_coeef}
 \tilde{T}^{PGI}_{\lambda_1',\lambda_2';\lambda_1,\lambda_2} =& S^{PG}_{\lambda_{i}} F_{S}^I(s,t) + P^{PG}_{\lambda_{i}} F_P^I(s,t) + V^{PG}_{\lambda_{i}} F_V^I(s,t) \nonumber \\ &+ A^{PG}_{\lambda_{i}} F_A^I(s,t) + T^{PG}_{\lambda_{i}} F_T^I(s,t),
\end{align}
where $\lambda_i$ represents $\lambda_1$, $\lambda_2$, $\lambda_1'$, $\lambda_2'$, 
which are the helicities of the initial and final particles, and 
$S^{PG}_{\lambda_{i}},P^{PG}_{\lambda_{i}},V^{PG}_{\lambda_{i}},A^{PG}_{\lambda_{i}},T^{PG}_{\lambda_{i}}$ are obtained from the Dirac algebra, 
\begin{align}
%scaler
 S^{PG}_{\lambda_{i}} &=  \frac{p_{t}^2}{m^2}\frac{1}{4}(1+P)^{2}(1+G)^2 \eta_{\lambda_1}\eta_{\lambda_1'} \delta_{\lambda_1,\lambda_2} \delta_{\lambda_1',\lambda_2'}, \\
%Psuedo-scaler
P^{PG}_{\lambda_{i}}  &= -\frac{E_{t}^2}{m^2}\frac{1}{4}(1-P)^2(1-G)^2 \delta_{\lambda_1,\lambda_2} \delta_{\lambda_1',\lambda_2'}, \\
%Vector
V^{PG}_{\lambda_{i}}  &= -\frac{1}{4}(1-P)^2 \left(1-\eta_{\lambda_1}\eta_{\lambda_2}G\right)\left(1-\eta_{\lambda_1'}\eta_{\lambda_2'}G\right)
\left( \frac{E_{t}+m}{2m} \right)^{2} \left(1-4\lambda_1 \lambda_2 \frac{E_{t}-m}{E_{t}+m}\right) 
\nonumber \\ & \times \left(1-4\lambda_1' \lambda_2' \frac{E_{t}-m}{E_{t}+m}\right)  
 \left( \chi^{\dag}_{\lambda_2}(\hat{p_{t}})\vec{\sigma} \chi_{\lambda_1}(\hat{p_{t}}) \right) \cdot \left( \chi^{\dag}_{\lambda_2'}(\hat{p_{t}}')\vec{\sigma} \chi_{\lambda_1'}(\hat{p_{t}}') \right)^{\ast} , 
 \\
%axial-vector
A^{PG}_{\lambda_{i}} &= \bigg[ \frac{1}{4}(1-P)^2(1+G)^2 \delta_{\lambda_1,\lambda_2} \delta_{\lambda_1',\lambda_2'}  \nonumber \\ 
& - \frac{1}{4}(1+P)^2(1-G)^2 \frac{p_{t}^2}{m^2}   \eta_{\lambda_1}\eta_{\lambda_1'} \delta_{\lambda_1,-\lambda_2}  \delta_{\lambda_1',-\lambda_2'} \nonumber \\ 
&\times \left( \chi^{\dag}_{\lambda_2}(\hat{p_{t}})\vec{\sigma} \chi_{\lambda_1}(\hat{p_{t}}) \right) \cdot \left( \chi^{\dag}_{\lambda_2'}(\hat{p_{t}}')\vec{\sigma} \chi_{\lambda_1'}(\hat{p_{t}}') \right)^{\ast}  \bigg] ,  
%tensor
\\
T^{PG}_{\lambda_{i}} &= \bigg[-\frac{1}{2} (1-P)^2\left(1-\eta_{\lambda_1}\eta_{\lambda_2}G\right)\left(1-\eta_{\lambda_1'}\eta_{\lambda_2'}G\right)\left( \frac{E_{t}+m}{2m} \right)^{2} \left(1+4\lambda_1 \lambda_2 \frac{E_{t}-m}{E_{t}+m}\right)  
\nonumber \\
&\times \left(1+4\lambda_1' \lambda_2' \frac{E_{t}-m}{E_{t}+m}\right) 
+ \frac{p_{t}^2}{m^2}\frac{1}{2}(1+P)^{2}(1-G)^2 \eta_{\lambda_1}\eta_{\lambda_1'} \delta_{\lambda_1,\lambda_2}  \delta_{\lambda_1',\lambda_2'} \bigg] 
\nonumber \\
&\times \left( \chi^{\dag}_{\lambda_2}(\hat{p_{t}})\vec{\sigma} \chi_{\lambda_1}(\hat{p_{t}}) \right) \cdot \left( \chi^{\dag}_{\lambda_2'}(\hat{p_{t}}')\vec{\sigma} \chi_{\lambda_1'}(\hat{p_{t}}') \right)^{\ast} .
\end{align}

Selecting specific $P$ and $G$ values,
\begin{equation}
\tilde{T}^{++I}_{\lambda_1',\lambda_2';\lambda_1,\lambda_2} = 4F^{I}_{S}(s,t)\frac{p_{t}^2}{m^2} \eta_{\lambda_1}\eta_{\lambda_1'} \delta_{\lambda_1,\lambda_2}\delta_{\lambda_1',\lambda_2'},
\end{equation}
\begin{align}
\tilde{T}^{+-I}_{\lambda_1',\lambda_2';\lambda_1,\lambda_2} &= \bigg[ -4 F^{I}_{A}(s,t) \eta_{\lambda_1}\eta_{\lambda_1'} \delta_{\lambda_1,-\lambda_2} \delta_{\lambda_1',-\lambda_2'} + 8F^{I}_{T}(s,t)\frac{p_{t}^2}{m^2} \eta_{\lambda_1}\eta_{\lambda_1'} \delta_{\lambda_1,\lambda_2} \delta_{\lambda_1',\lambda_2'} \bigg] \nonumber \\
& \times \left( \chi^{\dag}_{\lambda_2}(\hat{p_{t}})\vec{\sigma} \chi_{\lambda_1}(\hat{p_{t}}) \right) \cdot \left( \chi^{\dag}_{\lambda_2'}(\hat{p_{t}}')\vec{\sigma} \chi_{\lambda_1'}(\hat{p_{t}}') \right)^{\ast},
\end{align}
\begin{align}
\tilde{T}^{-+I}_{\lambda_1',\lambda_2';\lambda_1,\lambda_2} &=  \bigg[ -4 F^{I}_{V}(s,t)\frac{E_{t}^2}{m^2} \delta_{\lambda_1,-\lambda_2} \delta_{\lambda_1',-\lambda_2'} +8F^{I}_{T}(s,t)\delta_{\lambda_1,-\lambda_2} \delta_{\lambda_1',-\lambda_2'}\bigg] \nonumber 
\\ &\times \left( \chi^{\dag}_{\lambda_2}(\hat{p_{t}})\vec{\sigma} \chi_{\lambda_1}(\hat{p_{t}}) \right) \cdot \left( \chi^{\dag}_{\lambda_2'}(\hat{p_{t}}')\vec{\sigma} \chi_{\lambda_1'}(\hat{p_{t}}') \right)^{\ast} + 4 F^{I}_{A}(s,t)\delta_{\lambda_1,\lambda_2} \delta_{\lambda_1',\lambda_2'},
\end{align}
\begin{align}
\tilde{T}^{--I}_{\lambda_1',\lambda_2';\lambda_1,\lambda_2} &= -4 F^{I}_{P}(s,t)\frac{E_{t}^2}{m^2}\delta_{\lambda_1,\lambda_2} \delta_{\lambda_1',\lambda_2'} \nonumber \\ 
& - \bigg[ 4 F^{I}_{V}(s,t)\delta_{\lambda_1,\lambda_2} \delta_{\lambda_1',\lambda_2'} +8F^{I}_{T}(s,t) \frac{E_{t}^2}{m^2}\delta_{\lambda_1,\lambda_2} \delta_{\lambda_1',\lambda_2'}  \bigg] 
\nonumber \\ & \times \left( \chi^{\dag}_{\lambda_2}(\hat{p_{t}})\vec{\sigma} \chi_{\lambda_1}(\hat{p_{t}}) \right) \cdot \left( \chi^{\dag}_{\lambda_2'}(\hat{p_{t}}')\vec{\sigma} \chi_{\lambda_1'}(\hat{p_{t}}') \right)^{\ast}.
\end{align}

Finally, selecting specific helicity values allows one to solve for each Fermi invariant. Because the invariants have no explicit spin dependence there are many redundant equations allowing one to solve by selecting only helicity combinations of $(++;++)$ and $(++;--)$, where this notation refers to the $t$-channel cm frame helicities $(\lambda_1,\lambda_2;\lambda_1',\lambda_2')$. It is important to note that while only these two helicity configurations are selected, it does not mean that the end result will only have these selections. The final $s$-channel cm frame result will have all five combinations of helicity values. These choices, however, will prove beneficial when performing the Regge analysis, and is applicable since the Fermi invariants have no explicit spin dependence. Selecting these contributions yields
\begin{align}\label{eq:FItoAMP1}
F^{I}_{S}(s,t) &= \frac{m^2}{8|\vec{p}_{t}|^2}\left(\tilde{T}^{++I}_{++;++} - \tilde{T}^{++I}_{++;--} \right),   \\
F^{I}_{V}(s,t) &= \frac{-1}{8\cos(\theta_t)}\left(\tilde{T}^{--I}_{++;++} - \tilde{T}^{--I}_{++;--} \right) -2\frac{E_{t}^2}{m^2}F^{I}_{T}(s,t),  \\
 F^{I}_{T}(s,t) &= \frac{m^2}{8|\vec{p}_{t}|^{2}\cos(\theta_t)}\left(\tilde{T}^{+-I}_{++;++} - \tilde{T}^{+-I}_{++;--} \right), \\
F^{I}_{P}(s,t) &= \frac{m^2}{8E_{t}^2}\left(\tilde{T}^{--I}_{++;++} + \tilde{T}^{--I}_{++;--} \right), \\
F^{I}_{A}(s,t) &= \frac{1}{8} \left(\tilde{T}^{-+I}_{++;++} + \tilde{T}^{-+I}_{++;--} \right).  \label{eq:FItoAMP2} 
\end{align}

\subsection{Relationship Between Fermi Invariants and Partial Waves}
Now, in order to Reggeize, a partial wave expansion is set up with definite parity and $G$-parity, in the $t$-channel cm frame,
\begin{align} \label{eq:PWE}
    \tilde{T}^{PGI}_{\lambda_1',\lambda_2';\lambda_1,\lambda_2} &= 
   \sum_J(2J+1) [f^{JIG}_{\lambda_{1}', \lambda_{2}' ; \lambda_1, \lambda_2}(E_t)
  - P(-1)^{J+\lambda_1'-\lambda_1}f^{JIG}_{\lambda_{1}', \lambda_{2}' ; -\lambda_1, -\lambda_2}(E_t) ] 
  \nonumber \\
  & \times d^J_{\lambda_1 - \lambda_2,\lambda_1' - \lambda_2'}(\theta_t),
\end{align}
where $E_t$ and $\theta_t$ are the $t$-channel center of momentum energy and scattering angle, 
and the $f^{GIJ}_{\lambda_{1}', \lambda_{2}' ; \lambda_1, \lambda_2}(E_t)$ 
correspond to partial wave coefficients from expanding on to the Wigner-$d$ functions. 
Plugging (\ref{eq:PWE}) into (\ref{eq:FItoAMP1}) - (\ref{eq:FItoAMP2}) results in the partial wave contributions to the Fermi invariants,
\begin{align}\label{eq:FItoPW1i}
F^{I}_{S}(s,t) &= \frac{m^2}{8|\vec{p}_{t}|^2}\sum_{J}(2J+1) \left[ f^{JI+}_{++;++}(E_{t})  -  f^{JI+}_{++;--}(E_{t}) \right]\left[1+(-1)^{J}\right]P_J(z_{t}) \\
F^{I}_{V}(s,t) &= \frac{-1}{8z_{t}}\sum_{J}(2J+1) \left[ f^{JI-}_{++;++}(E_{t})  -  f^{JI-}_{++;--}(E_{t}) \right]\left[1-(-1)^{J}\right]P_J(z_{t})  
				\nonumber \\ &- 2\frac{E_{t}^2}{m^2}F^{I}_{T}(s,t)  \\
 F^{I}_{T}(s,t) &= \frac{m^2}{8|\vec{p}_{t}|^{2}z_{t}}\sum_{J}(2J+1) \left[ f^{JI+}_{++;++}(E_{t})  +  f^{JI+}_{++;--}(E_{t}) \right]\left[1-(-1)^{J}\right]P_J(z_{t}) \\
F^{I}_{P}(s,t) &= \frac{m^2}{8E_{t}^2}\sum_{J}(2J+1) \left[ f^{JI-}_{++;++}(E_{t})  +  f^{JI-}_{++;--}(E_{t}) \right]\left[1+(-1)^{J}\right]P_J(z_{t})  \\
F^{I}_{A}(s,t) &= \frac{1}{8} \sum_{J}(2J+1) \left[ f^{JI+}_{++;++}(E_{t})  +  f^{JI+}_{++;--}(E_{t}) \right]\left[1+(-1)^{J}\right]P_J(z_{t}) \label{eq:FItoPW1f}
\end{align}
Because of the choice of helicity configurations, the Wigner-$d$ functions reduce to Legendre polynomials $d^J_{00}(\theta_t) = P_J(z_t)$, where $z_t = \cos(\theta_t)$, and the Reggeization procedure will reduce to the spinless case. Each Fermi invariant trivially crosses back to the $s$-channel, (\ref{eq:Tpp}) and (\ref{eq:Tpn}), 
and none of the complications that are associated with a typical Regge analysis of particles with spin need to be explicitly dealt with. 
Now that the Fermi invariants are in terms of Legendre polynomials it will be shown in the next section how an analysis of these amplitudes in terms of continuous complex $J$ will result in a Regge approximation to the invariants. 

\section{Regge Analysis} \label{sec:ReggeAnalysis}
\begin{figure} \centering
	\includegraphics[height=6cm]{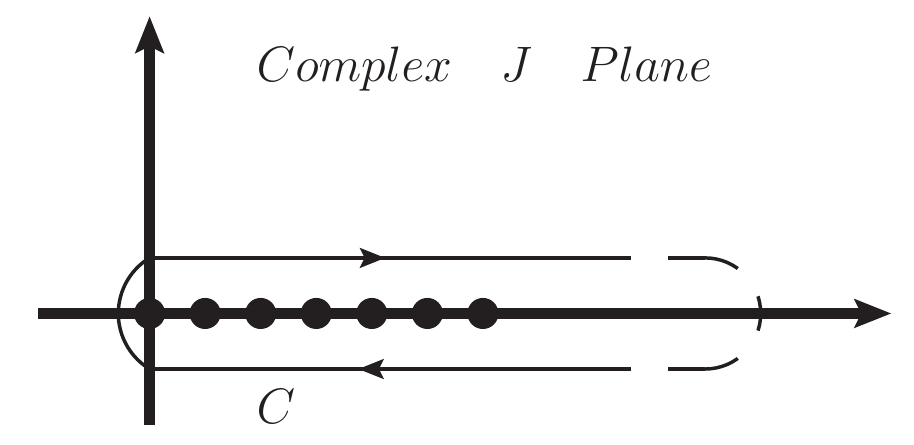} 
	\caption{Contour integral which represents the partial wave sum for continuous, complex $J$.}
	\label{fig:closed_contour}
\end{figure}

%The next step is to Reggeize spinless amplitudes, 
It is now time to determine the Regge contribution to the Fermi invariants. Note again that all spin dependence is dealt with explicitly with the spinors and gamma matrices, therefore typical methods to Reggeize spinless amplitudes \cite{martin,perl,collins} can be followed. Each Fermi invariant is in terms of combinations of amplitudes of the form,
\begin{equation}
 F^{I}_{i}(s,t) \propto R^{I}(s,t) = \sum_{J}(2J+1)f_i^J(E_{t})(1+\xi_i(-1)^J)P_J(z_{t}),
\end{equation}
where $i$ labels the type of invariant ($S$,$V$,$T$,$P$,$A$), and $\xi_i$ is $\pm1$ corresponding to the invariant determined by Eqs. (\ref{eq:FItoPW1i}) - (\ref{eq:FItoPW1f}) . The coefficient $f_i^J(E_{t})$ is the combination of partial waves also given by Eqs. (\ref{eq:FItoPW1i}) - (\ref{eq:FItoPW1f}) . 
The first step to Reggeize is to allow $J$ to be continuous and complex allowing this summation to be rewritten as a contour integral,
\begin{equation}\label{eq:contour1}
 R^{I}(s,t) = \frac{-1}{2i} \oint_C \frac{(2J+1)f_i^J(E_{t})[P_J(-z_{t}) + \xi_iP_J(z_{t})]}{\sin{\pi J}}dJ,
\end{equation}
where the contour is shown in Fig. \ref{fig:closed_contour}. The property that $(-1)^JP_J(z_{t})=P_J(-z_{t})$ has also been used.
 %\begin{center} 
 %\end{center}
There are a few things that need to be verified, however, before moving on. These are discussed thoroughly in \cite{martin, collins}, so details will be omitted, but the main issue is to determine that the inverse transform is still defined now that $J$ is continuous and complex. 
The standard inverse transform,
\begin{equation}
 f_i^J(E_{t})(1+\xi_i(-1)^J) = \frac{1}{2} \int_{-1}^{1} dz P_J(z_{t}) R^{I}(s(z_{t}),t),
\end{equation}
will not suffice here, as it is now, because eventually it will be required to take $|J| \rightarrow \infty $, and $P_J(z)$ blows up everywhere except the real axis. This implies that this is not a proper analytical continuation to complex $J$, and that this continuation is not unique. There is, however, a way out of this dilemma by examining $R^{I}(s(z_{t}),t)$ further. 

Assume that $R^{I}(s,t)$ satisfies a fixed $t$ dispersion relation, and for simplicity ignore any subtractions,
\begin{equation}
 R^{I}(s,t) |_{t-\mathrm{fixed}} = \int_{s_0}^{\infty} \frac{D_s(s',t)}{s'-s} ds',
\end{equation}
where $D_s(s,t) = \frac{1}{2i}\left[ R^I(s+i\epsilon,t) - R^I(s-i\epsilon,t) \right]$ is the discontinuity of $R^I(s,t)$ across the $s$-channel cut, and $D_s(s,t) = \Im [R^I(s,t)]$ for real values of s. More details regarding the $s$-channel cut, and analytic properties of the amplitudes can be found in \cite{martin}. 
Now this can be rewritten in terms of $z_{t}$, using 
\begin{align}
 s' - s &= -2\vec{p_{t}}^2 ( 1 + z_{t}' -1 - z_{t} ) =  -2\vec{p_t}^2(z_{t}' - z_{t}) \\
ds' &= -2\vec{p_{t}}^2 dz_{t}'
\end{align}
So the dispersion relation becomes,
\begin{equation}
 R^{I}(s(z_t),t) |_{t-fixed} = \int_{z_{t0}}^{\infty} \frac{D_s(s'(z_{t}'),t)}{z_{t}'-z_{t}} dz_{t}'.
\end{equation}
If this is true then the inverse transform can be taken in terms of this dispersion relation, and appropriate asymptotic behavior can be ensured. So the coefficient becomes,
\begin{align}
 f_i^J(E_{t})(1+\xi_i(-1)^J) &= \frac{1}{2} \int_{-1}^{1} dz_{t} P_J(z_{t}) R^{I}(s(z_{t}),t) \nonumber \\
                         &= \frac{1}{2} \int_{-1}^{1} dz_t P_J(z_{t}) \int_{{z_{t}}_0}^{\infty} \frac{D_s(s'(z_{t}'),t)}{z_{t}'-z_{t}} dz_{t}' \nonumber \\ \label{eq:coefficient}
                         &= \int_{{z_{t}}_0}^{\infty} D_s(s'(z_{t}'),t) Q_J(z_{t}') dz_{t}'.
\end{align}
In the last line Neumann's formula was used to do the integral over $z_{t}$,
\begin{equation}
 Q_J(z') =  \frac{1}{2} \int_{-1}^{1} dz \frac{P_J(z)}{z'-z}.
\end{equation}
Now the coefficients are defined over the entire complex plane because $Q_J(z_{t}) \rightarrow 0$ for $|J| \rightarrow \infty$, and one can return to Reggeizing the amplitudes using (\ref{eq:contour1}).

\begin{figure} \centering
    \includegraphics[height=10cm]{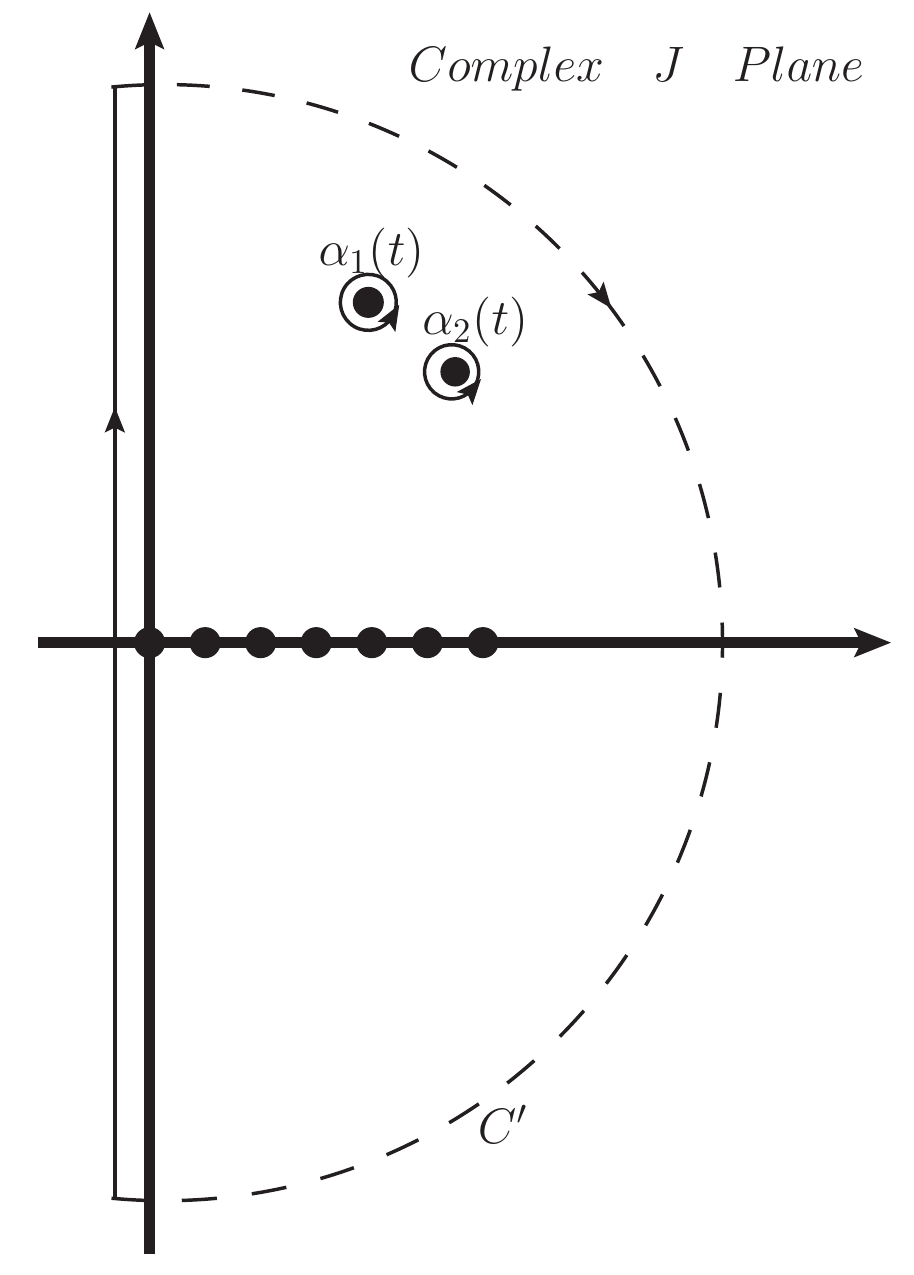} 
    \caption{Complex $J$ plane with contour opened up. The $\alpha_i$ represent the location of Regge poles. Only two poles are shown here although in principle there can be many.}
\label{fig:open_contour}
\end{figure}

So now that the coefficients are properly defined the contour can be opened up to include the entire right hand side of the complex $J$ plane. When doing this the assumption is that the only contributions picked up are simple poles in the upper right plane. This assumption is motivated from the non-relativistic case where one can show that this is true. In the relativistic case this assumption is not actually true since there can also be cuts, however, this possibility is ignored. Historically many models ignore Regge cuts with good success \cite{irving_Regge_phenom,irvingNN}, and since the goal of this work is to develop an effective parametrization scheme an assumption of simple poles is justified. The new contour is shown in figure \ref{fig:open_contour}. 

The invariant then becomes
\begin{align}\label{eq:contour2}
 R^{I}(s,t) =& \frac{-1}{2i} \int_{-\frac{1}{2}-i\infty}^{-\frac{1}{2}+i\infty} \frac{(2J+1)f_i^J(E_{t})[P_J(-z_{t}) + \xi_iP_J(z_{t})]}{\sin{\pi J}}dJ \nonumber \\ 
                 &- \sum_{k} \frac{\pi(2\alpha_k(t)+1)\beta_k(t)[P_{\alpha_k(t)}(-z_{t}) + \xi_iP_{\alpha_k(t)}(z_{t})]}{\sin{\pi \alpha_k(t)}},
\end{align}
where $\alpha_k(t)$ is the position of the pole, and $\beta_k(t)$ is the residue of $f_i^J(E_{t})$ at that pole. The sum over $k$ represents that, in general, many Regge poles can contribute. It should be noted that all the $z_{t}$ dependence, and therefore all the $s$ dependence, is still in the Legendre polynomial. Also, the large semi-circle at infinity is discarded because, from the earlier work, (\ref{eq:coefficient}), the coefficients will go to zero for $|J| \rightarrow \infty$. In  (\ref{eq:contour2}), the first term is referred to as the background integral, and the second is the Regge poles that were picked up.

Now, since the background integral goes as $z_{t}^{-\frac{1}{2}}$, if the limit $z_{t} \rightarrow \infty$ is taken it does not contribute. In this limit only the Regge contribution is relevant,
\begin{equation}
R^{I}(s,t)   = - \sum_{k} \frac{\pi(2\alpha_k(t)+1)\beta_k(t)[P_{\alpha_k(t)}(-z_{t}) + \xi_iP_{\alpha_k(t)}(z_{t})]}{\sin{\pi \alpha_k(t)}}.
\end{equation}
Taking this limit is the technical reason for performing the Regge analysis in the crossed $t$-channel.
From (\ref{eq:mandelstam_s_tchannel}) and (\ref{eq:mandelstam_t_tchannel}), $z_t = -1 - \frac{2s}{t- 4m^2}$. 
In the $N \bar{N}$, $t$-channel cm frame where this analysis has been performed $z_t$ is bounded by $-1 \leq z_t \leq 1$, and taking $z_t \rightarrow \infty$ is unphysical. 
However, in the the $NN$ $s$-channel cm frame $1 \leq z_t \leq \infty$, which means that if the analysis result is analytically continued back to the $s$-channel cm frame the approximation that has been found is in a physical region. In other words, the $s$-channel cm frame amplitudes can be approximated by $t$-channel cm frame Regge poles.
%
%This large $z_t$ limit is completely unphysical here, however, it is a physical limit back in the original, uncrossed, $s$-channel. 
%If we recall that $z_t = -1 - \frac{2s}{t- 4m^2}$, then we see that $z_{t} \rightarrow \infty$ is simply a large $s$, small $t$ approximation.
%In the $s$-channel cm frame this limit is a high energy, small angle approximation. 
%%Had we tried to directly ``Reggeize'' in the $s$-channel we would have been forced to take this unphysical limit and everything would have been useless (we would actually have found an approximation for the crossed channel). 
%So even though we have done this whole analysis here in the crossed channel, this final Reggeized result is an approximation to the Fermi invariant back in the original channel. 

This is essentially the result. The scattering amplitudes are in terms of the Fermi invariants, and these are now parametrized in terms of Regge poles. 
It is now convenient to clean things up a bit, which is necessary in order to pragmatically utilize this analysis. 
Taking the approximation $P_{\alpha}(-z_{t}) \approx e^{-i\pi\alpha}P_{\alpha}(z_{t})$ the Legendre polynomial can be factored out,
 \begin{equation}
R^{I}(s,t)   = - \sum_{k} \frac{\pi(2\alpha_k(t)+1)\beta_k(t)(e^{-i\pi\alpha_k(t)} + \xi_{i})P_{\alpha_k(t)}(z_{t})}{\sin{\pi \alpha_k(t)}}.
\end{equation}
Since the large $z_{t}$ limit has been taken the Legendre polynomials can be simplified somewhat. The Legendre polynomials can be written as a sum \cite{edmunds1996angular},
\begin{equation}
 P_J(z) = \sum_{r=0}^{\nu} (-1)^r \frac{(2J-2r)! z^{J-2r} }{2^{J}r!(J-r)!(J-2r)!},
\end{equation}
where $\nu = \frac{J}{2}$ or $=\frac{J-1}{2}$, whichever is an integer. So for very large $z$ the $r=0$ term is sufficient,
\begin{equation}
 P_J(z) \mathop {\sim } \limits_{z \rightarrow \infty}   \frac{\Gamma(2J+1) z^J}{2^J \Gamma(J+1)^2}.
\end{equation}
Using this expansion and $z_{t} =( -1+\frac{2s}{4m^2 - t})$, the invariants can be written with the typical Regge dependence,
\begin{equation}
 R^{I}(s,t)   = - \sum_{k}\pi(2\alpha_k(t)+1)\beta_k(t)\frac{e^{-i\pi\alpha} + \xi_{i}}{\sin{\pi \alpha_k(t)}}\frac{\Gamma(2\alpha_k(t)+1)}{2^{\alpha_k(t)} \Gamma(\alpha_k(t)+1)^2} \left(-1+\frac{2s}{4m^2-t}\right)^{\alpha_k(t)}.
\end{equation}
It is also convenient to simplify the phase factor,
\begin{align}
\frac{e^{-i\pi\alpha} \pm 1}{\sin{\pi \alpha_k(t)}} = 
\frac{e^{-\frac{i\pi\alpha}{2}}}{2\sin(\frac{\pi\alpha}{2})\cos(\frac{\pi\alpha}{2})} \left(e^{-\frac{i\pi\alpha}{2}} \pm e^{\frac{i\pi\alpha}{2}}\right) 
=
\left\{ 
\begin{array}{c}
 \frac{e^{-i(\pi\alpha(t)/2)}}{\sin(\frac{\pi\alpha}{2})} \quad  + \\ \\ \frac{-ie^{-i(\pi\alpha(t)/2)}}{\cos(\frac{\pi\alpha}{2})} \quad  - 
\end{array} \right.  
\end{align}

Assuming all extra $t$ dependence, including the $\Gamma$ functions and the $\sin(\pi\alpha/2)$ or $\cos(\pi\alpha/2)$ in the phase function, can be absorbed into the residue and label it with isospin, parity, and G-parity, $\beta \rightarrow \beta^{IPG}$. The phase function can be defined as,
\begin{equation}\label{eq:signature} %\frac{e^{-i\pi\alpha(t)} \pm 1}{\sin{\pi \alpha(t)}} \approx 
\xi_{\pm}(t) = \left\{ \begin{array}{c}
 e^{-i(\pi\alpha(t)/2 + \delta)} \quad  + \\ \\ -ie^{-i(\pi\alpha(t)/2 + \delta)} \quad  - 
\end{array}  \right. ,
\end{equation}
where an additional phase $\delta$ has been introduced, which accounts for the various approximations that have been made, and the fact that all extra $t$ dependence has been absorbed into the residue. Ultimately it provides an extra degree of freedom which is convenient when fitting certain Reggeons. Finally for reasons discussed more thoroughly in Section \ref{sec:res_traj_params}, it is necessary to multiply by an overall large angle cutoff factor $\zeta(s,t)$. The final expression can now be written as, 
\begin{equation}
 R^{I}(s,t)   = \zeta(s,t) \sum_{k}\xi_{k\pm}(t)\beta_{k}^{IPG}(t)\left(-1+\frac{2s}{4m^2 - t}\right)^{\alpha_k(t)}.
\end{equation}

\section{Summary of Calculation}
This section is to summarize the results of the calculation, and discuss the parameters entering into the Regge exchanges. The Fermi invariants are given in terms of Regge exchanges by,
\begin{equation} \label{eq:FI_to_RP_t}
\left( \begin{array}{c}
F^{I}_{S}(s,t)  \\ \\
F^{I}_{V}(s,t) \\ \\
F^{I}_{T}(s,t) \\ \\
F^{I}_{P}(s,t) \\ \\
F^{I}_{A}(s,t)      
\end{array} \right)
=
\left( \begin{array}{ccccc}
\frac{m^2}{2(t-4m^2)} & 0 & 0 & 0 & 0 \\ \\
0 & \frac{t-4m^2}{8\left(2s+t-4m^2 \right)} & \frac{t}{8(2s+t-4m^2)} & 0 & 0 \\ \\
0 & 0 & -\frac{m^2}{4(2s+t-4m^2)} & 0 & 0 \\ \\
0 & 0 & 0 & \frac{-m^2}{2t} & 0 \\ \\
0 & 0 & 0 & 0 & \frac{1}{8} \\ \\
\end{array} \right)
\left( \begin{array}{c}
R^{I++}_{+1}(s,t)  \\ \\
R^{I--}_{-2}(s,t) \\ \\
R^{I+-}_{-3}(s,t) \\ \\
R^{I--}_{+4}(s,t) \\ \\
R^{I-+}_{+5}(s,t)      
\end{array} \right),
\end{equation}
where $m = .93895$ (GeV) is the nucleon mass, and the right-most vector is defined by a sum of Regge exchanges,
\begin{equation}\label{eq:ReggeExchange}
 R^{IPG}_{\pm j}(s,t)= \zeta(s,t)\sum_{k}\xi_{k\pm}(t)\beta_{k}^{IPG}(t) \left(-1+\frac{2s}{4m^2-t}\right)^{\alpha_k(t)},
\end{equation}  
where $\beta(t)$ and $\alpha(t)$ correspond to the residue and the trajectory of the Regge pole and are discussed in the following section, 
$\zeta(s,t)$ is a cutoff factor also discussed in the following section, and
$j$ is simply the position of the Regge exchange in the vector, which is referred to as the ``type'' of the exchange. 
Also note that while Reggeons with $PG = --$ enter into two different positions in (\ref{eq:FI_to_RP_t}), 
the residues of any contributing poles in these positions are not necessarily the same. 

The $u$-channel exchanges of Fig. \ref{fig:stat_wFI}(b) are taken into account by a simple substitution of $t \rightarrow u$ in (\ref{eq:FI_to_RP_t}). 
An additional factor of $\frac{t}{4m^2}$ is utilized for type 4 exchanges, guaranteeing that the amplitude $d = 0$ at $t = 0$, which is required by conservation of angular momentum. 
In addition, type 5 exchanges are multiplied by a factor of $\frac{4m^2}{s}$, which is assumed can be factored from $F_{A}(s,t)$. 
This is necessary in order to cancel with an additional factor of $s$ in the matrix $C^{t}_{ij}$, and prevents amplitude $e$ from blowing up at large $s$. 
This seems to be a general problem with expressing the amplitudes in terms of the Fermi invariants at large $s$, and $F_{A}(s,t)$ should either always be 
redefined or parameterized to explicitly cancel this factor of $s$ in order to avoid this problem.  
Now that the Fermi invariants are parametrized in terms of Regge exchanges this result can be plugged into (\ref{eq:Tpp}) and (\ref{eq:Tpn}) for a Regge approximation of the $s-$channel helicity amplitudes.

\subsection{Residue and Trajectory} \label{sec:res_traj_params}
Linear Regge trajectories are utilized, $\alpha(t) = \alpha_0+\alpha_1t$. 
These are obtained from the well established meson masses available from the Particle Data Group\cite{PDG} and are shown in Fig. \ref{fig:trajectories}.
In addition to the mesonic trajectories, it was also necessary to utilize ``effective'' trajectories, which are discussed in more detail in section \ref{sec:Fit_Details}. These were included so that the fit to the low $s$ data could use the same form as the Regge parameterization.

Three different parametrizations are used for the residues,
\begin{align} \label{EQ:residues}
\beta_{I}(t)   &= \beta_0e^{\beta_1t} \nonumber \\
\beta_{II}(t)  &= \left(1 - e^{\gamma t}\right)\beta_0e^{\beta_1t}  \\  
\beta_{III}(t) &= \frac{t}{4m^2}\beta_0e^{\beta_1t} \nonumber
\end{align}
where $\beta_0$ and $\beta_1$ are fit parameters. $\gamma$ was set by hand and is used in two exchanges in order to reproduce the diffraction minimum in the high energy 
proton-proton data. 
Utilization of the different types of residues for various Regge exchanges was based on trial and error.

Equation (\ref{eq:ReggeExchange}) differs from the usual expression in that the full expression for $\cos(\theta_t)$ was kept. 
Generally the Regge limit assumes that $\cos(\theta_t)\gg1$, which implies that $s \gg 4m^2 - t$. In extrapolating from the region where the SAID partial wave analysis
has been performed to higher $s$, this condition is violated in two respects. First, data where $s$ is of the same order of magnitude as $4m^2$ are included. 
Second, in the same region there are significant data for $4m^2 - s < t <0$. So at backward angles $t$ is of the same order of magnitude as $s$. 
For this reason the exact expression for $\cos(\theta_t)$ is kept.

A practical problem associated with fitting at low $s$ is that the $u$ channel contributions necessarily overlap those from the $t$ channel.
Fitting to data near $\theta = 0^\circ$, where $t=0$ and $u = 4m^2 - s$, and near $\theta = 180^\circ$, where $u = 0$ and $t = 4m^2 - s$,
can be affected substantially by the tail of the crossed channel. This can cause the fitting procedure to become very sensitive, if not unstable.
As a result it was found useful to introduce a cutoff factor,
\begin{equation}
	\zeta(s,t) = \left(1 - e^{20\left(\frac{t}{4m^2 - s} - 1\right)}\right),
\end{equation}
to decouple the $t$ and $u$ channel contributions at the endpoints in order to simplify the fitting procedure. 
This has no effect at large $s$ where
the two channels have no significant overlap, but is extremely useful for smaller values of $s$.
\section{Electromagnetic Effects}
In order to properly describe the proton-proton interaction electromagnetic effects should be taken into account. These effects take place for very small $-t$ values or small angles, and show up in the high energy data. The region where these effects are noticeable is referred to as the Coulomb region, whereas the region which is dominated by the strong interaction is referred to as the hadronic region. The transition area between these two is the interference region. 

The full proton vertex is used,
\begin{align} \label{eq:EM_vertex}
\Gamma^{\mu}_{EM} = F_1(t) \gamma^{\mu} - \frac{F_2(t)}{2m} & i \sigma^{\mu \nu}q_{\nu},
\end{align}
where $q_{\nu}$ is the four momentum of the photon, and $F_1(t)$ and $F_2(t)$ are the Dirac form factors of the proton and are related to the electric ($G_E(t)$) and magnetic ($G_M(t)$) form factors by,
\begin{align}
  F_1(t) &= \frac{G_E(t) - G_M(t) t/4m^2}{1 - t/4m^2} \\ 
  F_2(t) &= \frac{G_M(t) - G_E(t)}{1 - t/4m^2}. 
\end{align} 
A typical parametrization of the form factors, which is sufficient for this work is,
\begin{equation}
G_E = G_M/2.79 = (1 - t/.71)^{-2}.
\end{equation}

The amplitudes for a one photon exchange are then, 
\begin{align}\label{eq:EMampa}
a_{EM}(s,t) &=  \frac{4\pi/137}{2t(4m^2 - s)(-4m^3 + mt)^2} \nonumber \\   
       &\times(-8G_{E}G_{M}{m^2}stu- G_{M}^2t(32m^6 + s^2t + 2m^2t(s + t) - 8m^4(s + 2t)) \nonumber  \\
       &- 8G_{E}^2m^4(16m^4 + 2s^2 + 3st + t^2 - 4m^2(3s + 2t))),
\end{align}
\begin{align}\label{eq:EMampb}  
b_{EM}(s,t) &= -\frac{4\pi/137}{2mt(-4m^2 + t)^2}\sqrt{\frac{stu}{(s - 4m^2)^2} } \nonumber \\
       &\times\left( (s-u)(4m^2G_{E}^2 + G_{M}^2 t) + 2G_{E}G_{M}(16m^4 - st - 4m^2(s + t) )   \right)  ,
\end{align}
\begin{align}\label{eq:EMampc}
c_{EM}(s,t) &= -\frac{(4\pi/137)u}{2t(4m^2 - s)(-4m^3 + mt)^2} \nonumber \\
       &\times (8G_{E}^2m^4(u-s) + 8G_{E}G_{M}m^2st + G_{M}^2t(-8m^4 + 2m^2t - st)),
\end{align}
\begin{align}\label{eq:EMampd}
d_{EM}(s,t) &=  \frac{4\pi/137}{(s - 4m^2)(t-4m^2)^2}  
       \times (4G_{E}G_{M}su + G_{E}^2s(s-u) \nonumber \\
       & + G_{M}^2(16m^4 + 2s^2 + 3st + t^2 - 4m^2(3s + 2t))) ,
\end{align}
\begin{align}\label{eq:EMampe}
e_{EM}(s,t) &=  -\frac{4\pi/137}{(s-4m^2)(t-4m^2)^2} \times(4G_{E}G_{M}su + G_{E}^2s(s-u) \nonumber \\
	&+ G_{M}^2(16m^4 + 2s^2 + 3st + t^2 - 4m^2(3s + 2t))).
\end{align}
where $u = 4m^2 -s -t$.

In order to account for higher order effects a helicity-dependent constant and phase are utilized. 
Since the electromagnetic contribution is dominated by ``no flip'' (amplitudes $a$ and $c$) and ``single flip'' (amplitude $b$) contributions, the one photon exchange amplitudes are redefined  as follows,
\begin{align} \label{eq:EM_phase_amp}
 a'_{EM}(s,t)  = \beta_ae^{i\delta_a}a_{EM}(s,t) \\
 b'_{EM}(s,t)  = \beta_be^{i\delta_b}b_{EM}(s,t) \\
 c'_{EM}(s,t)  = \beta_ce^{i\delta_c}c_{EM}(s,t),
\end{align}
where $\beta_a$, $\beta_b$, $\beta_c$, $\delta_a$, $\delta_b$, and $\delta_c$  are fit to available polarization and differential cross section data. 
Utilizing this approach allows one to keep the electromagnetic effects under control, and smoothly fit through the Coulomb, interference, and hadronic regions. 
\chapter{Nuceon-Nucleon System}
\label{sec:NNresults}
\section{Fit Details} \label{sec:Fit_Details}
Presented here is the solution of the fit to available data of $NN$ scattering observables \cite{FVO_Reggemodel}. 
The $\chi^2$ values are given in Table \ref{ta:chi^2}, 
and the parameter values are given in Table \ref{ta:params_pol}. 
%In comparison with the SAID analysis, there is overlap between the two models at lower energies, 
%and we expect the SAID solution to be more precise, i.e. lower $\chi^2$.
%The emphasis on the fit is the ability to extrapolate to higher energies. 
%Using a Regge model over the entire angular region may give less precise results than the SAID parameterization, but it does allow this extrapolation.
%A further test of the results will be when we implement them into the electrodisintegration of the deuteron process, as it will allow us to see 
%how much the results vary in comparison with the use of the SAID amplitudes through the kinematic region of overlap.
%
The data set was assembled from the SAID analysis \cite{SAIDdata}, the Durham database \cite{Durham}, 
the Cudell dataset \cite{Cudell:2005sg}, and the Particle Data Group \cite{PDG}. 
The original experimental papers are referenced here 
\cite{BARASHENKOV,Schwaller:1979eu,MARSHALL,SUTTON,SCHWALLER,GUZHAVIN,MESHCHERYAKO,DZHELEPOV,DZHELEPOV2,ELIOFF,SMITH,IGO,Shimizu:1982dx,Jaros:1977it,Chen:1956zz,Hart:1962zz, COLETTI,COLETTI2,Longo:1962zz,Blue:1962zz,PARKS,Diddens:1962zz,TAYLOR,Ginestet:1969zx,Jenni:1977kv, BELLETTINI1, BELLETTINI2,Almeida:1969bv,CZAPEK,BLOBEL,BREITENLOHNE,Ashmore:1960zz,Jabiol:1977ku,Apokin:1976zu,Bushnin:1973yv,Ammosov:1972cx,Brick:1982dy,Brenner:1981kf,Bartenev:1972nf,Barish:1974ga,Kafka:1978pp,Firestone:1974pd,Dao:1972jb,Bromberg:1973fi,Amaldi:1972uw,Amos:1981dr,Amos:1985wx,Amaldi:1976yf,Baksay:1978sg,Carboni:1984sg,Amaldi:1978vc,Eggert:1975bd,Ambrosio:1982gq,Honda:1992kv,Baltrusaitis:1984ka,Shapiro:1965zza,CARVALHO,LAW,Bugg:1966zz,Badier:1972ui,Galbraith:1965jk,Carroll:1974yv,Denisov:1971jb,Carroll:1975xf,Carroll:1978vq,KRUCHININ,DALKHAZAV,AZIMOV,Albers:2004iw,Shimizu1982445,Garcon:1986ni,Albrow:1970hn,Williams:1972gk,Kammerud:1971ac,Jenkins:1979nc,Fujii:1962zz,Eisner:1965zz,Ankenbrandt:1968zz,Rust:1970wp,Ambats:1974is,Clyde:1966zz,Preston:1960zz,Cork:1957zz,Alexander:1967zz,Rubinstein:1984kf,Akerlof:1976gk,Nagy:1978iw,Amaldi:1979kd,Breedon:1988kd,Faissler:1980fk,Breakstone:1984te,Bizard:1974hc,Terrien:1987jt,Perl:1969pg,Palevsky:1962zz,Miller:1971yc,Friedes:1965zz,Kreisler:1966kf,Altmeier:2004qz,Kobayashi:1991am,Bystricky:1986nj,Perrot:1987pv,Garcon:1986qp,DallaTorreColautti:1989nk,Andreev:2004bs,Cozzika:1967fz,Neal:1967jv,Marshak:1978wh,Ball:1987bh,Bell:1980rp,Diebold:1975yu,Miller:1977pm,Makdisi:1980gf,Zhurkin:1978rr,Ball:1999yy,Grannis:1966zz,Lin:1978xu,Parry:1973fj,BAREYRE,Allgower:1999ad,Allgower:1998ma,Allgower:1999ac,Ball:1999cn,Arvieux:1997vg,Deregel:1976jx,Abshire:1974ed,Klem:1977xq,Rust:1975kv,Aschman:1977rf,Borghini:1971mq,Kramer:1977pf,Corcoran:1980ew,Crabb:1977mg,Gaidot:1976kj,Okada:2006dd,Snyder:1978gs,Fidecaro:1980ee,Fidecaro:1978rb,Fidecaro:1981dk,Akchurin:1993xd,Ball:1992qn,Sakuda:1982eg,deLesquen:1999yz,Robrish:1970jw,Abolins:1973dy,Crabb:1979nh,Crabb:1982eh,Bauer:2004aw,Allgower:2000pv,Allgower:2001qs,Perrot:1988tw,Ball:1994uz,Bystricky:1985bz,Lehar:1988tx,Auer:1978zp,Auer:1983vm,Fontaine:1989ak,Auer:1982cv,Ball:1988pc,Abshire:1975bp,Ball:1994bp,Lac:1989fx,Lac:1989aj,Allgower:1998wf,Allgower:1998dy,Binz}. 
The collected dataset that the fit is based on is intended to be made available to the community, and can currently be obtained by contacting the author. 
Where both the statistical and systematic errors are available, the larger of the two was used in fitting the model to the data.

%All observables were fit simultaneously. 
%and \ref{ta:params_unpol} for the polarized and unpolarized solutions respectively. 

$\chi^2$ is calculated as the sum of each observable,
\begin{align}
\chi^2   &= \sum_k w_k \chi_k^2 \\
\chi_k^2 &= \sum_i \left( \frac{O_k(\vec{\beta},s_i,t_i) - E_{k,i}}{\sigma_i} \right)^2,
\end{align}
where $k$ denotes the various observables given in Table \ref{ta:chi^2}, $i$ denotes the data point, $E_{k,i}$ is the experimental value, sigma is the experimental error, $O_k(\vec{\beta},s_i,t_i)$ is the model value dependent on the parameters represented by $\vec{\beta}$, and $w_k$ is a weight factor discussed in the following paragraph.

The number of data points for each observable is given in Table( \ref{ta:chi^2}), 
and it should be noted that for every observable the number of points for proton-neutron data is considerably lower than the proton-proton case. 
Besides this asymmetry between proton-proton and proton-neutron data there are varying amounts of data for the various observables.
In order to avoid the largest data sets dominating the fit, 
weights were implemented, $w_k$, in order to keep all observables on the same footing.
The weights are given in Table \ref{ta:chi^2}. In order to calculate the reported, non biased $\chi^2$ the weights are set equal to 1.

\begin{figure}
    \includegraphics[width=15cm]{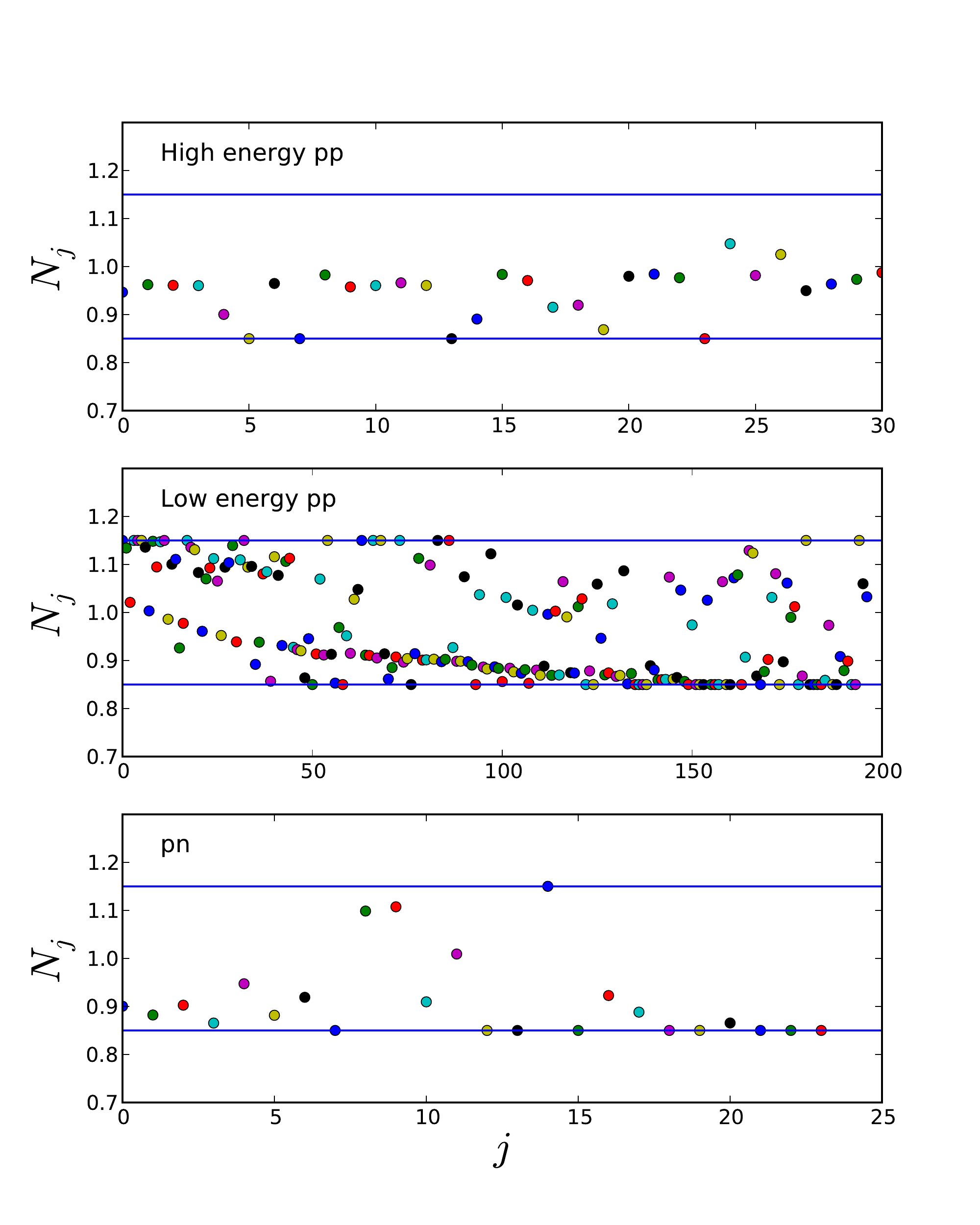}  %\textwidth
    \caption{Normalization factors which allow the experimental values for the differential cross sections to float plotted vs. $j$, an integer corresponding to number of data sets and ordered from low to high $s$. The solid lines represent the minimum and maximum values ($\pm$ 15\%) that the data was allowed to shift by.}
    \label{fig:norm_floats}
\end{figure}

Because the data set includes differential cross section data from many sources, there is a potential problem with normalization. 
In order to correct for this the fit was made to the shape of the differential cross section data, allowing the overall magnitude of each differential cross section data set to float. This is accomplished by fitting to a modified $\chi^2$. The data is organized into sets based on the energy, $s$, and the data source. Defining, 
\begin{align}
		\chi^2_{d\sigma/dt} &= \sum_j \tilde{\chi_j}^2 \\
		\tilde{\chi_j}^2    &=  \sum_i \left(\frac{O(\vec{\beta},s_j,t_i) - N_j E_i}{\sigma_i}  \right)^2,
\end{align}
where $j$ denotes the data set, and $N_j$ which is a normalization which is introduced allowing the data to float. $N_j$ needs to be calculated for each data set, and is determined by finding the minimum of $\tilde{\chi_j}^2$ as a function of $N_j$. To calculate this the above is expanded,
\begin{align}
		\tilde{\chi_j}^2 &= A N_j^2 - B N_j + C, \label{eq:dsg_chi}\\
		A                &= \sum_i \left( \frac{E_i}{\sigma_i} \right)^2, \\
		B				 &= \sum_i 2\frac{O_i E_i}{\sigma^2}, \\ 
		C				 &= \sum_i \left( \frac{O_i}{\sigma_i} \right)^2.		
\end{align}
Taking the derivative with respect to $N_j$ and setting equal to zero,
\begin{equation}\label{eq:N_j}
		N_j = \frac{B}{2 A}
\end{equation}
Ideally if all the data has a consistent normalization then this value will be $1$, however in practice this is not the case. When fitting $N_j$ is calculated for each iteration of the fitting routine and fit to (\ref{eq:dsg_chi}), however $N_j$ was only allowed to vary by plus or minus 15\%. The normalization values are presented in Fig. \ref{fig:norm_floats}. 

In order to perform the fit, programs were written independently in Fortran and Python. The Fortran code used the amoeba minimization subroutine from numerical recipes \cite{Press:2007:NRE:1403886}. The Python code utilized the pyminuit module \cite{pyminuit}, which gives low level Minuit \cite{Minuit} functionality to Python functions. 

In order to fit to such a large data set a hierarchy was developed in the fitting procedure. This is also represented in the order of the observables in Table \ref{ta:chi^2} . The total cross sections were fit first as these are calculated at $t = 0$, therefore fewer parameters are required, and this is well within the Regge limit. 
Next the high energy proton-proton data was fit, which is available for differential cross sections and polarization. 
Again this is well within the Regge limit, and enables one to have confidence when extrapolating results. 
Then fitting was performed for the lower energy differential cross sections since all other observables require us to divide by these. 
Finally was the polarization, and then the double polarization observables. 

Since the differential cross sections, polarization, and double polarization variables span the entire angular range the entire data sets were not introduced all at once. Instead the data was introduced at the endpoints and then the angular region was slowly increased from both ends. This was necessary since there was no knowledge of how many Regge exchanges would be required to fit all the data. As the amount of data was increased to the fit additional Regge exchanges would be added as necessary in order to get an acceptable solution. The procedure was iterative and introducing new Regge exchanges was based on trial and error.

The trajectories labeled $X_i$ in Table \ref{ta:params_pol} are effective trajectories in that they do not correspond to the meson spectrum. These trajectories were introduced to obtain a fit to the data primarily at low energies and over all angles. 
Trajectories $X_1$ and $X_2$ are introduced to reproduce the diffractive structure in the proton-proton differential cross sections at large values of $s$.
Several of the remaining effective trajectories have relatively large negative values of $\alpha_0$. 
These are required to provide more rapid variation of observables at small values of $s$.
The need for this is most obvious for the total cross sections, which reach a peak and then decrease in value with decreasing $s$.
Although less obvious, these are also important for other observables at low $s$.

The parameters of the fit are generally quite tightly constrained for all trajectories. 
However, both $\beta_1$ and $\alpha_1$ contribute to the $t$ dependence of the amplitudes; 
$\beta_1$ through the exponential factor, and $\alpha_1$ by changing the exponent of $\left(-1+\frac{2s}{4m^2-t}\right)$.
If $\beta_1$ is large, it controls the $t$ dependence and $\alpha_1$ is not tightly constrained. 
%Examination of the uncorrelated errors indicates that,
%with the exception of the situation mentioned above, errors in the fitting parameters are less than one percent. 
The model would benefit greatly with a full error analysis, and this is planned as a future work.
\begin{table}[htbp] \centering
\caption{$\chi^2$ values per data, number of data points, and weights used for each observable. The fit has a total of 136 parameters.}
\begin{tabular}{llccccc}
\hline \hline
Observable 							  &  	 & $N$	& $\chi^2/N$	&   weight	\\ 	\hline
\multirow{2}{*}{$\sigma$}             & $pp$ & 181 	& 0.9 			&	50.0	\\ 
                                      & $pn$ & 69 	& 0.2 			&	50.0	\\ 	\hline
$\frac{d\sigma}{dt}(s>20(GeV^2))$     & $pp$ & 1635 & 1.7 			&	1.0		\\ 	\hline
\multirow{2}{*}{$\frac{d\sigma}{dt}$} & $pp$ & 3481 & 2.4 			&	4.0		\\ 
				     				  & $pn$ & 745 	& 2.6 			&	8.0		\\ 	\hline
\multirow{2}{*}{$P$ $(A_N)$}          & $pp$ & 3410 & 2.5 			&	3.0		\\ 
                                      & $pn$ & 508 	& 3.1 			&	12		\\ 	\hline
\multirow{2}{*}{$A_{YY}$}             & $pp$ & 1587 & 4.6 			&	2.0		\\ 
                                      & $pn$ & 117 	& 2.6 			&	30.0	\\  \hline
\multirow{2}{*}{$A_{ZX}$}             & $pp$ & 568 	& 5.6 			&	4.0		\\ 
                                      & $pn$ & 81 	& 1.2 			&	30.0	\\  \hline
\multirow{2}{*}{$A_{ZZ}$}             & $pp$ & 608 	& 5.8 			&	3.0		\\ 
                                      & $pn$ & 89 	& 2.6 			&	30.0	\\  \hline
$A_{XX}$                              & $pp$ & 276 	& 9.5 			&	4.0		\\  \hline
\multirow{2}{*}{$D$}                  & $pp$ & 188 	& 4.9 			&	20.0	\\ 
                                      & $pn$ & 37   & 3.0 			&	30.0	\\  \hline
\multirow{2}{*}{$D_{T}$}              & $pp$ & 281 	& 6.7 			&	3.0		\\ 
			              			  & $pn$ & 8 	& 0.4 			&	30.0	\\  \hline
total 								  &  	 & 13869 & 3.1 \\  %\slashbox{6111}
\hline \hline
\end{tabular}
\label{ta:chi^2}
\end{table}
\newpage
\section{Fit Results}
In this section, plots of the fit solution to the NN system are presented. All data that were fit to are displayed in order to show a qualitative result of the fit to all data, and to visualize that some of the data sets can exhibit an apparent lack of consistency with one another. This is most likely because the data sets were collected over a large time frame at a number of different facilities, therefore the quality of the data varies. 
The solid lines through the data represent the Regge model. Electromagnetic effects are turned off for all observables except the high energy proton-proton data where data are available in the Coulomb region. The $s$ values on all the plots are in units of $\mathrm{GeV}^2$. This was suppressed in the plots due to space constraints. All the differential cross section plots are shown with the normalization factors, $N_j$, given in (\ref{eq:N_j}).

\begin{figure}
    \includegraphics[width=15cm]{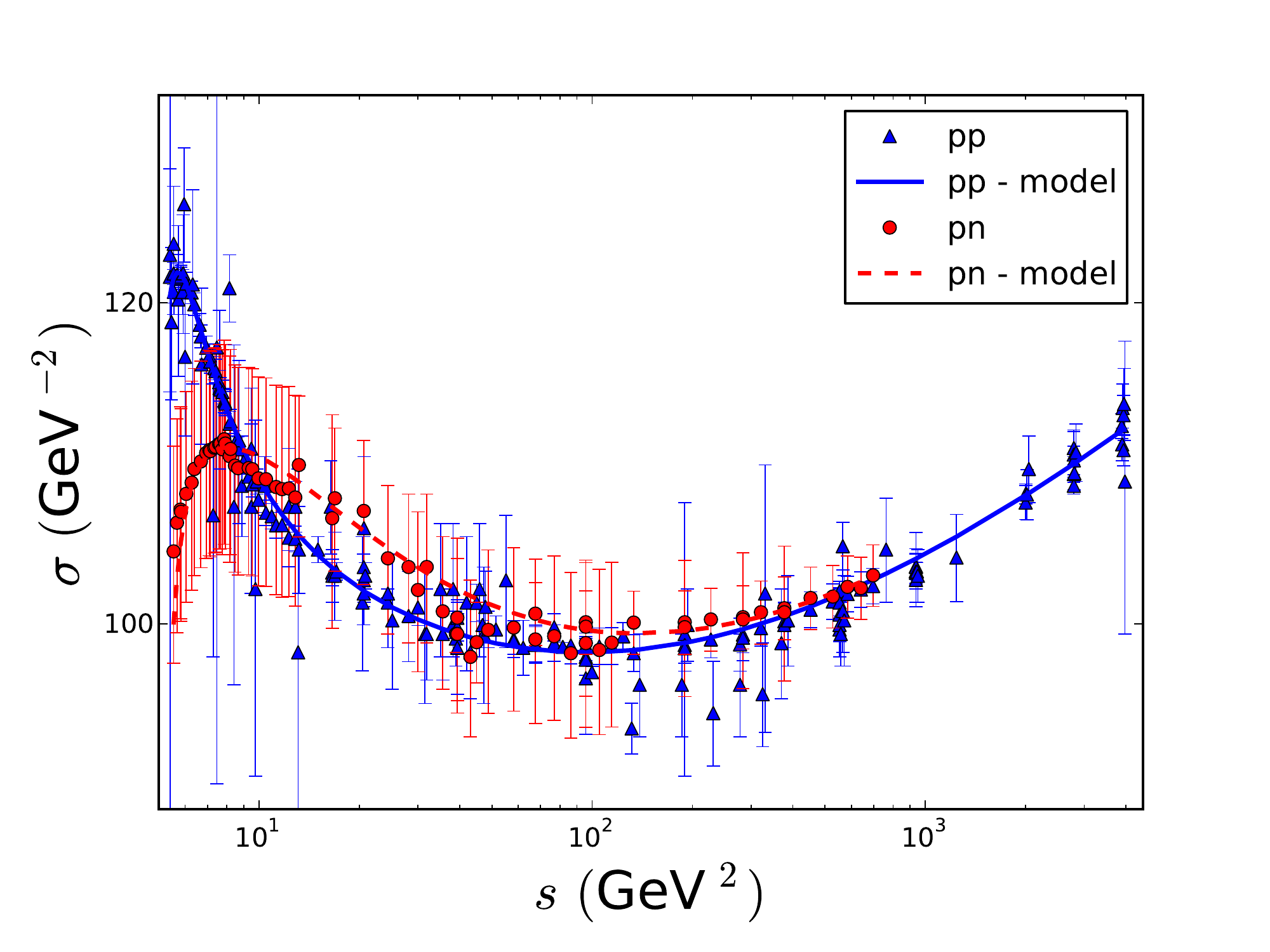}  %\textwidth
    \caption{Total cross sections for proton-proton and proton-neutron as a function of Mandelstam $s$.}
    \label{fig:total}
\end{figure}

The total cross sections for both proton-proton and proton-neutron are presented in Fig. \ref{fig:total}. 
As these are calculated at $t=0$, the Regge approximation works extremely well. 

\begin{figure}
    \includegraphics[width=15cm]{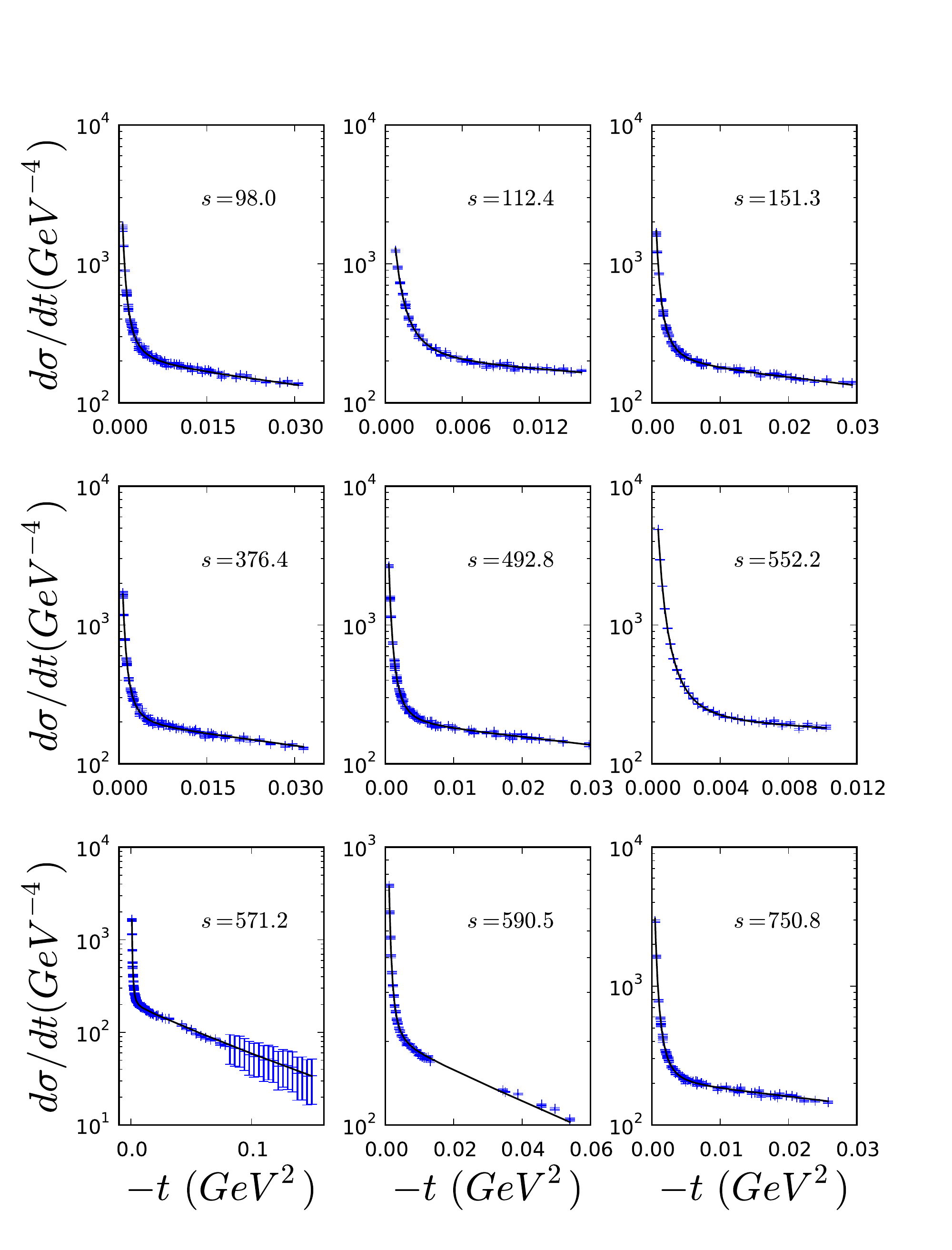} 
    \caption{High energy results for proton-proton differential cross sections as a function of $-t$ through the Coulomb region.}
    \label{fig:Highpp_coul1}
\end{figure}

\begin{figure}
    \includegraphics[width=15cm]{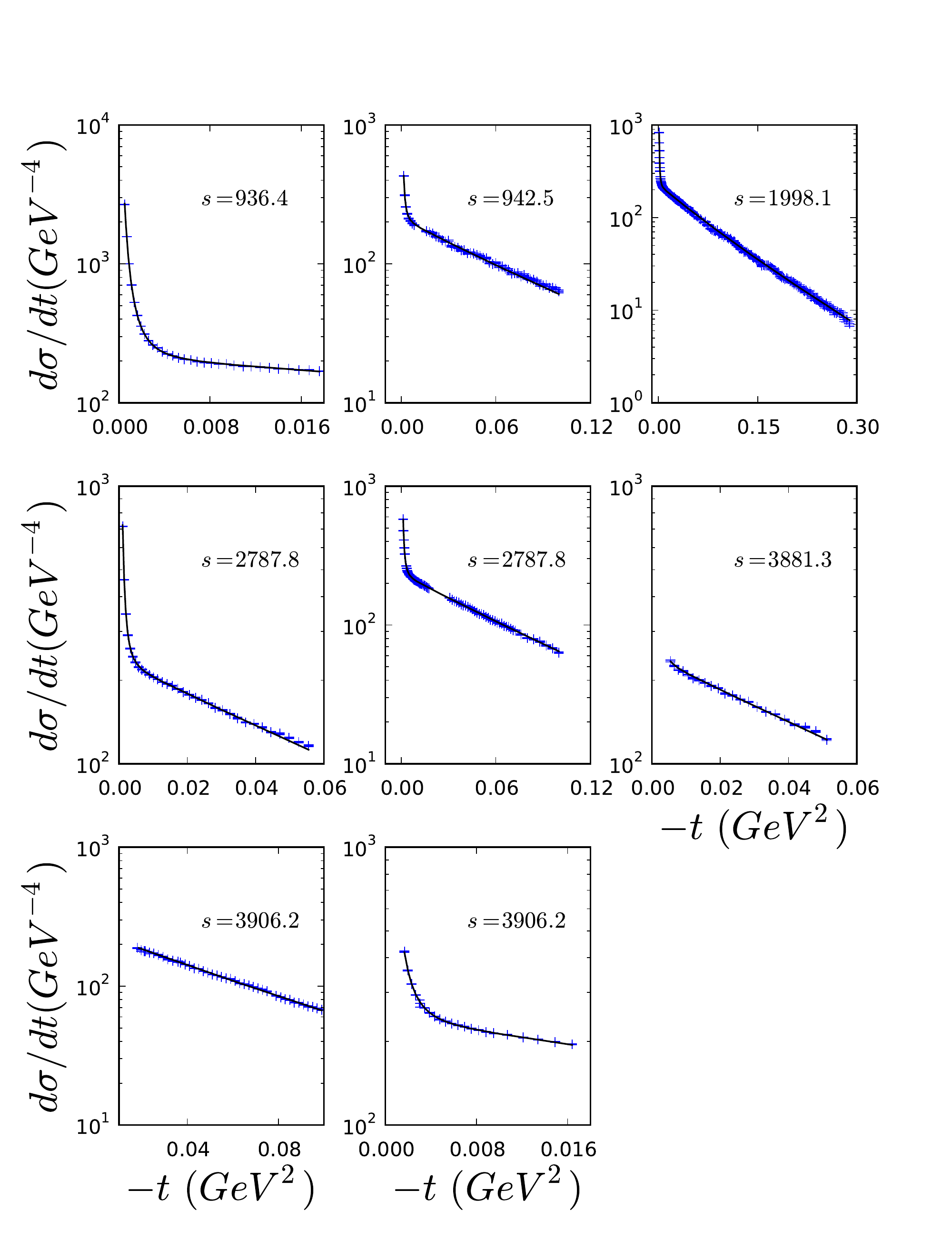} 
    \caption{High energy results for proton-proton differential cross sections as a function of $-t$ through the Coulomb region.}
    \label{fig:Highpp_coul2}
\end{figure}

\begin{figure}
    \includegraphics[width=15cm]{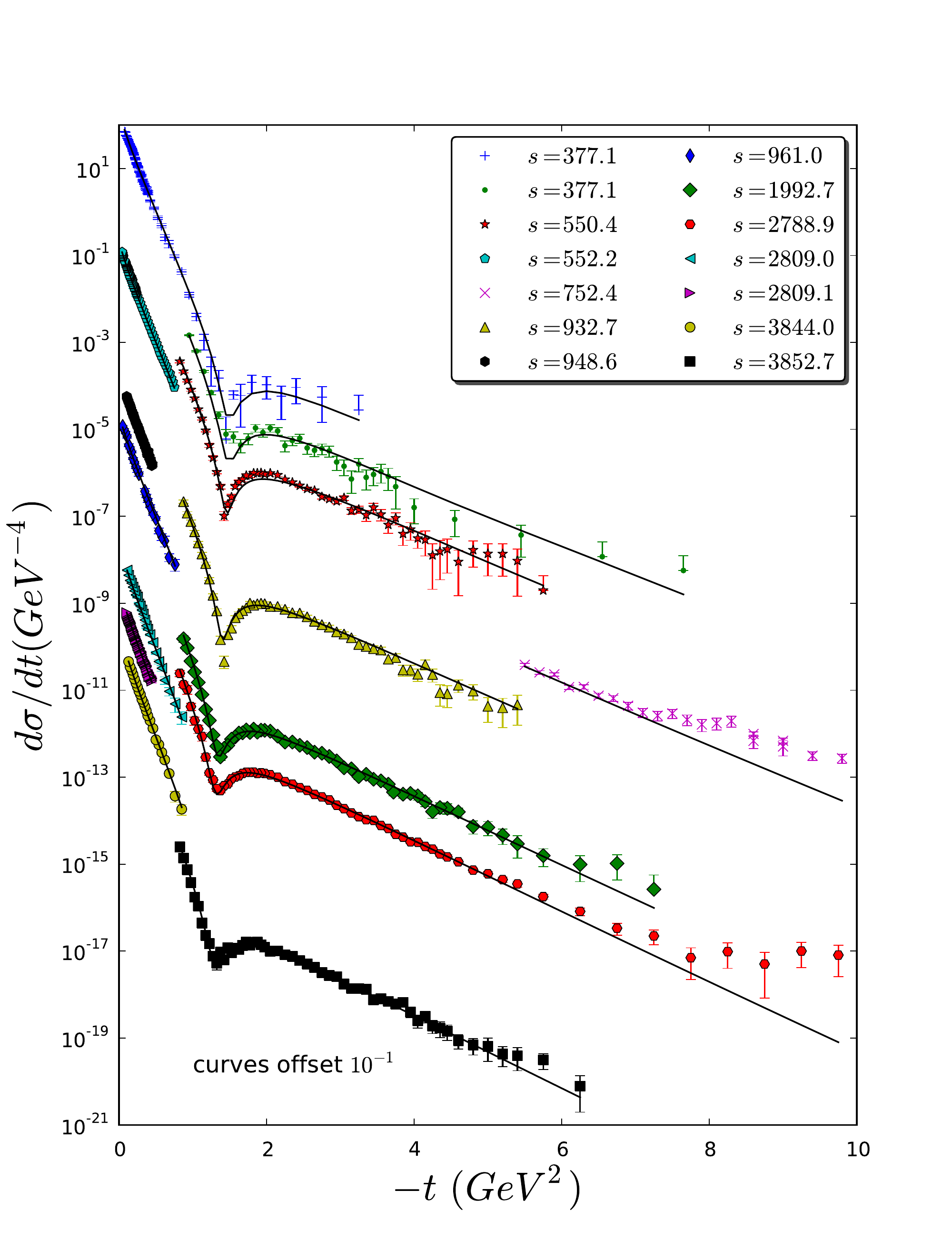} 
    \caption{High energy results for proton-proton differential cross sections as a function of $-t$ through the linear and dip regions. 
    This data was fit through $-t\leq 8 (\mathrm{GeV}^2)$. For error bars going negative only the upper bound is plotted.}
    \label{fig:Highpp_lindip}
\end{figure}

\begin{figure}
    \includegraphics[width=15cm]{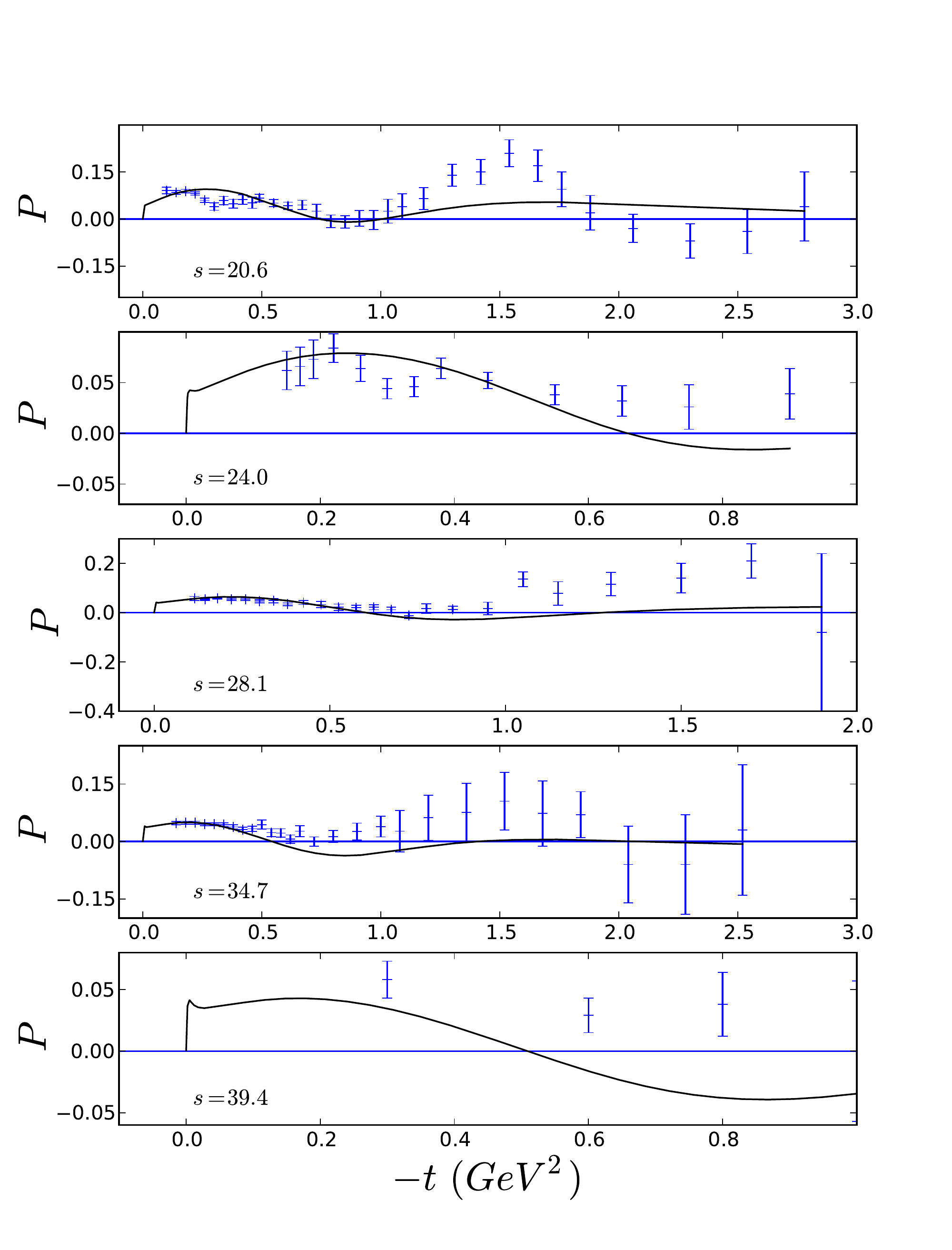} 
    \caption{High energy results for proton-proton polarization as a function of $-t$.}
    \label{fig:HighP1}
\end{figure}

\begin{figure}
    \includegraphics[width=15cm]{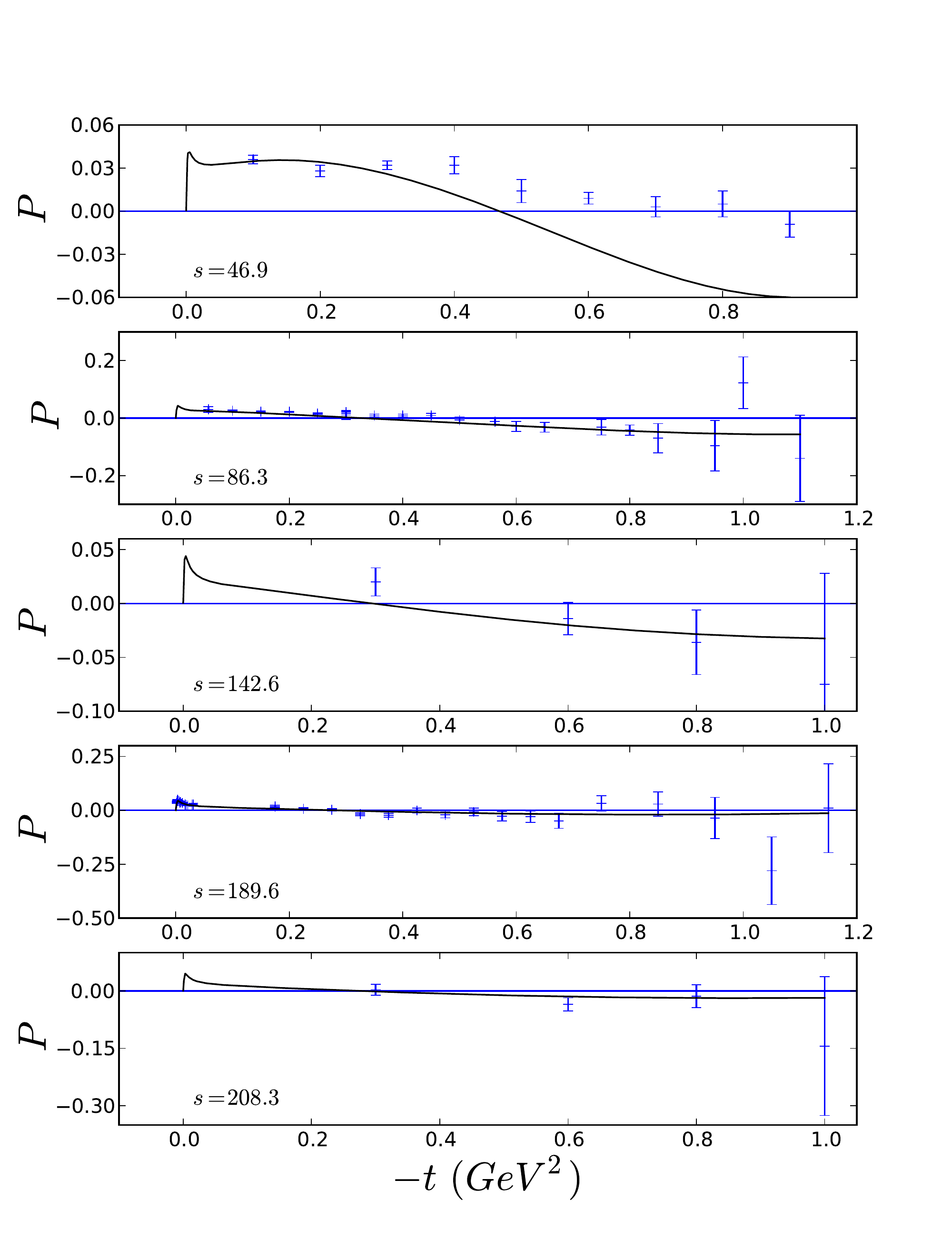} 
    \caption{High energy results for proton-proton polarization as a function of $-t$.}
    \label{fig:HighP2}
\end{figure}

\begin{figure}
\includegraphics[width=15cm]{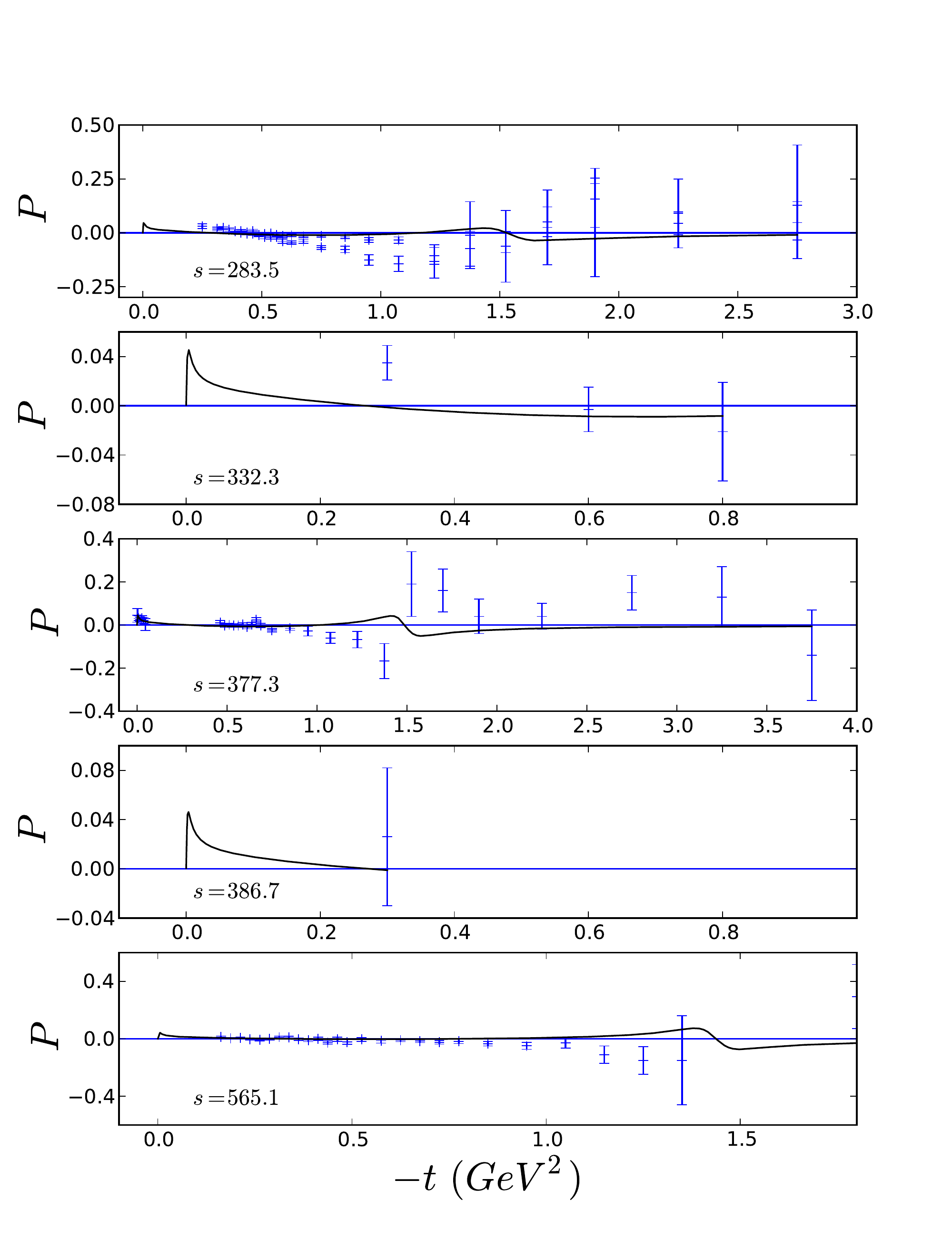} 
\caption{High energy results for proton-proton polarization as a function of $-t$.}
\label{fig:HighP3}    
\end{figure}

\begin{figure}
    \includegraphics[width=15cm]{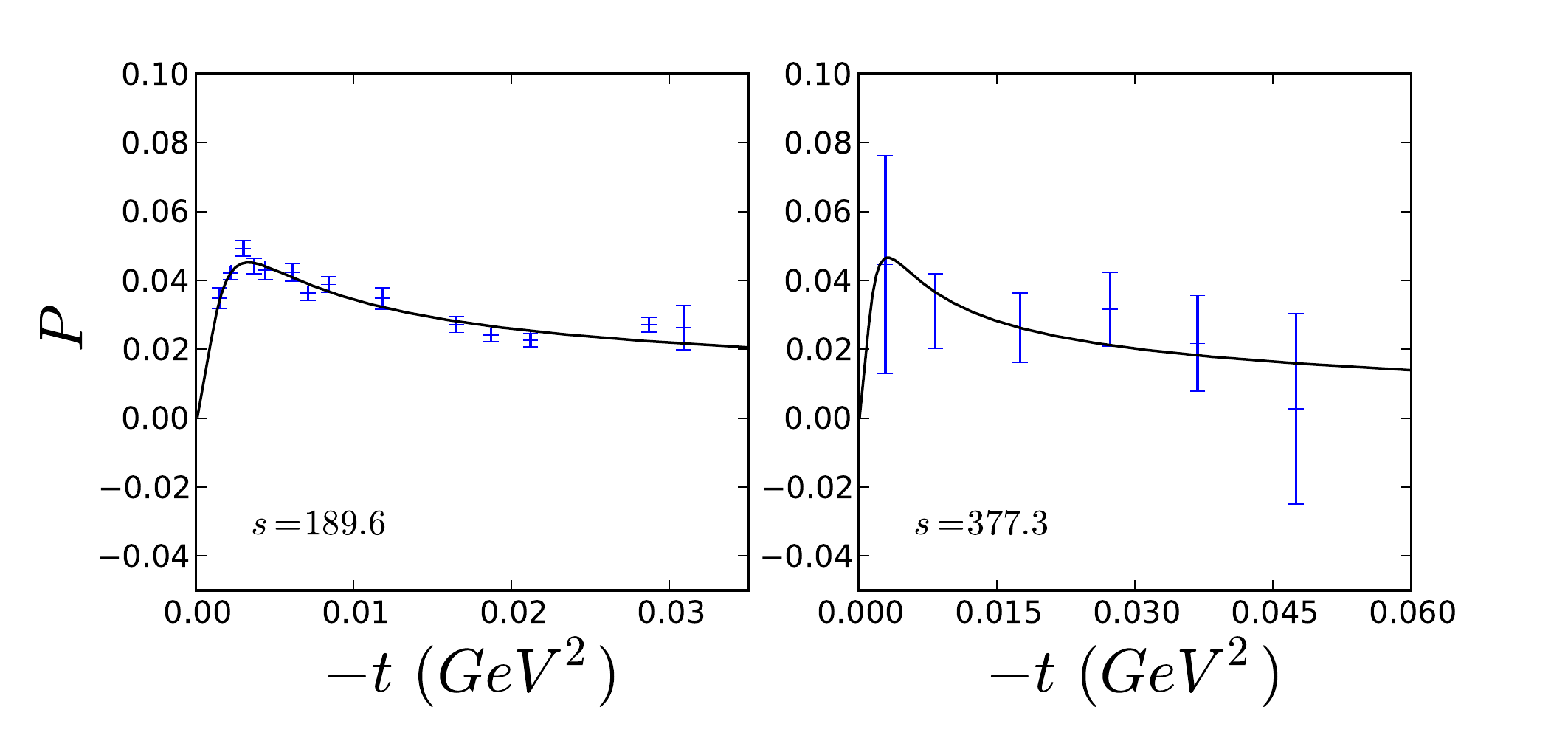} 
    \caption{High energy results for proton-proton polarization as a function of $-t$ zoomed to show features in the Coulomb region.}
    \label{fig:HighP_coul}
\end{figure}

In order to constrain the model at large $s$, fits were performed to high energy proton-proton data. 
Figures. \ref{fig:Highpp_coul1} - \ref{fig:Highpp_coul2} show the differential cross section data through the Coulomb region. Figure \ref{fig:Highpp_lindip} show the high energy proton-proton differential cross section through the linear-dip region. The fit for the high energy differential cross section was performed out to $-t = 8$ GeV$^2$, and excellent agreement is noted through the fit region, although less so at very high $|t|$ as expected from a Regge model.
There is also high energy polarization data available. This is presented in Figs. \ref{fig:HighP1} - \ref{fig:HighP3} . Figure \ref{fig:HighP_coul} shows zoomed in high energy polarization data in the Coulomb region, which was used along with the low $t$ differential cross section data to fit the phases of the electromagnetic interaction. 
These results illustrate the ability of the Regge model to scale to higher energies.

\begin{figure}
    \includegraphics[width=15cm]{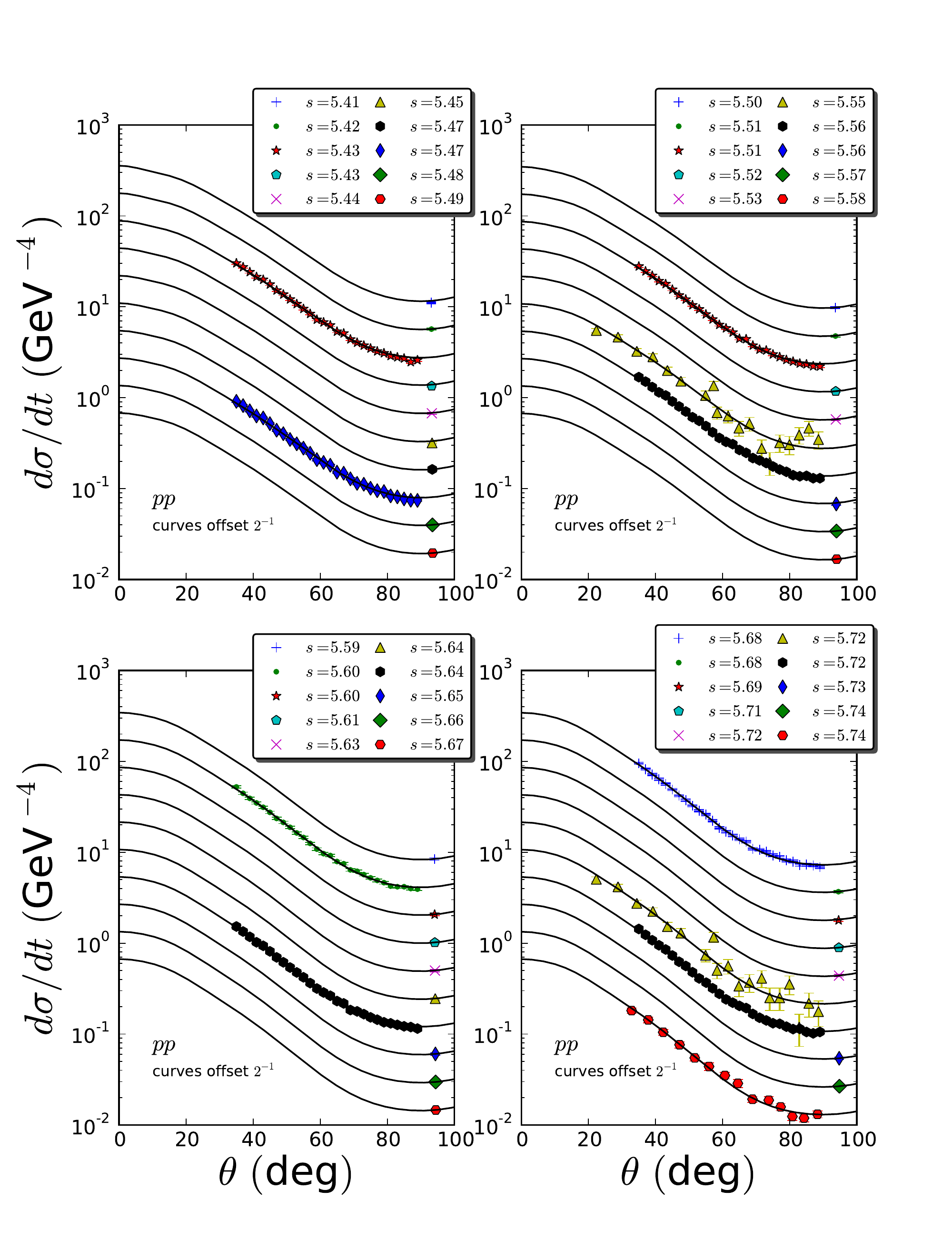} 
    \caption{ Proton-proton differential cross sections as a function of center of mass angle $\theta$. Each data set is offset by a factor of two.}
    \label{fig:DSGppL1}
\end{figure}

\begin{figure}
    \includegraphics[width=15cm]{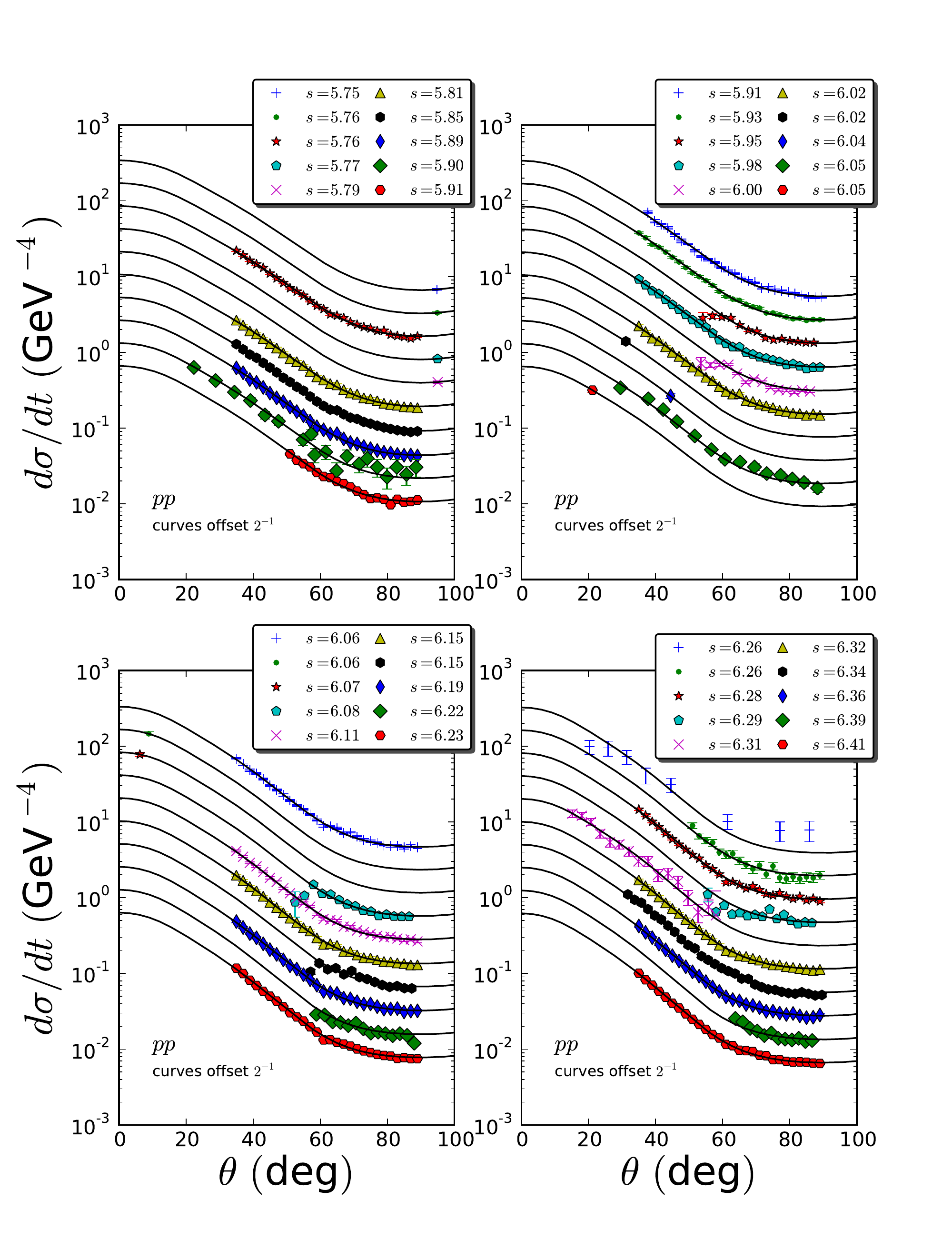} 
    \caption{Proton-proton differential cross sections as a function of center of mass angle $\theta$. Each data set is offset by a factor of two. (cont.)}
    \label{fig:DSGppL2}
\end{figure}

\begin{figure}
    \includegraphics[width=15cm]{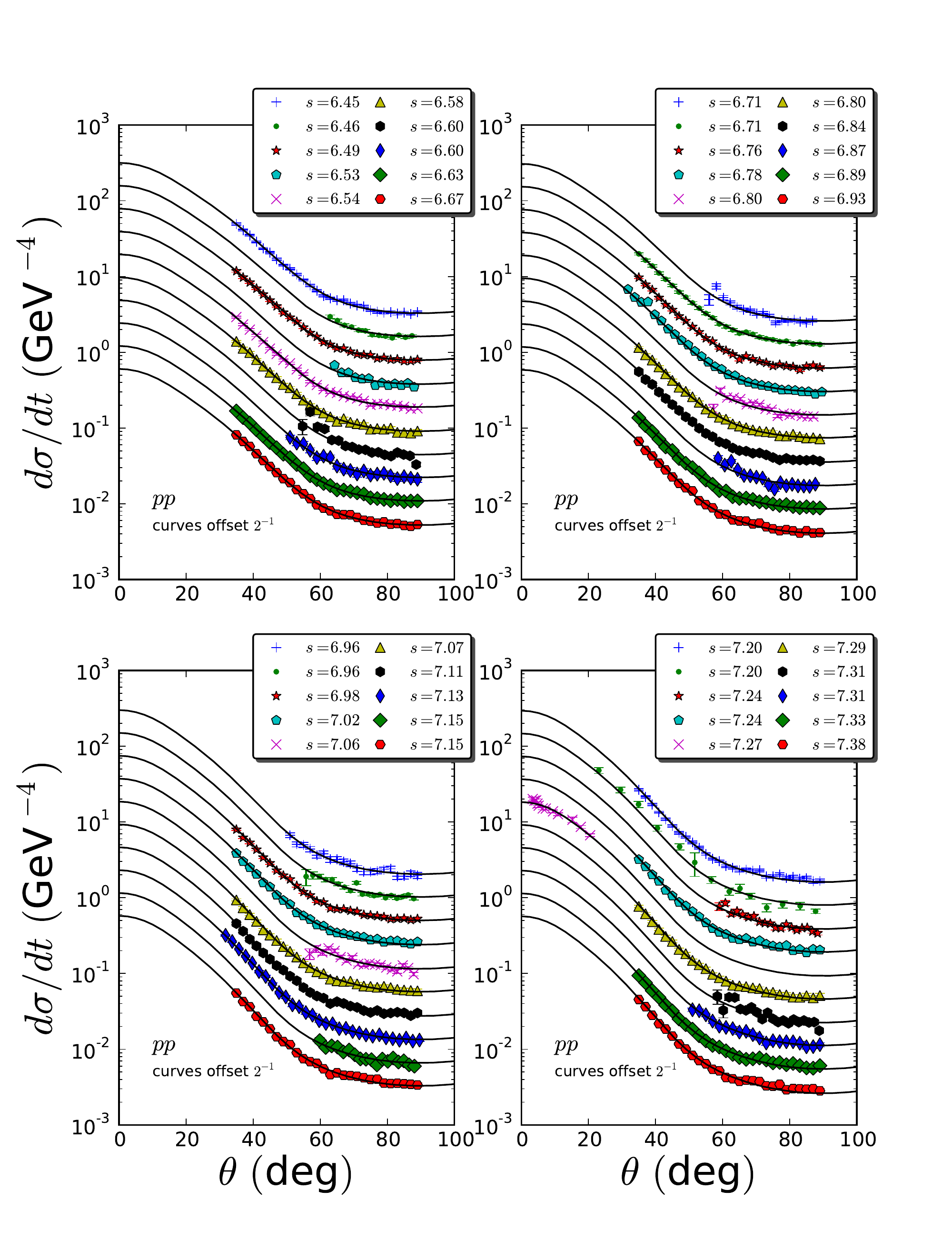} 
    \caption{Proton-proton differential cross sections as a function of center of mass angle $\theta$. Each data set is offset by a factor of two. (cont.)}
    \label{fig:DSGppL3}
\end{figure}

\begin{figure}
    \includegraphics[width=15cm]{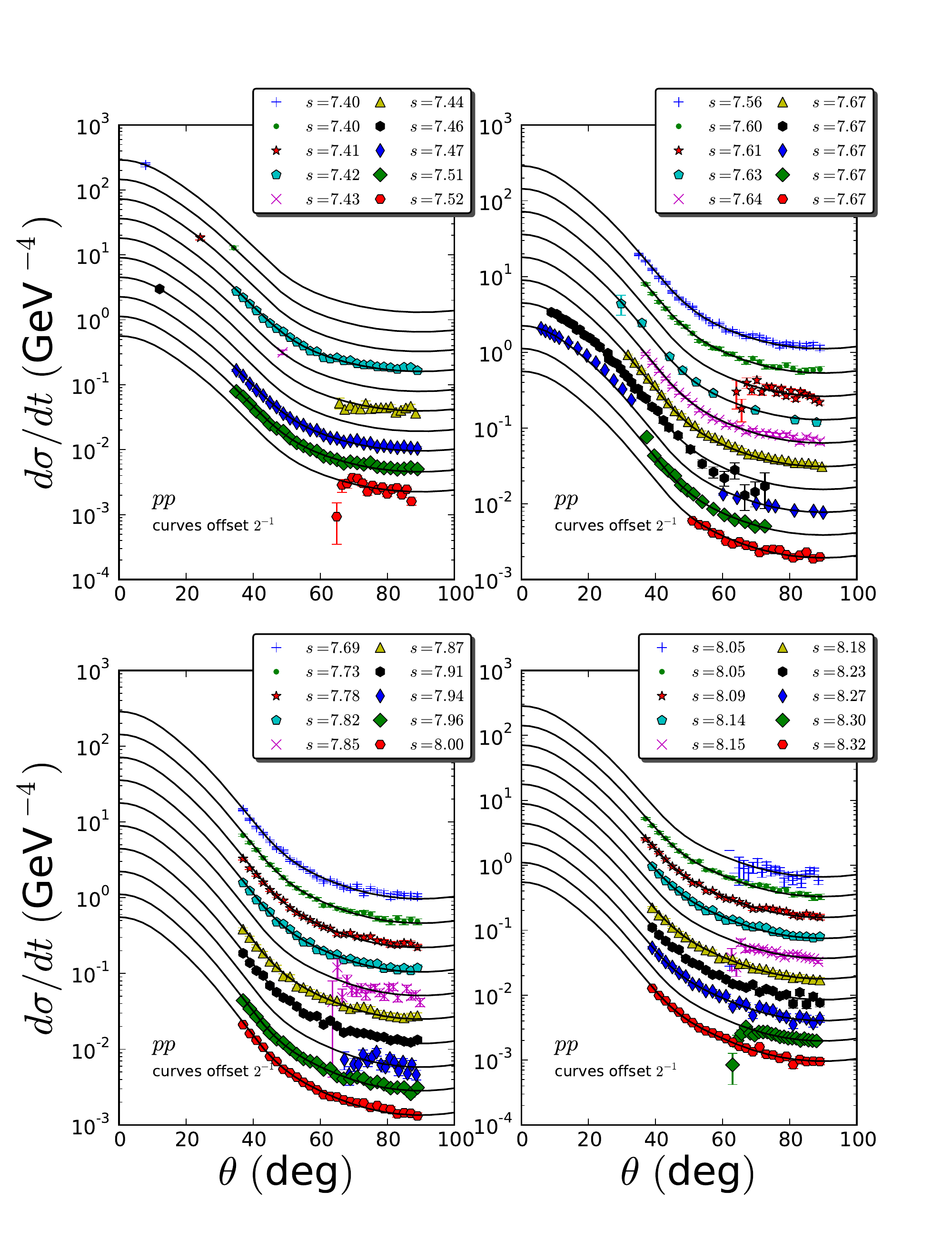} 
    \caption{Proton-proton differential cross sections as a function of center of mass angle $\theta$. Each data set is offset by a factor of two. (cont.)}
    \label{fig:DSGppL4}
\end{figure}

\begin{figure} 
    \includegraphics[width=15cm]{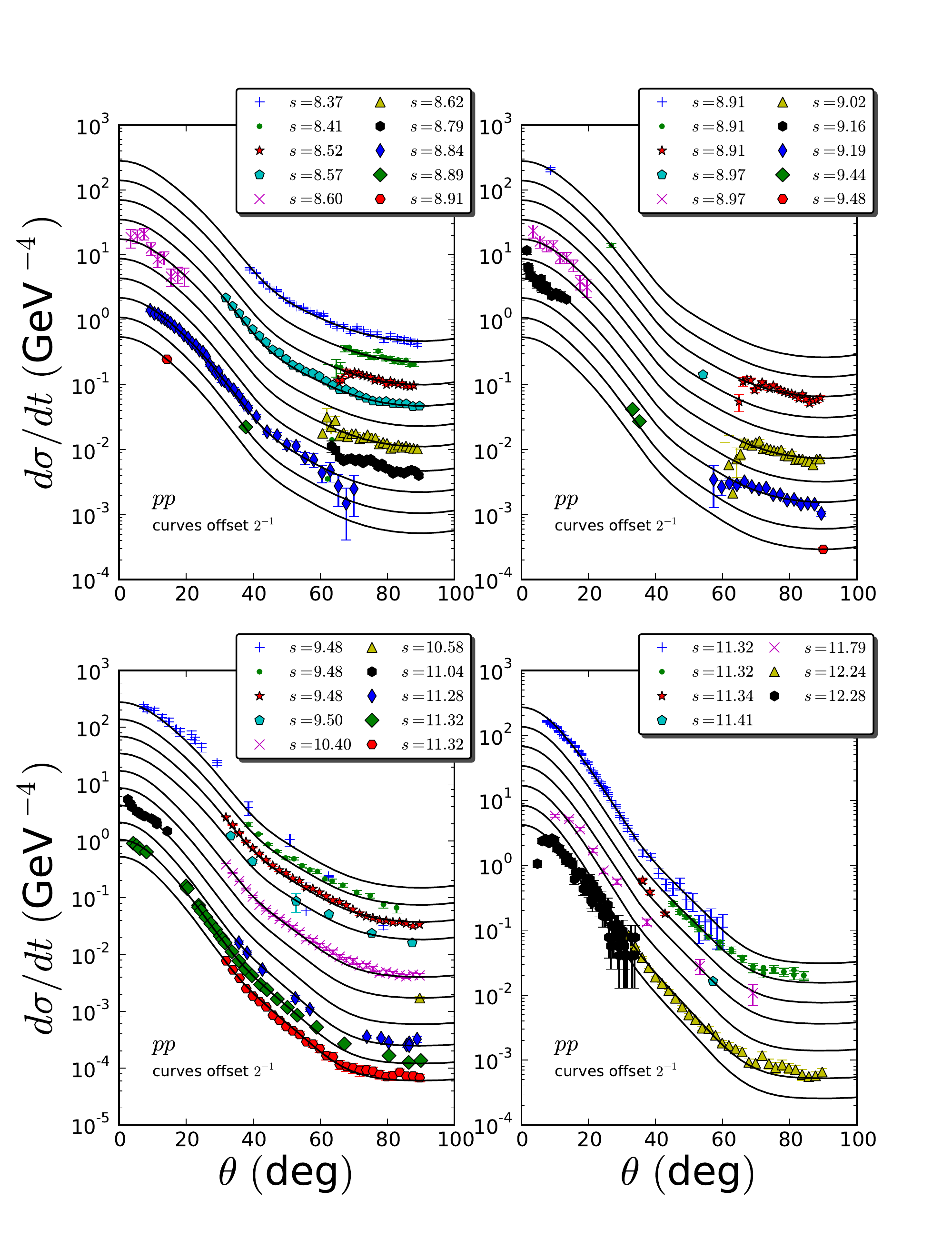} 
    \caption{Proton-proton differential cross sections as a function of center of mass angle $\theta$. Each data set is offset by a factor of two. (cont.)}
    \label{fig:DSGppL5}
\end{figure}

\begin{figure} 
    \includegraphics[width=15cm]{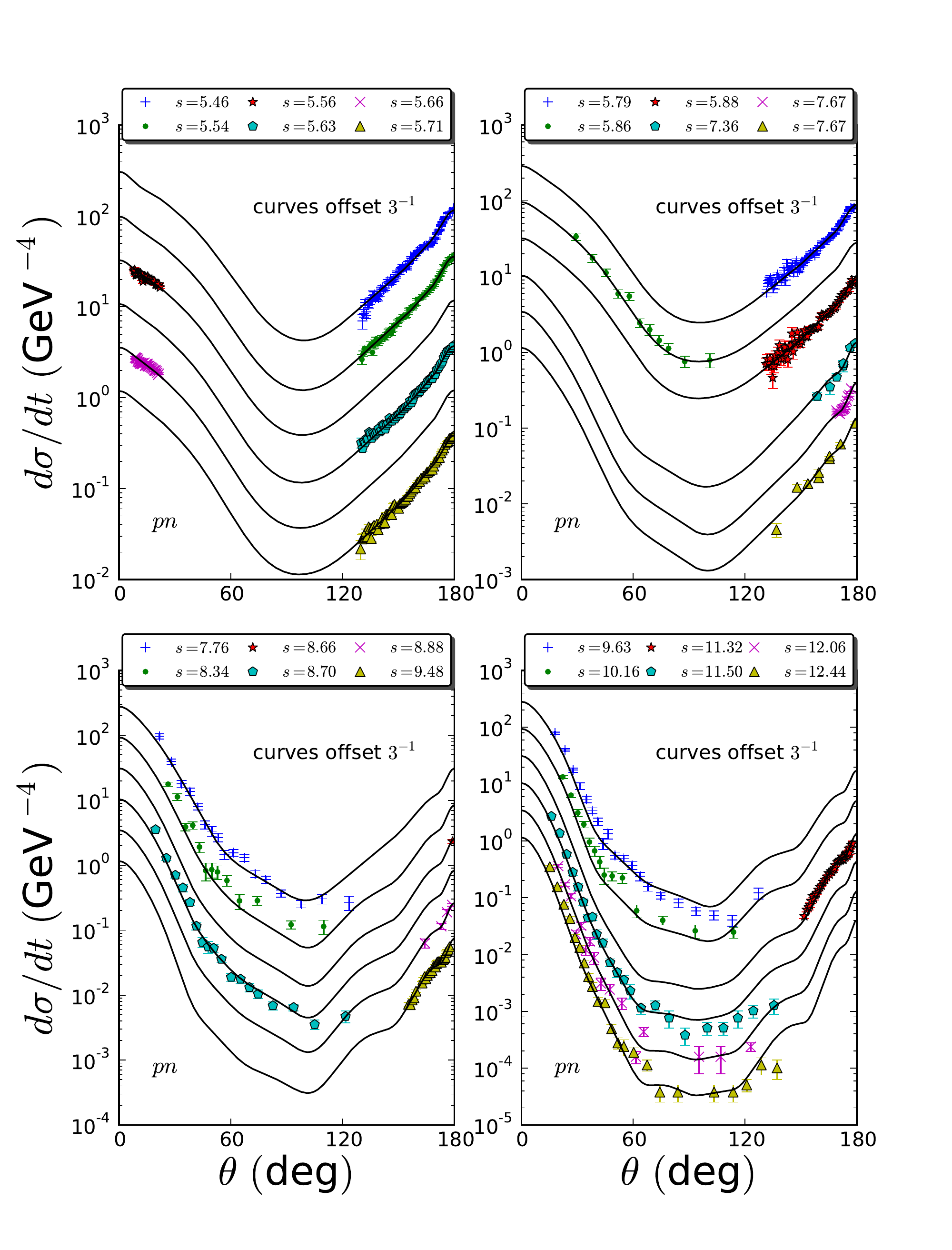} 
    \caption{Proton-neutron differential cross sections as a function of center of mass angle $\theta$. Each data set is offset by a division of three.}
    \label{fig:DSGpn}
\end{figure}

Low energy differential cross sections for proton-proton scattering are shown in Figs. \ref{fig:DSGppL1} - \ref{fig:DSGppL5} and proton-neutron in Fig. \ref{fig:DSGpn}. 
The model works very well, especially considering that the data are described over the entire angular region, and for relatively low $s$, 
well outside of where one would typically expect the Regge approximation to be valid. 

\begin{figure}
    \includegraphics[width=15cm]{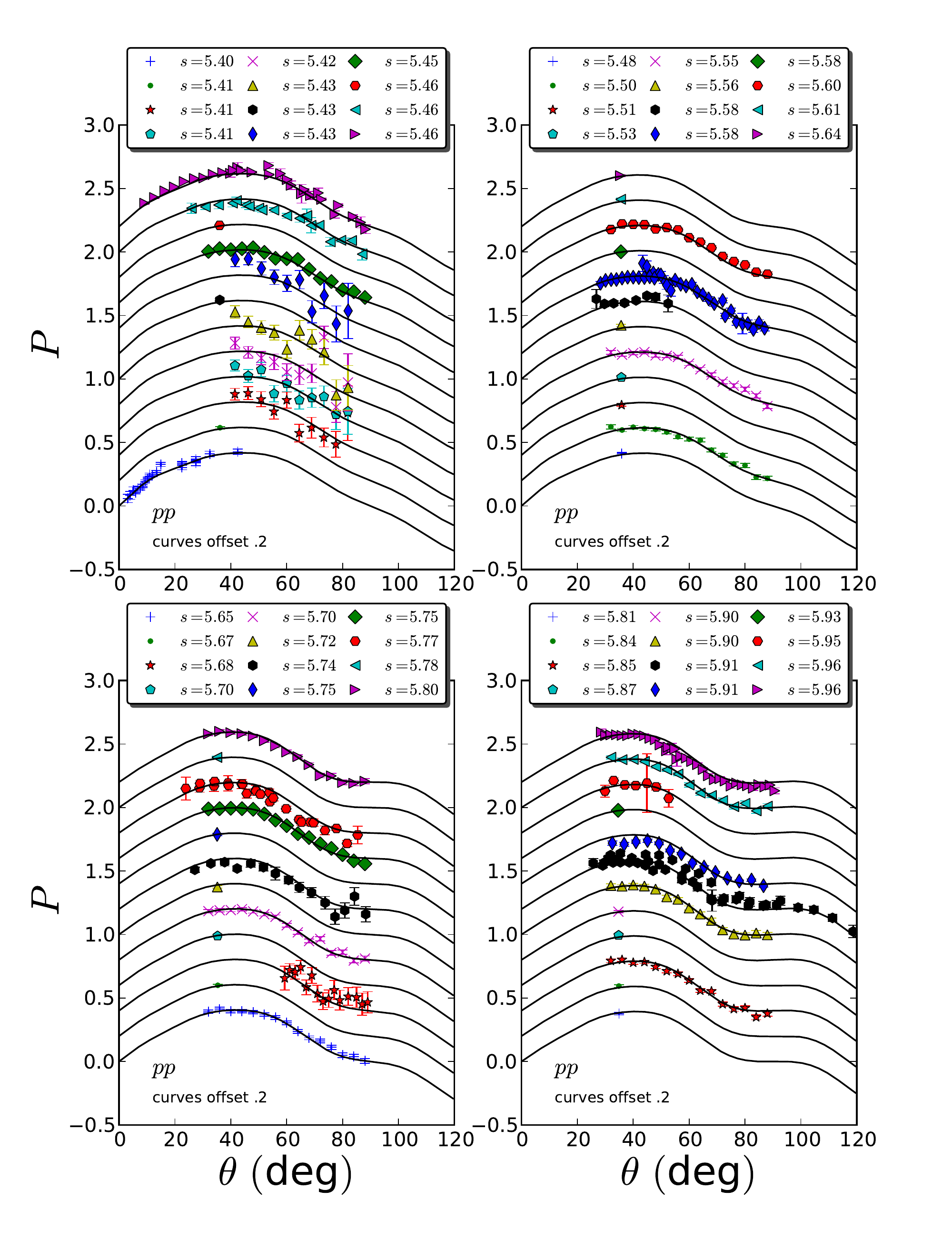} 
    \caption{Polarization for proton-proton as a function of center of mass angle $\theta$.}
    \label{fig:Ppp10}
\end{figure}

\begin{figure}
    \includegraphics[width=15cm]{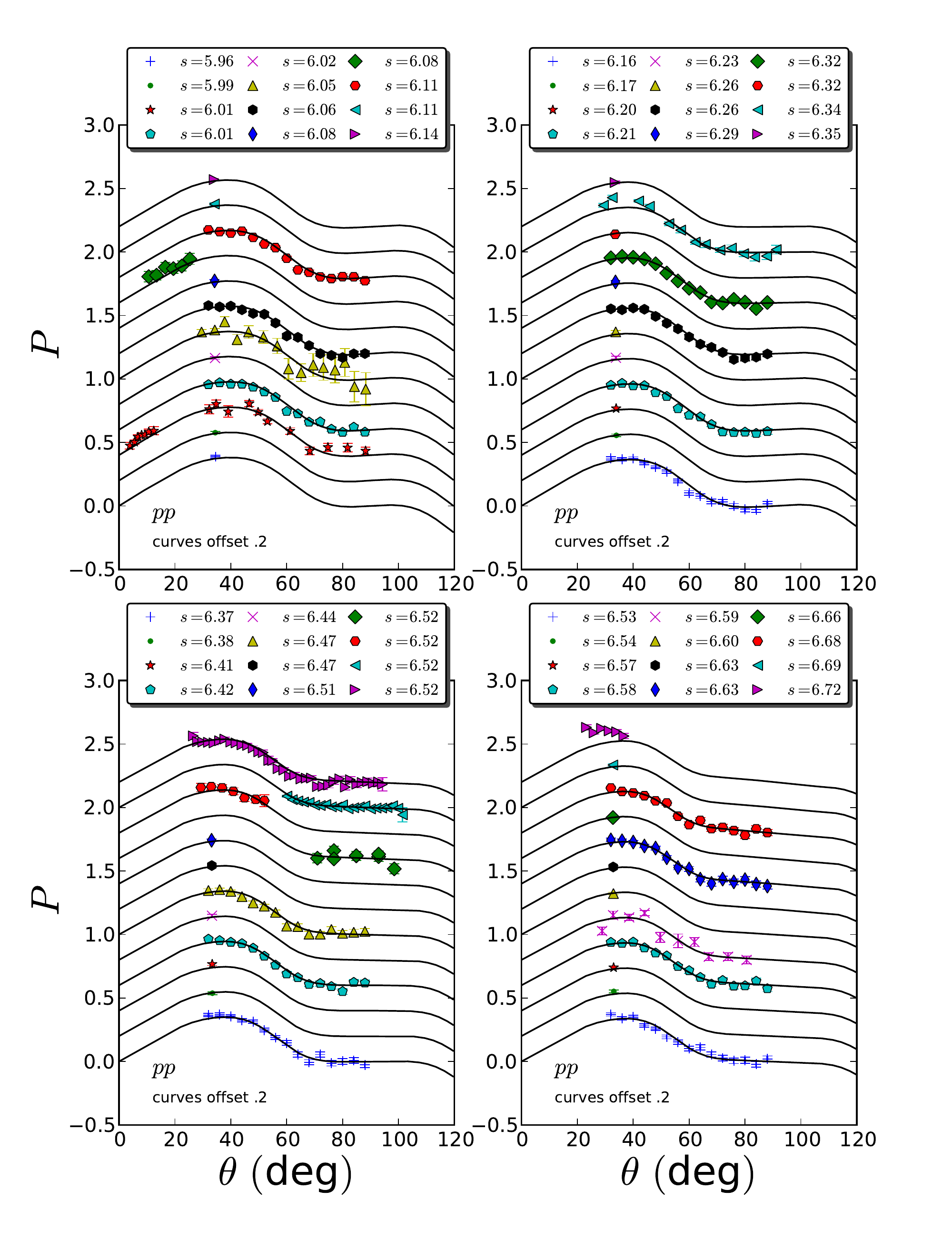} 
    \caption{Polarization for proton-proton as a function of center of mass angle $\theta$.}
    \label{fig:Ppp11}
\end{figure}

\begin{figure}
    \includegraphics[width=15cm]{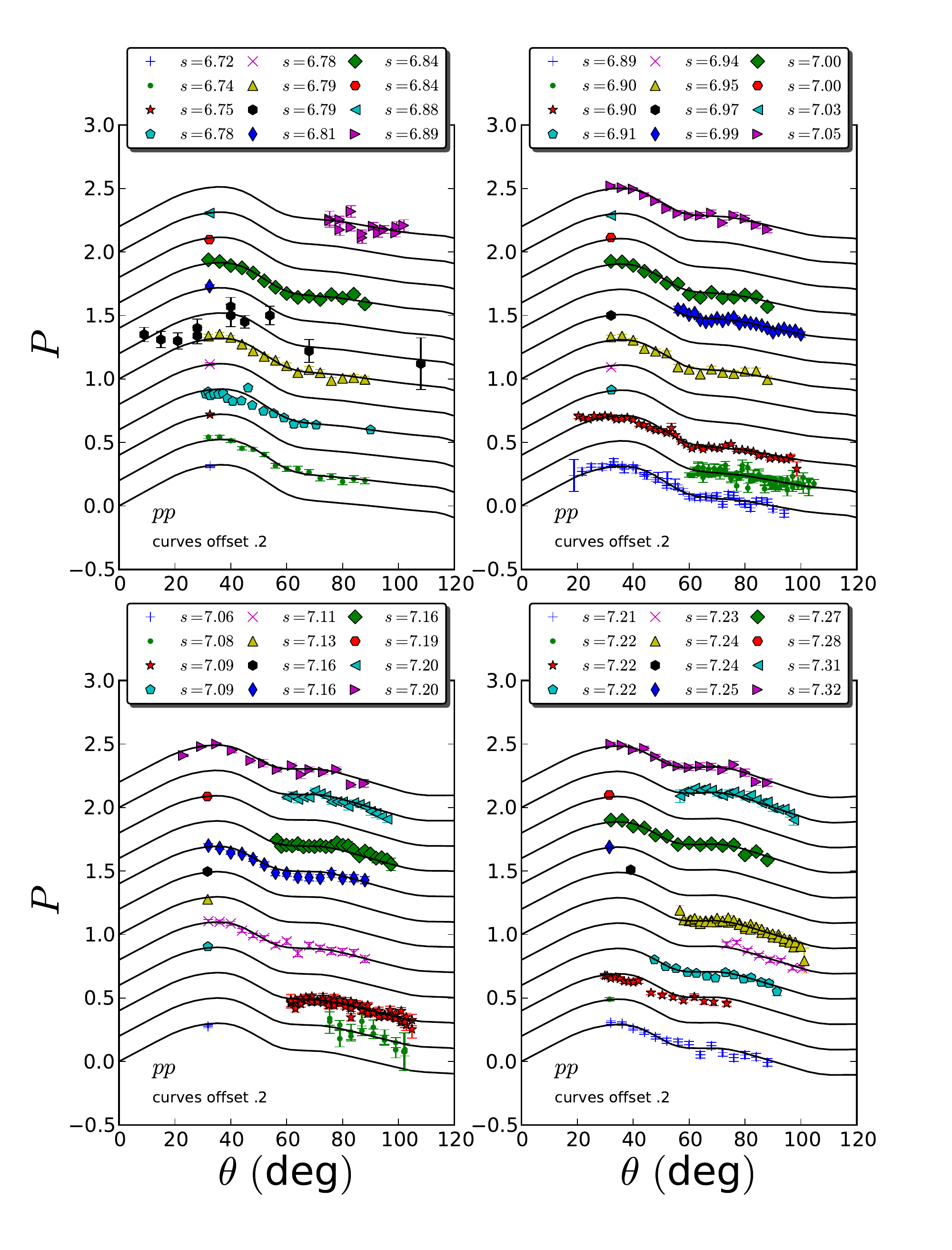} 
    \caption{Polarization for proton-proton as a function of center of mass angle $\theta$.}
    \label{fig:Ppp12}
\end{figure}

\begin{figure}
    \includegraphics[width=15cm]{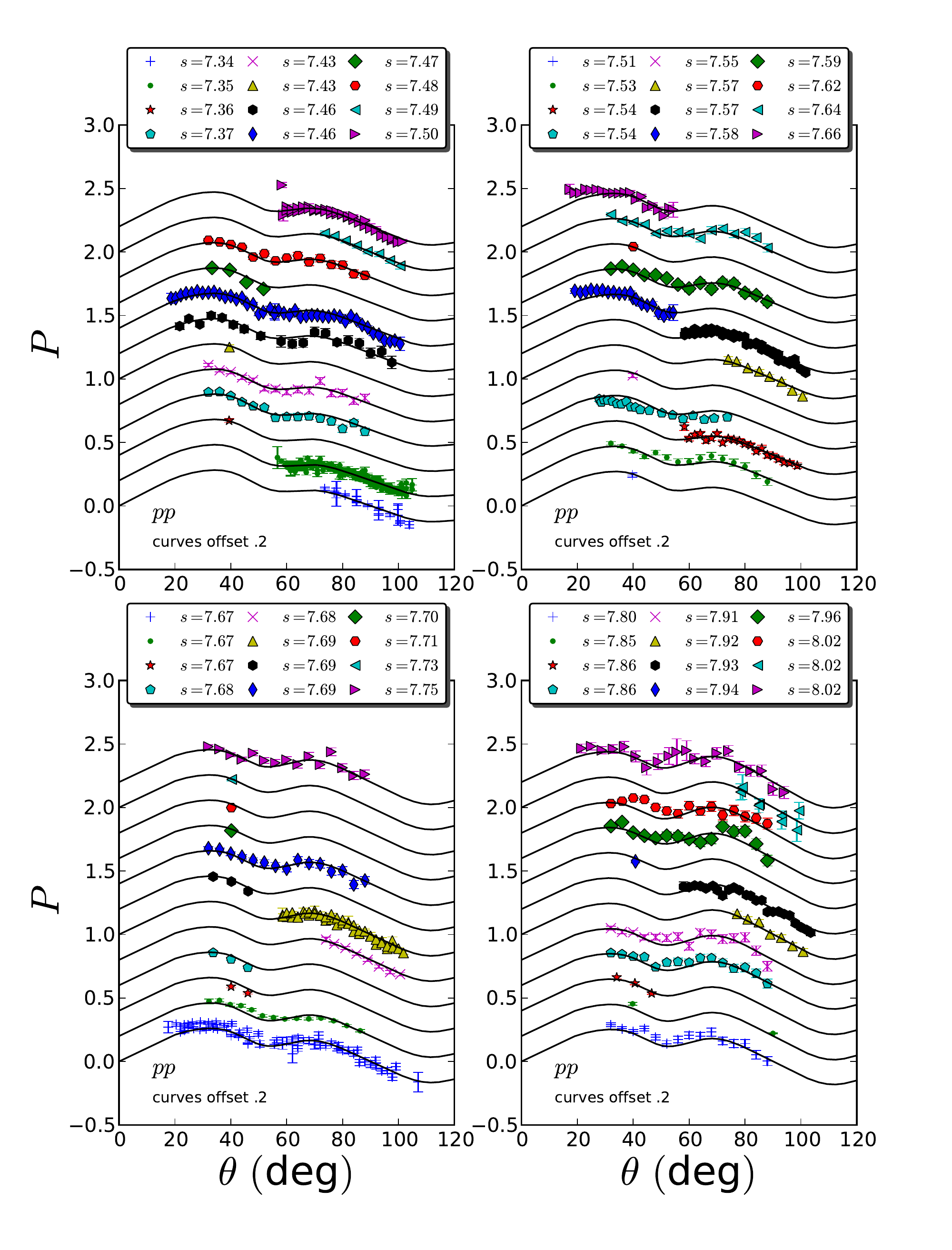} 
    \caption{Polarization for proton-proton as a function of center of mass angle $\theta$.}
    \label{fig:Ppp13}
\end{figure}

\begin{figure}
    \includegraphics[width=15cm]{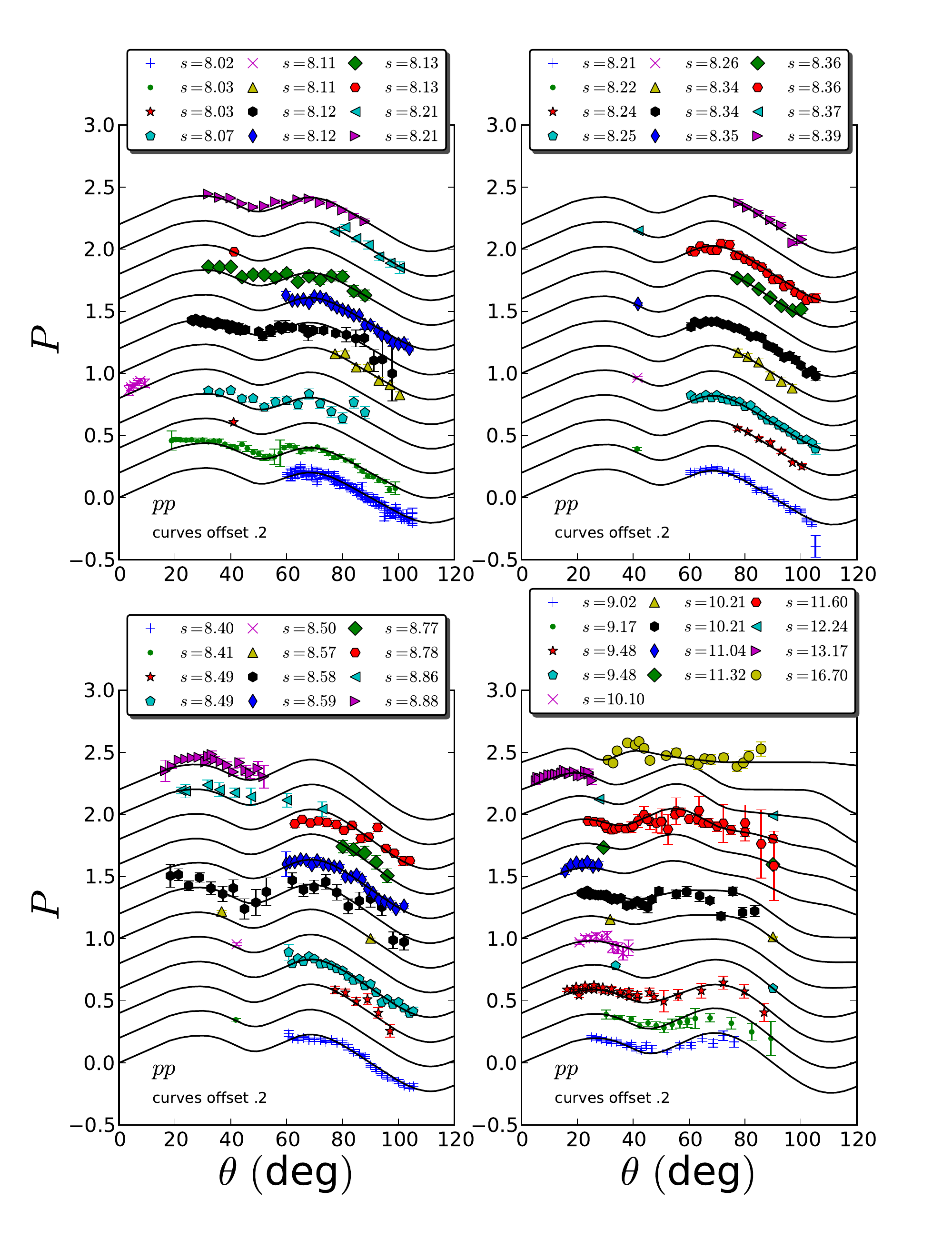} 
    \caption{Polarization for proton-proton as a function of center of mass angle $\theta$.}
    \label{fig:Ppp14}
\end{figure}

\begin{figure}
    \includegraphics[width=15cm]{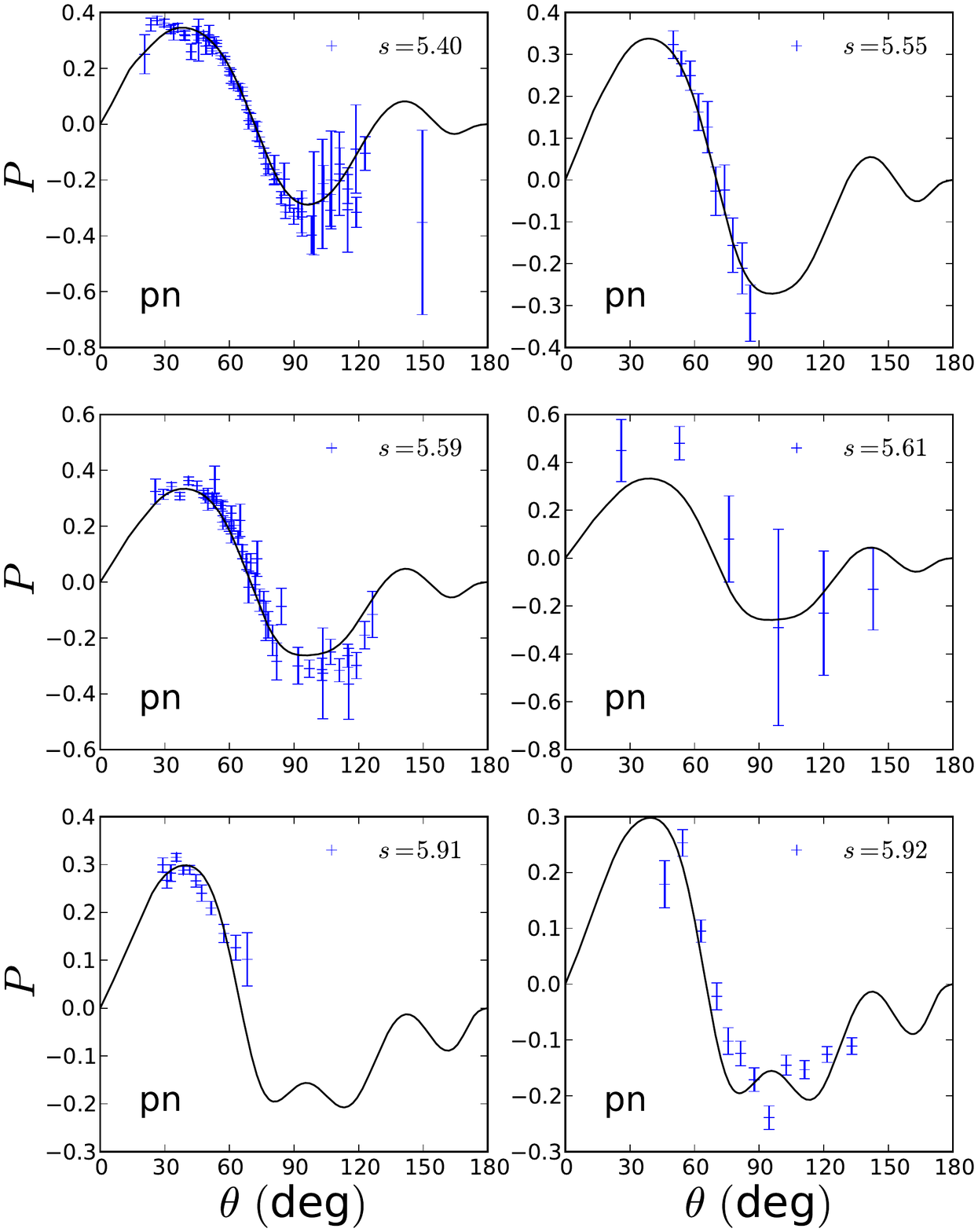} 
    \caption{Polarization for proton-neutron as a function of center of mass angle $\theta$.}
    \label{fig:Ppn1}
\end{figure}

\begin{figure}
    \includegraphics[width=15cm]{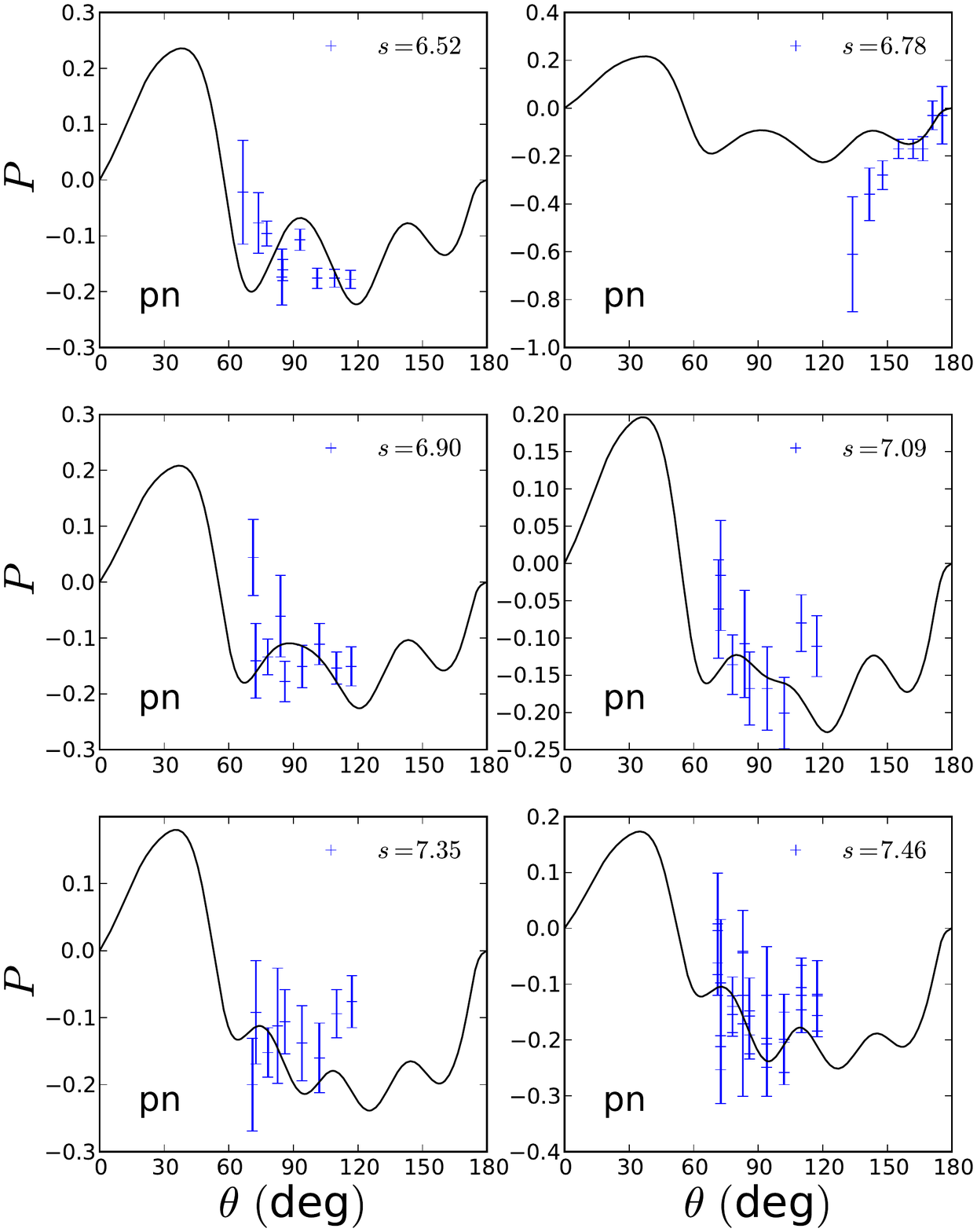} 
    \caption{Polarization for proton-neutron as a function of center of mass angle $\theta$.}
    \label{fig:Ppn2}
\end{figure}

\begin{figure}
    \includegraphics[width=15cm]{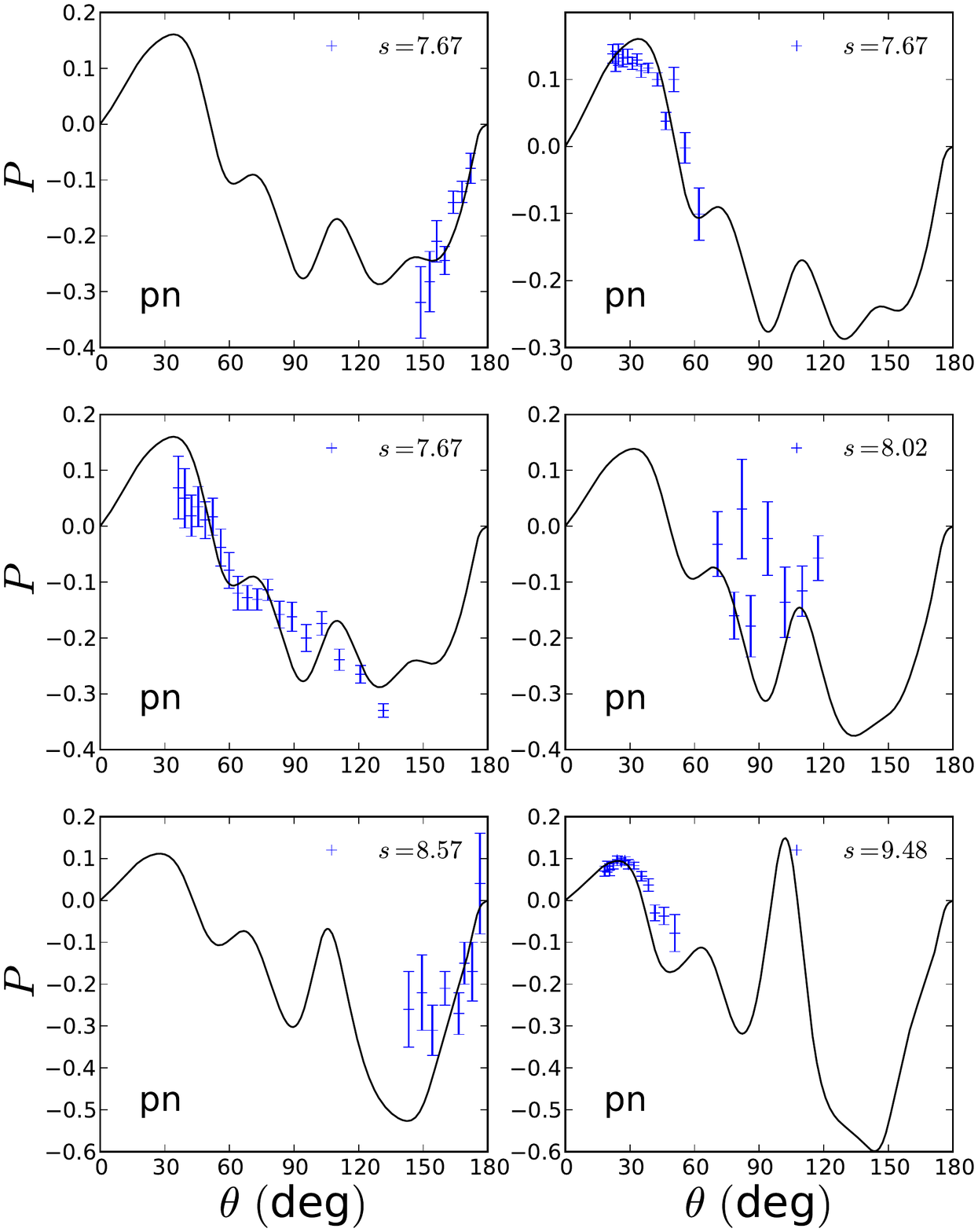} 
    \caption{Polarization for proton-neutron as a function of center of mass angle $\theta$.}
    \label{fig:Ppn3}
\end{figure}

\begin{figure}
    \includegraphics[width=15cm]{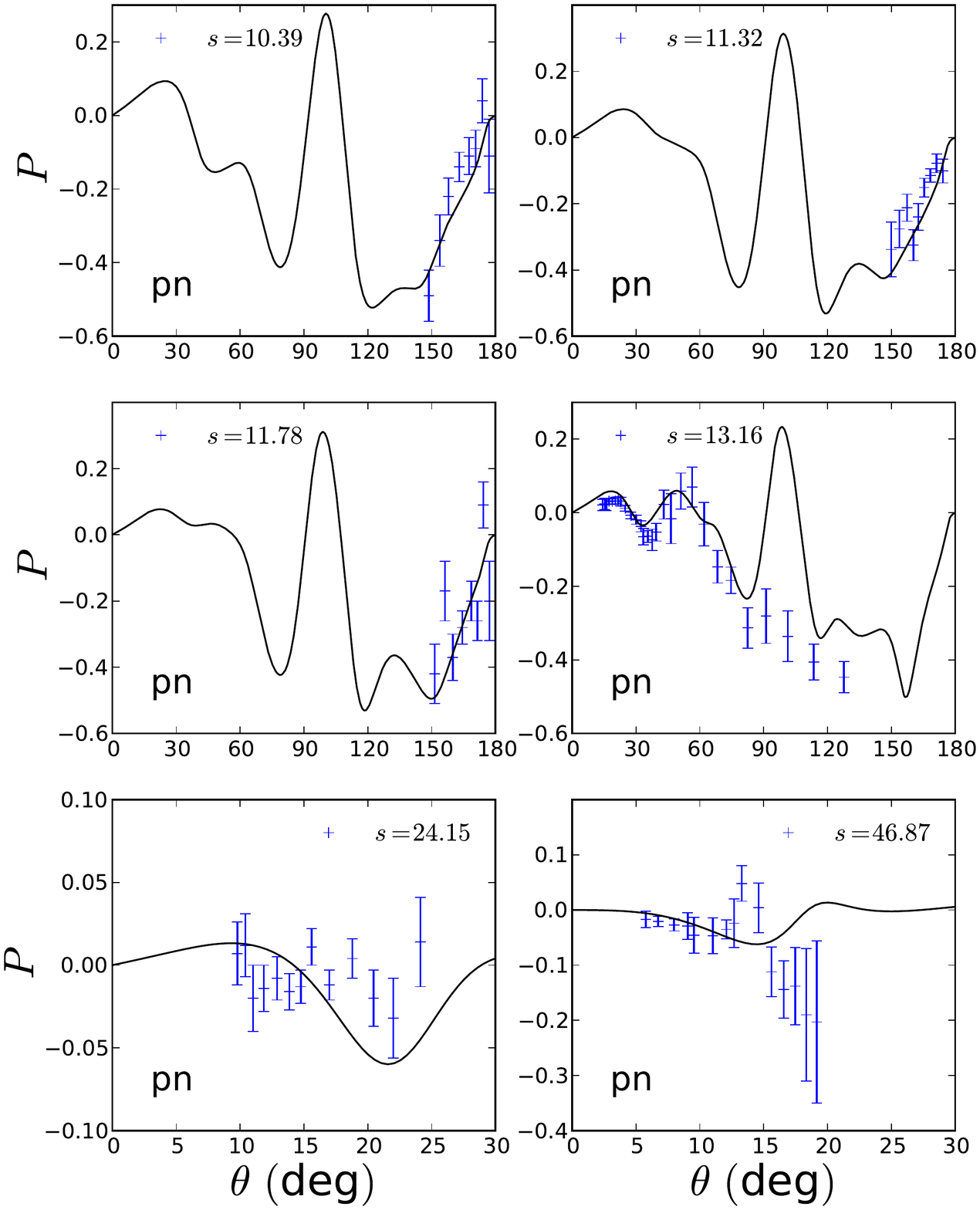} 
    \caption{Polarization for proton-neutron as a function of center of mass angle $\theta$.}
    \label{fig:Ppn4}
\end{figure}

Single polarization, or analyzing power, are presented for proton-proton, Figs. \ref{fig:Ppp10} -- \ref{fig:Ppp14}, and for proton-neutron Figs. \ref{fig:Ppn1} -- \ref{fig:Ppn4}. 
Again the model describes the data well, although more proton-neutron data would be useful to constrain the model further. 
One can see in the proton-neutron case that as the energy is increased there is insufficient data to constrain the model around $90^\circ$, and a large peak begins to form.

Finally the double-polarization observables are presented.
These were fit with minimal priority, due to the lack of data. In most cases the model works as intended and roughly describes the data, and in some cases very well. 

\begin{figure}
%\begin{center}
    \includegraphics[width=15cm]{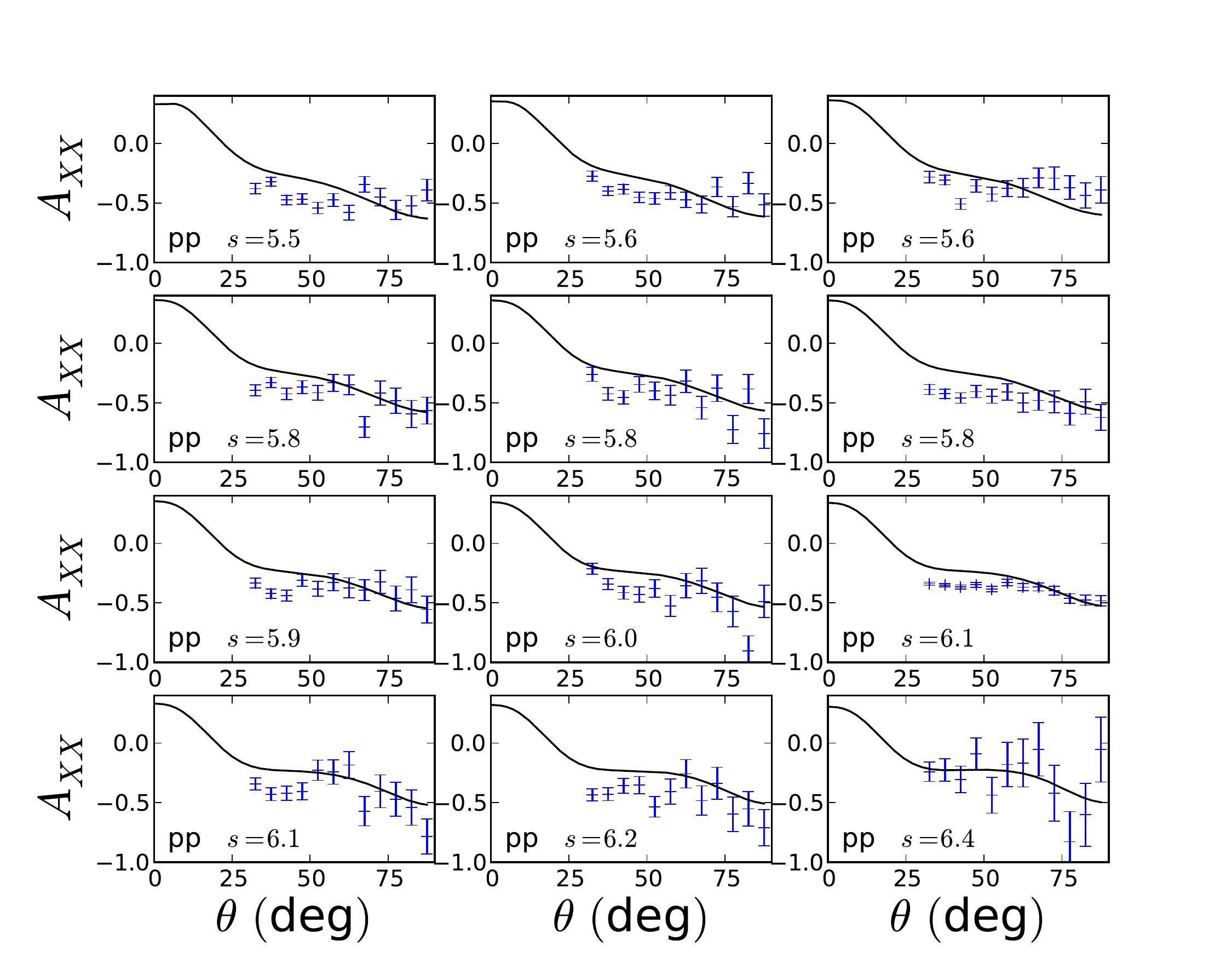} 
    \caption{Double polarization observable $A_{XX}$ for proton-proton scattering as a function of center of mass angle $\theta$.}
    \label{fig:AXX1}
%\end{center}
\end{figure}

\begin{figure}
%\begin{center}
    \includegraphics[width=15cm]{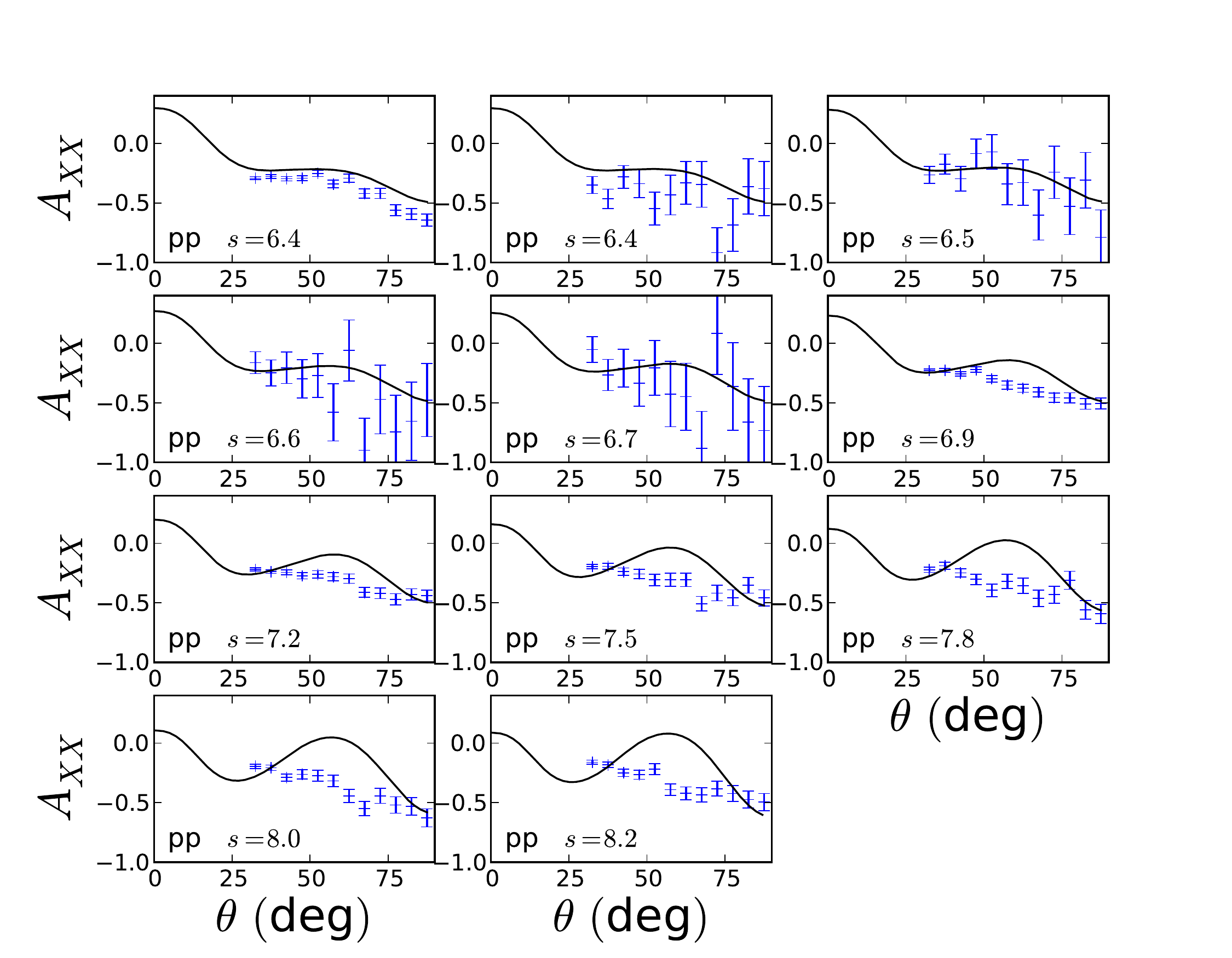} 
    \caption{Double polarization observable $A_{XX}$ for proton-proton scattering as a function of center of mass angle $\theta$.}
    \label{fig:AXX2}
%\end{center}    
\end{figure}

In Figs. \ref{fig:AXX1} - \ref{fig:AXX2} $A_{XX}$ is presented for proton-proton scattering. The model works well for the lower energy data, but does not do as well in the higher energy region with the more precise error. At the energies considered for this model there are no proton-neutron data for this observable. 

\begin{figure}
%\begin{center}
    \includegraphics[width=15cm]{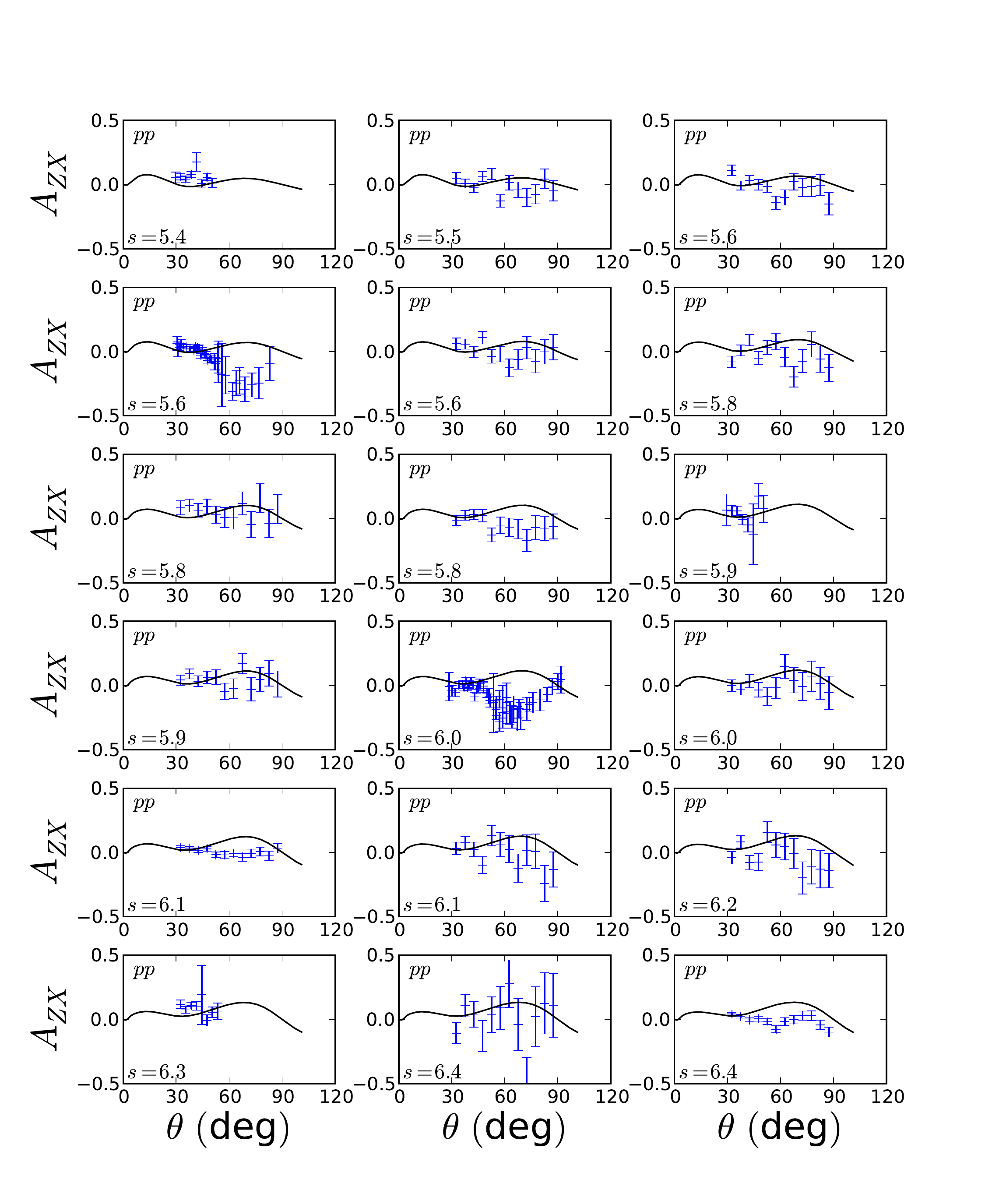} 
    \caption{Double polarization observable $A_{ZX}$ for proton-proton scattering as a function of center of mass angle $\theta$.}
    \label{fig:AZXpp1}
%\end{center}    
\end{figure}

\begin{figure}
%\begin{center}
    \includegraphics[width=15cm]{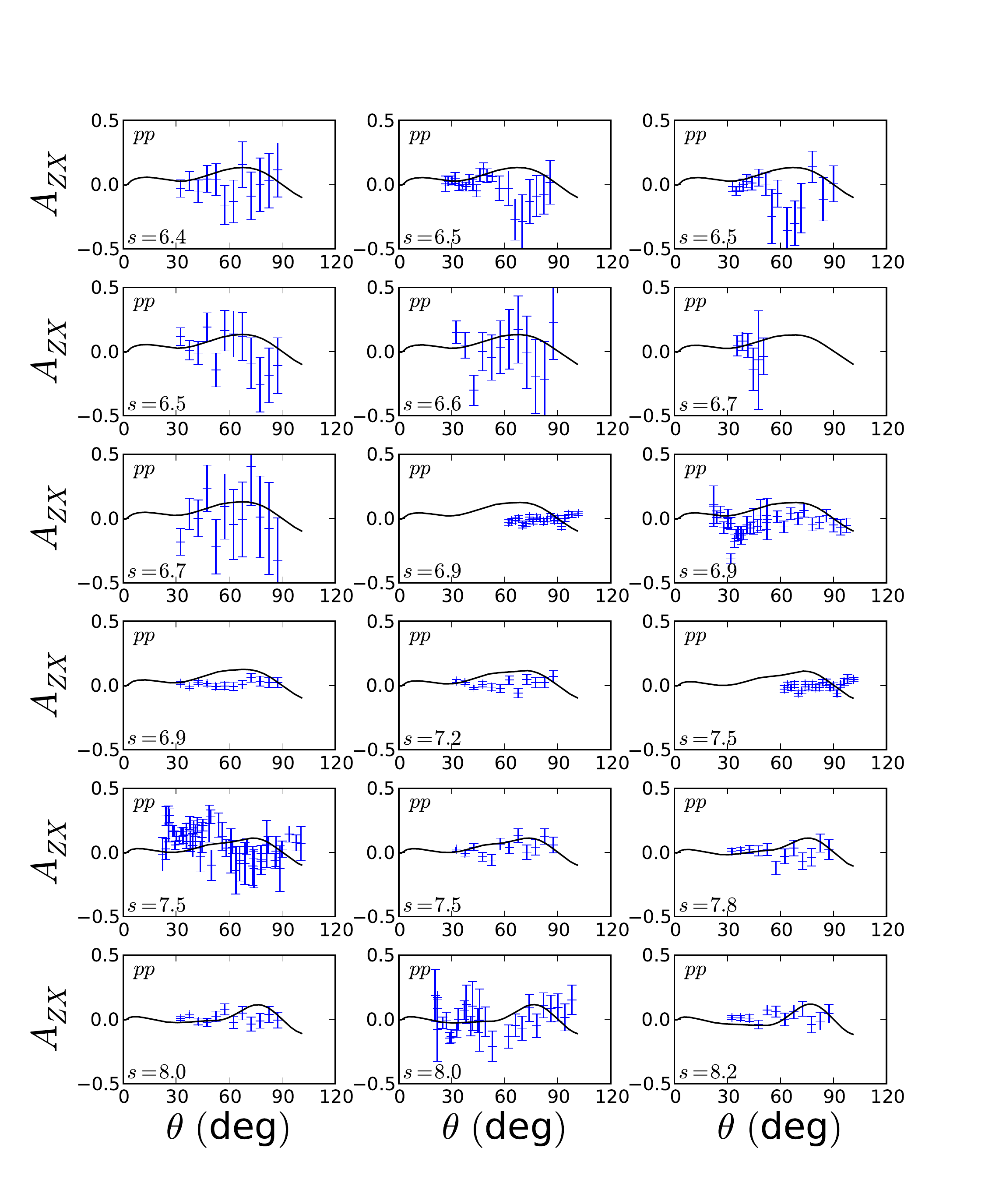} 
    \caption{Double polarization observable $A_{ZX}$ for proton-proton scattering as a function of center of mass angle $\theta$.}
    \label{fig:AZXpp2}
%\end{center}    
\end{figure}

\begin{figure}
    \includegraphics[width=15cm]{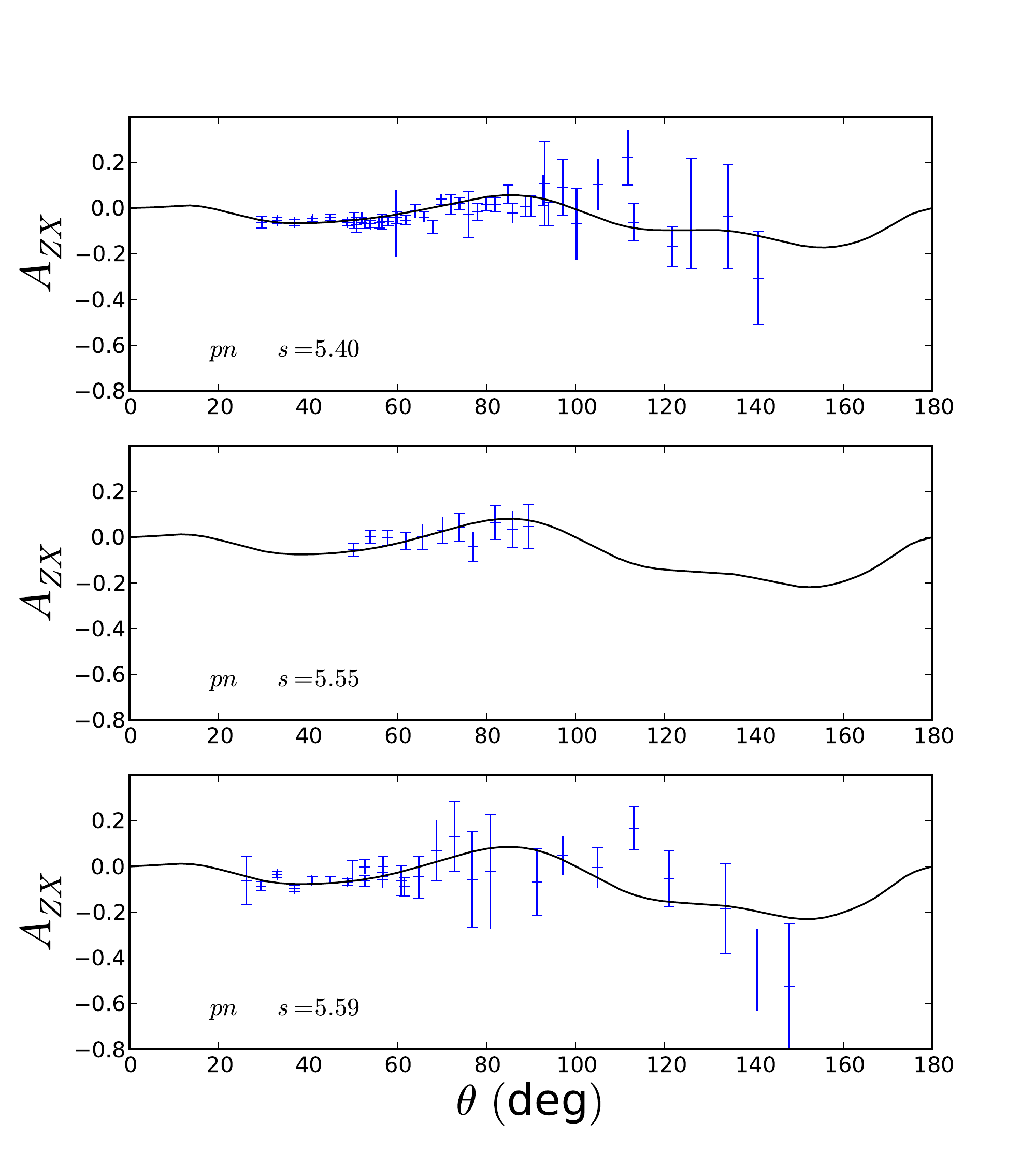} 
    \caption{Double polarization observable $A_{ZX}$ for proton-neutron scattering as a function of center of mass angle $\theta$.}
    \label{fig:AZXpn}   
\end{figure}

$A_{ZX}$ is presented in Figs. \ref{fig:AZXpp1} - \ref{fig:AZXpp2} and \ref{fig:AZXpn} for proton-proton and proton-neutron scattering respectively. The fit is acceptable and describes most of the data, although there are features in the data which the model fails to reproduce.

\begin{figure}
    \includegraphics[width=15cm]{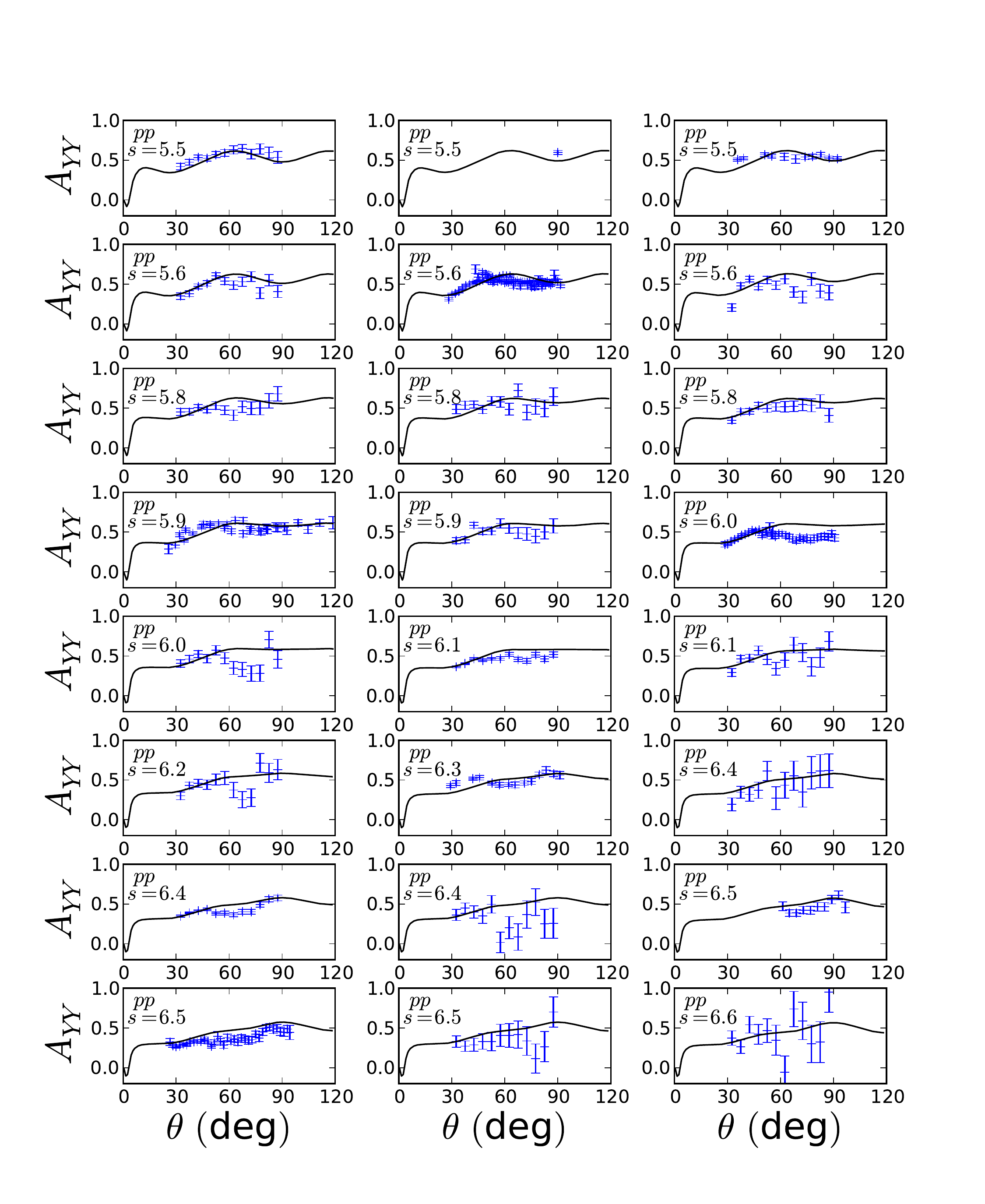} 
    \caption{Double polarization observable $A_{YY}$ for proton-proton scattering as a function of center of mass angle $\theta$.}
    \label{fig:AYYpp1}   
\end{figure}

\begin{figure}
    \includegraphics[width=15cm]{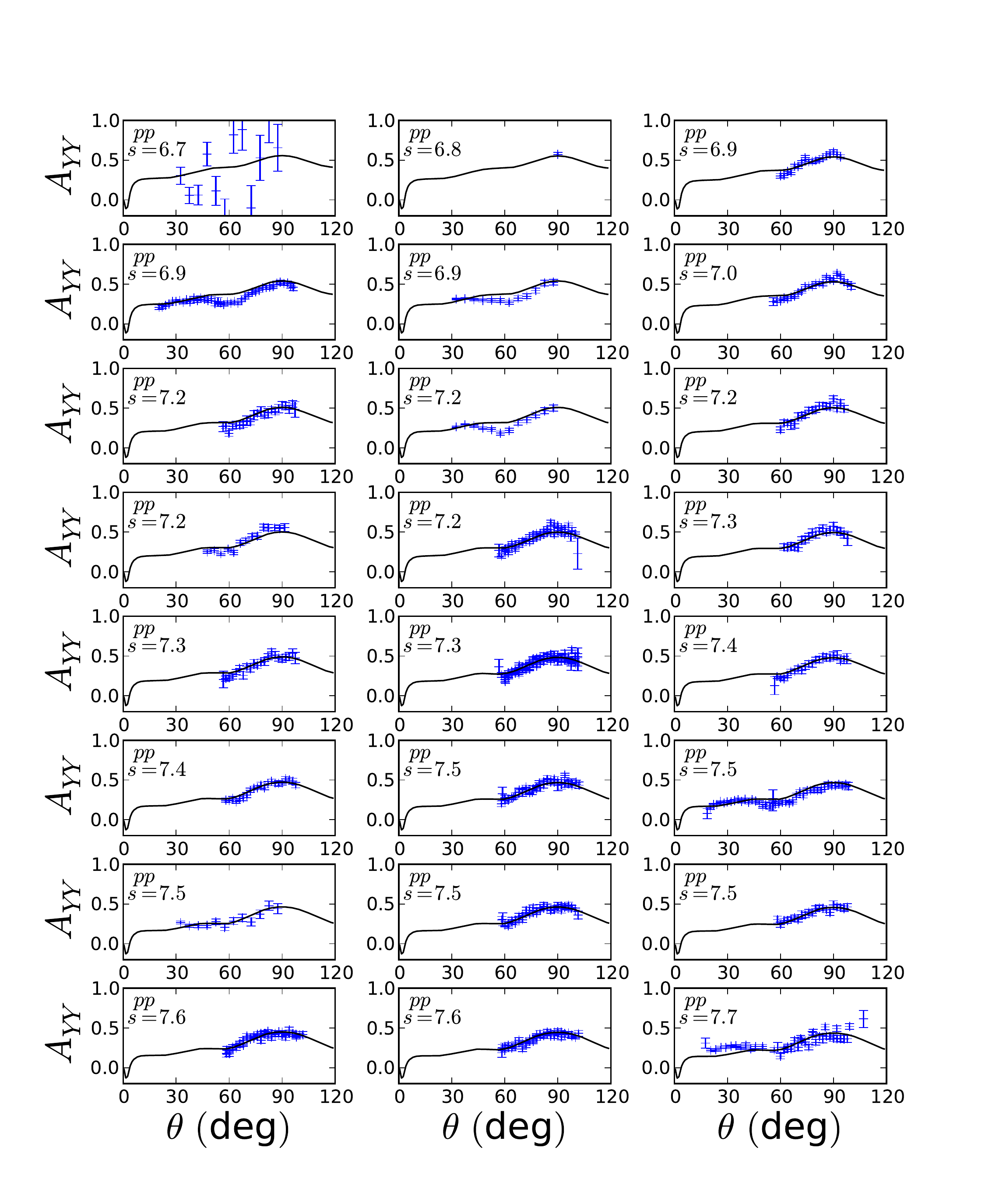} 
    \caption{Double polarization observable $A_{YY}$ for proton-proton scattering as a function of center of mass angle $\theta$.}
    \label{fig:AYYpp2}   
\end{figure}

\begin{figure}
    \includegraphics[width=15cm]{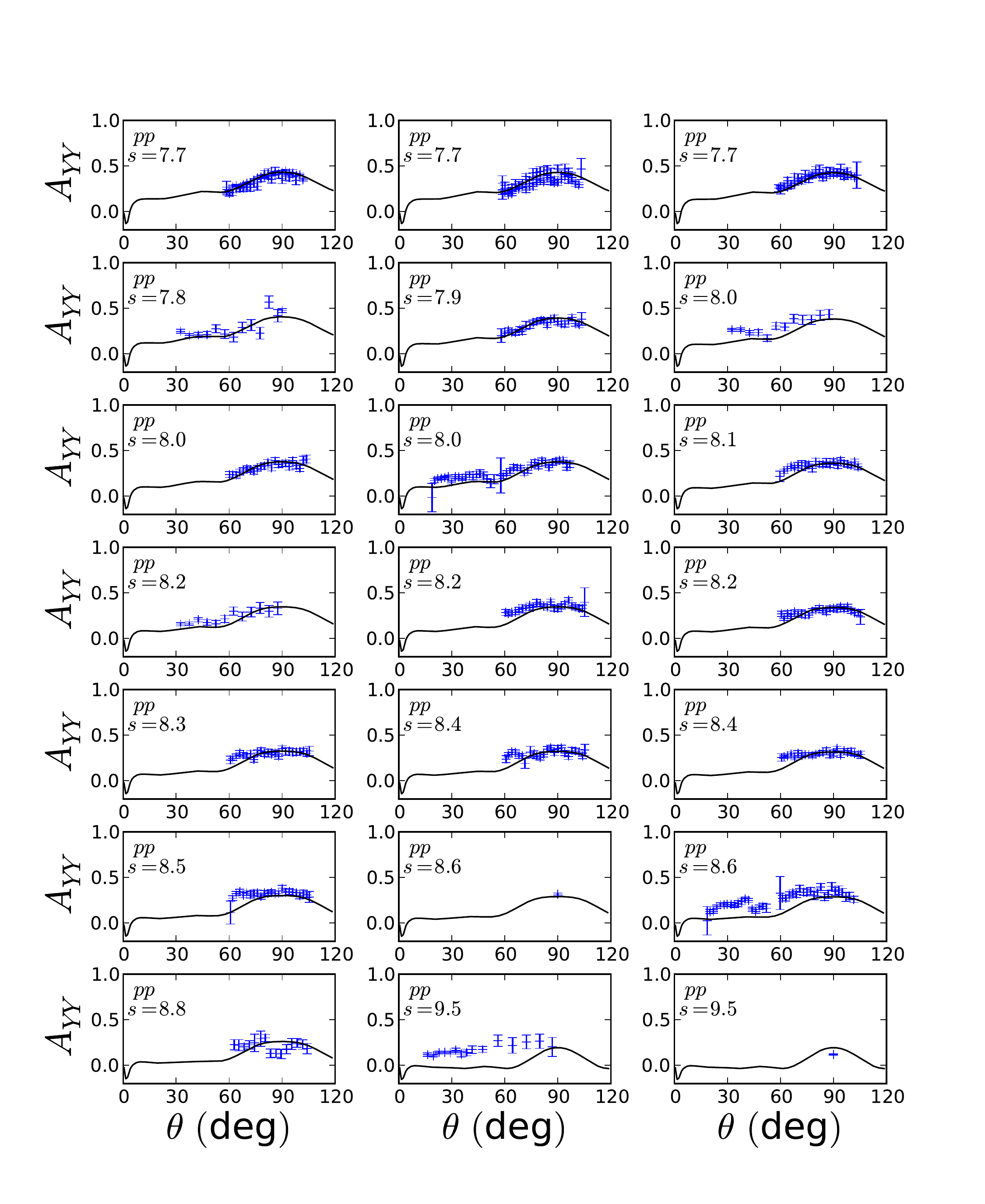} 
    \caption{Double polarization observable $A_{YY}$ for proton-proton scattering as a function of center of mass angle $\theta$.}
    \label{fig:AYYpp3}   
\end{figure}

\begin{figure}
    \includegraphics[width=15cm]{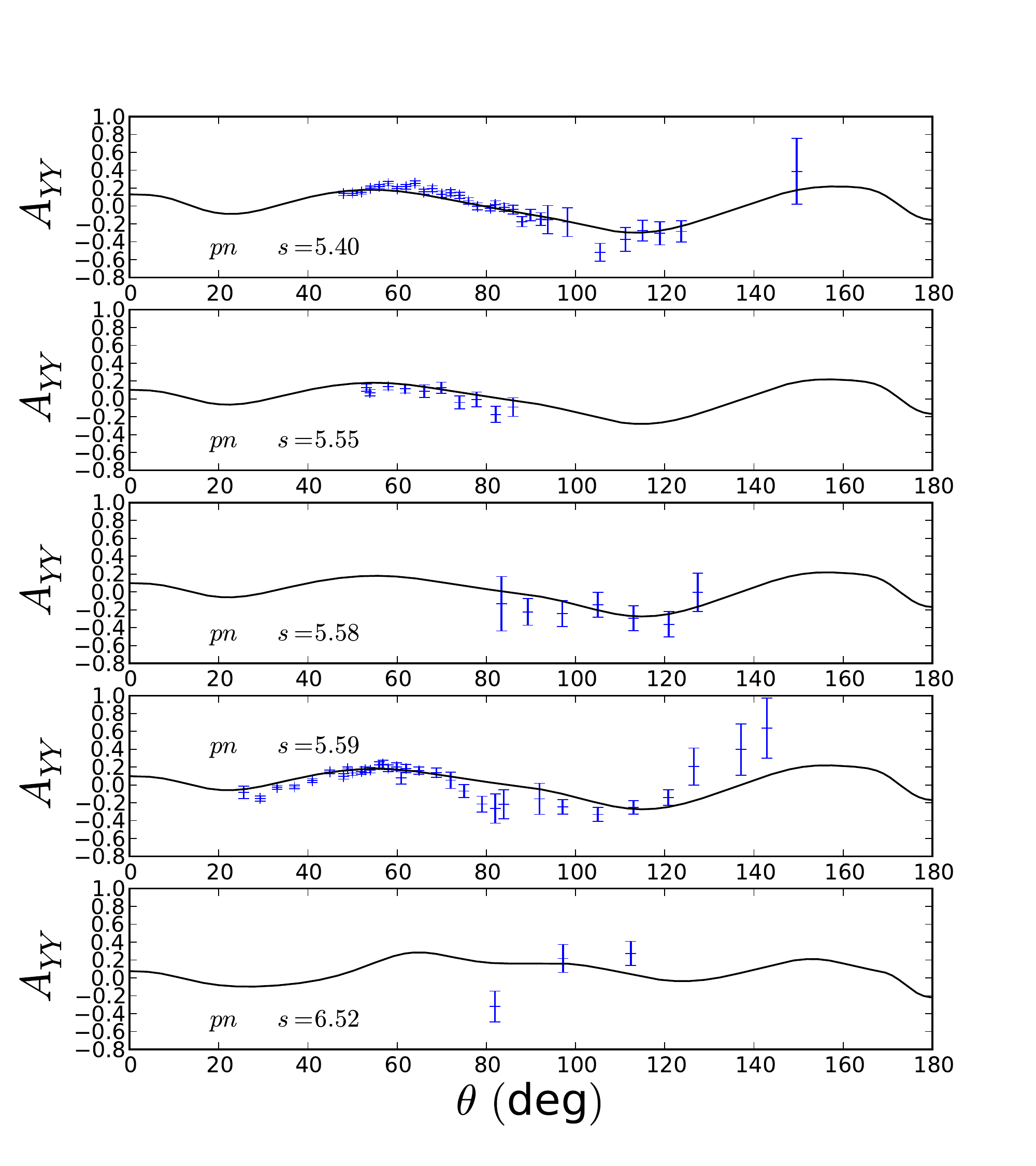} 
    \caption{Double polarization observable $A_{YY}$ for proton-neutron scattering as a function of center of mass angle $\theta$.}
    \label{fig:AYYpn}   
\end{figure}

$A_{YY}$ is shown in Figs. \ref{fig:AYYpp1} - \ref{fig:AYYpp3} and \ref{fig:AYYpn} for proton-proton and proton-neutron respectively. $A_{YY}$ has the most data points of all the double polarization observables, and it can be seen that with sufficient data the model is working well, reproducing various features in the data.

\begin{figure}
    \includegraphics[width=15cm]{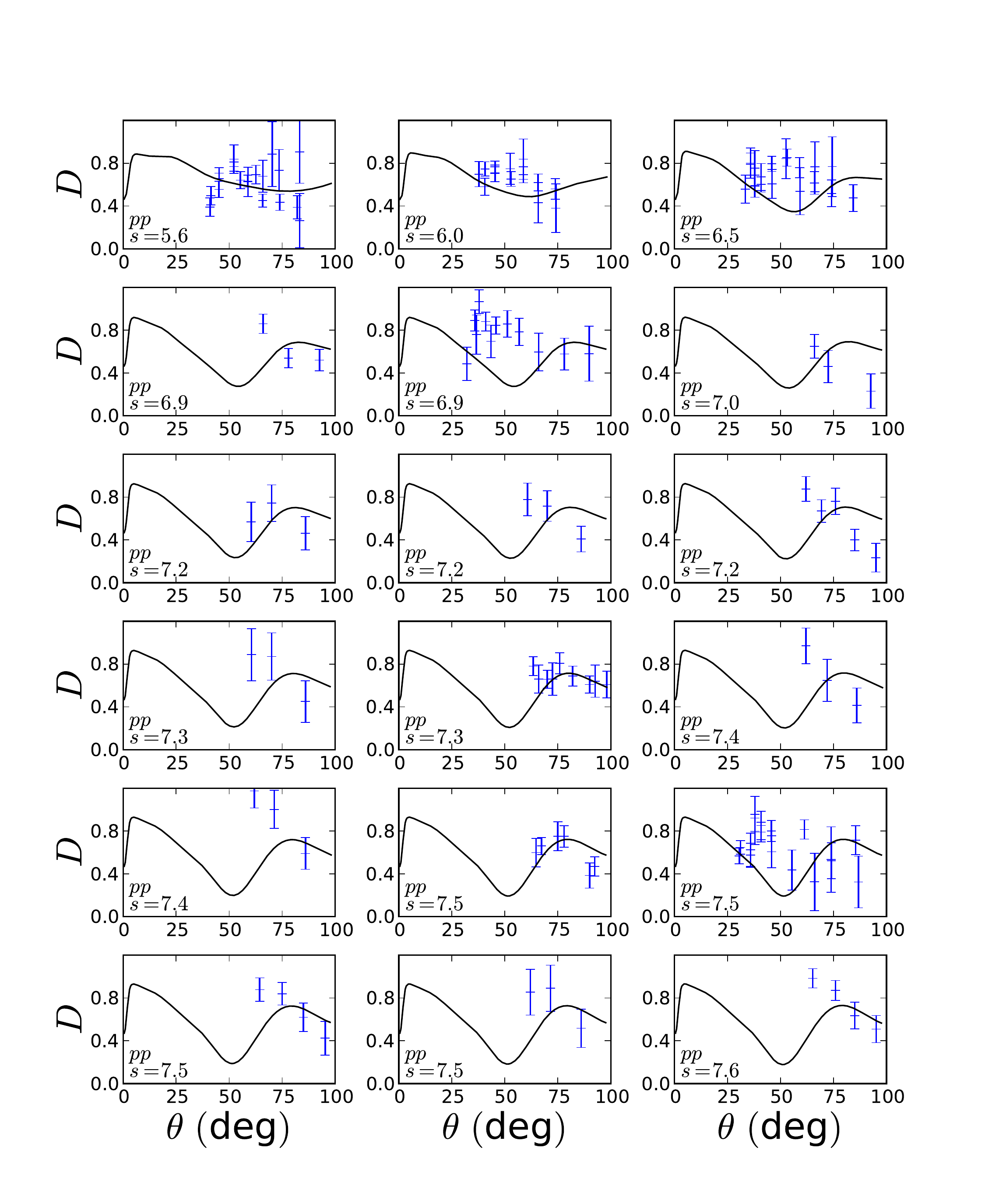} 
    \caption{Double polarization observable $D$ for proton-proton scattering as a function of center of mass angle $\theta$.}
    \label{fig:Dpp1}   
\end{figure}

\begin{figure}
    \includegraphics[width=15cm]{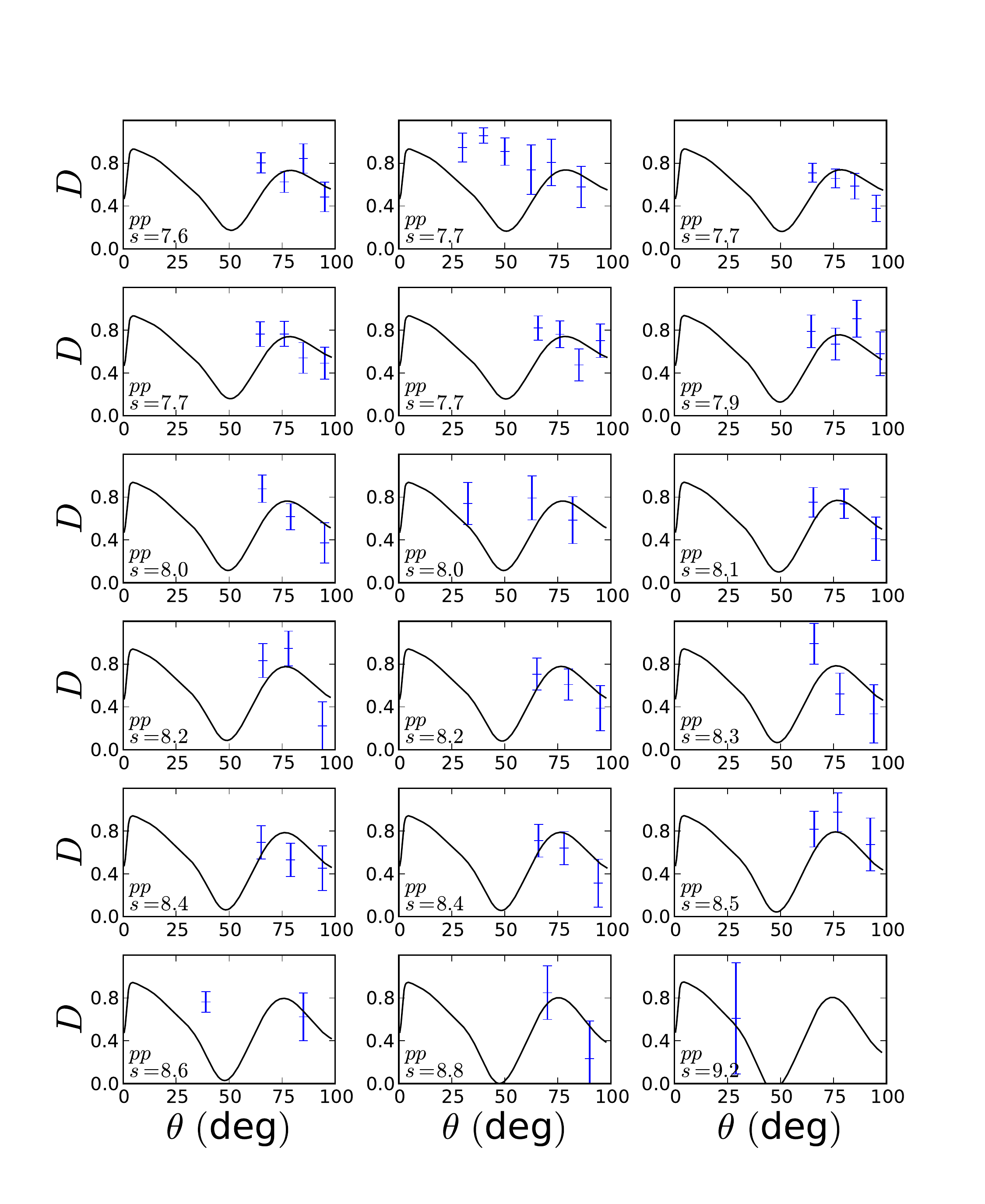} 
    \caption{Double polarization observable $D$ for proton-proton scattering as a function of center of mass angle $\theta$.}
    \label{fig:Dpp2}   
\end{figure}

\begin{figure}
    \includegraphics[width=15cm]{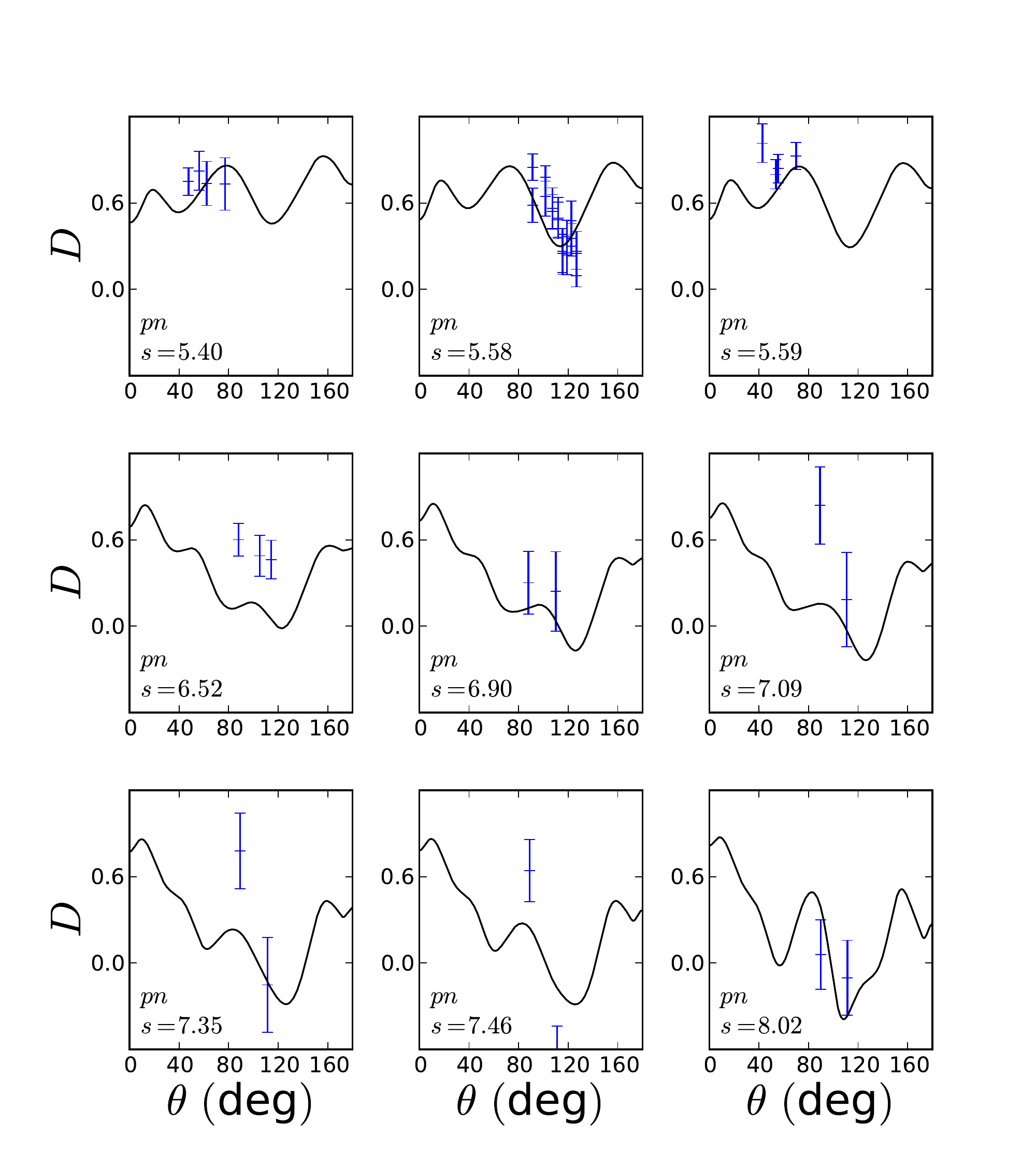} 
    \caption{Double polarization observable $D$ for proton-neutron scattering as a function of center of mass angle $\theta$.}
    \label{fig:Dpn}   
\end{figure}

In Figs. \ref{fig:Dpp1} - \ref{fig:Dpp2} and \ref{fig:Dpn} $D$ is shown for proton-proton and proton-neutron respectively. The model works as intended in the regions where data are available. While this observable has been measured over a large energy range notice that only a few angular points are available at each energy.

\begin{figure}
    \includegraphics[width=15cm]{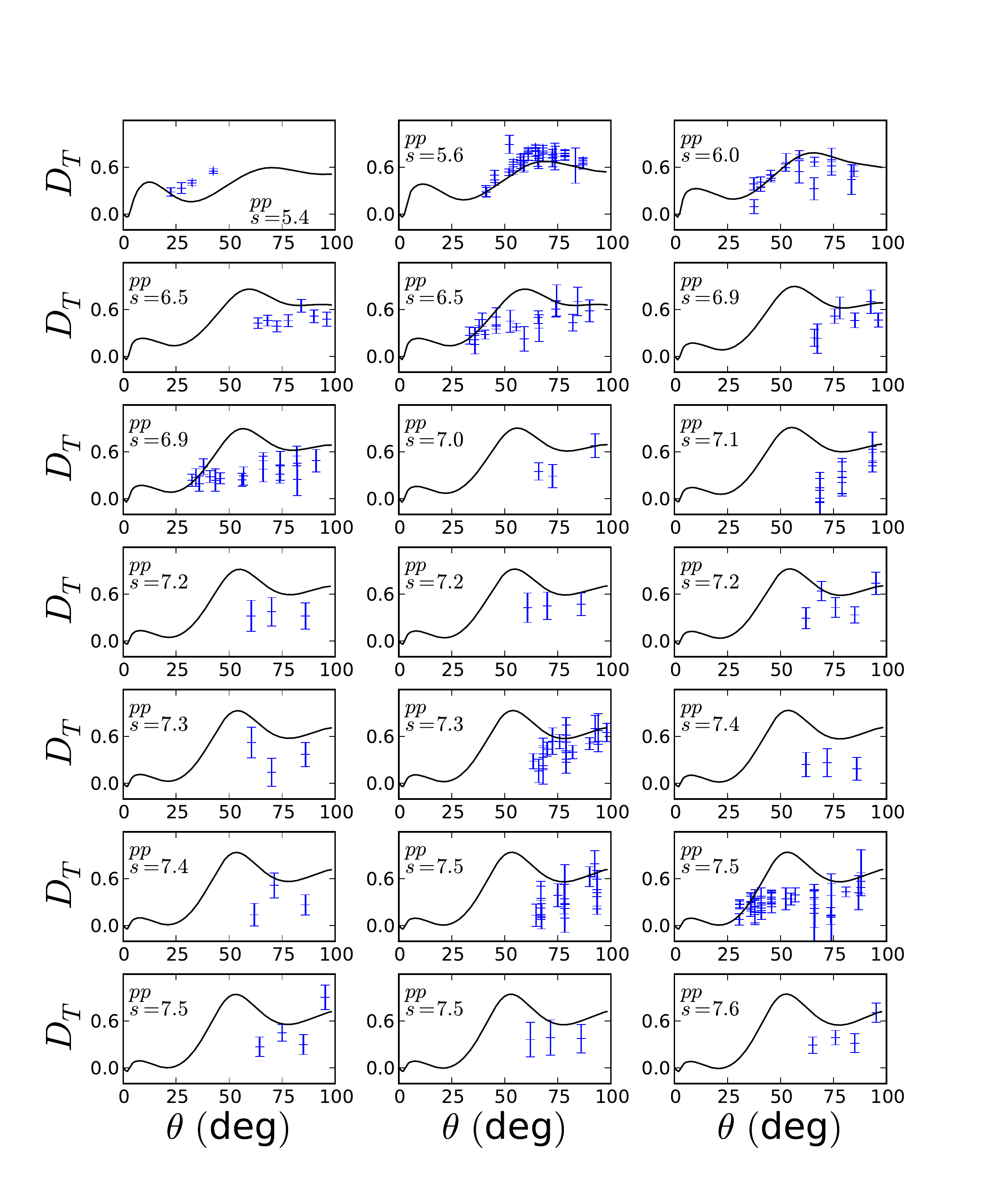} 
    \caption{Double polarization observable $D_T$ for proton-proton scattering as a function of center of mass angle $\theta$.}
    \label{fig:DTpp1}   
\end{figure}

\begin{figure}
    \includegraphics[width=15cm]{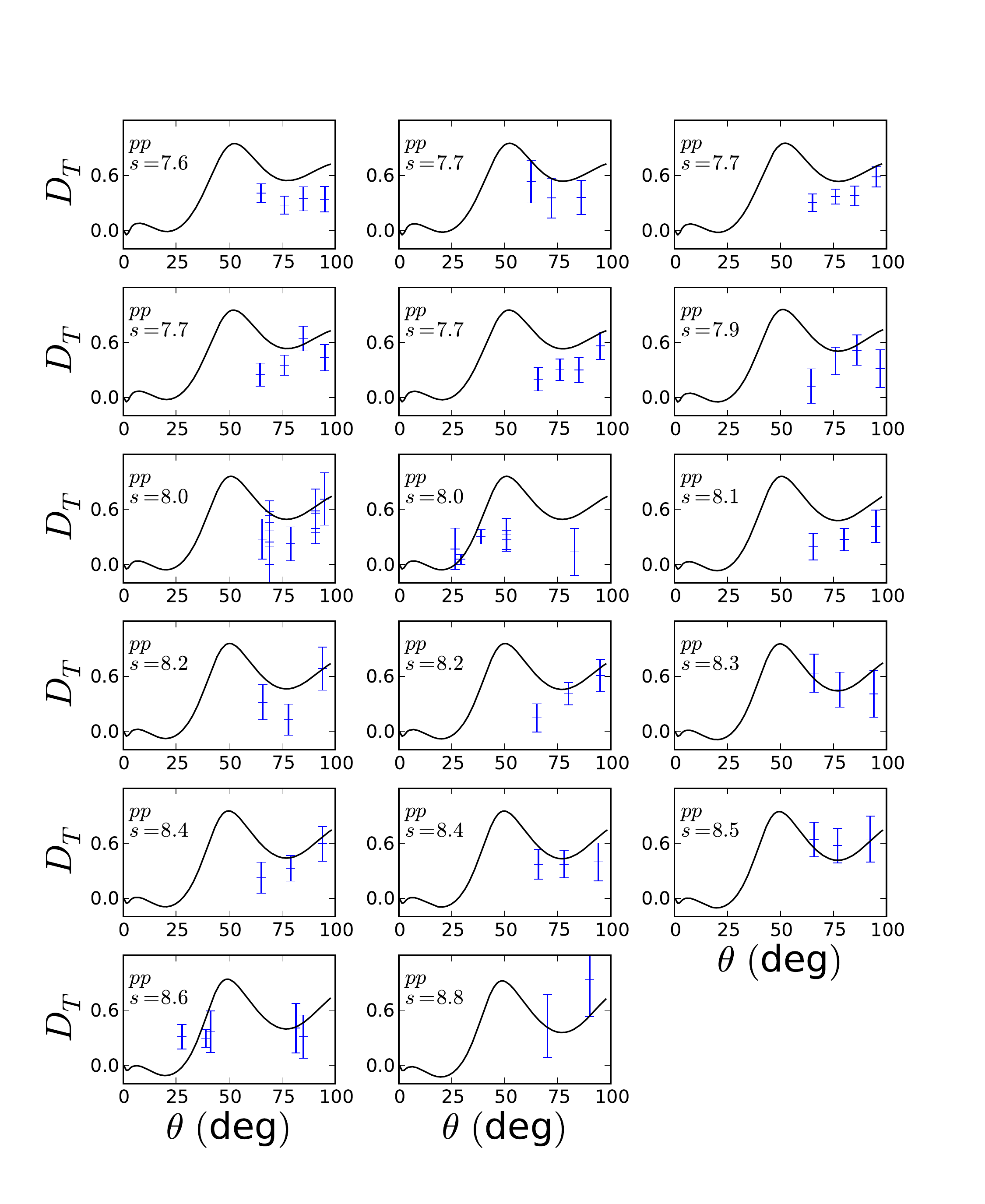} 
    \caption{Double polarization observable $D_T$ for proton-proton scattering as a function of center of mass angle $\theta$.}
    \label{fig:DTpp2}   
\end{figure}

\begin{figure}
    \includegraphics[width=15cm]{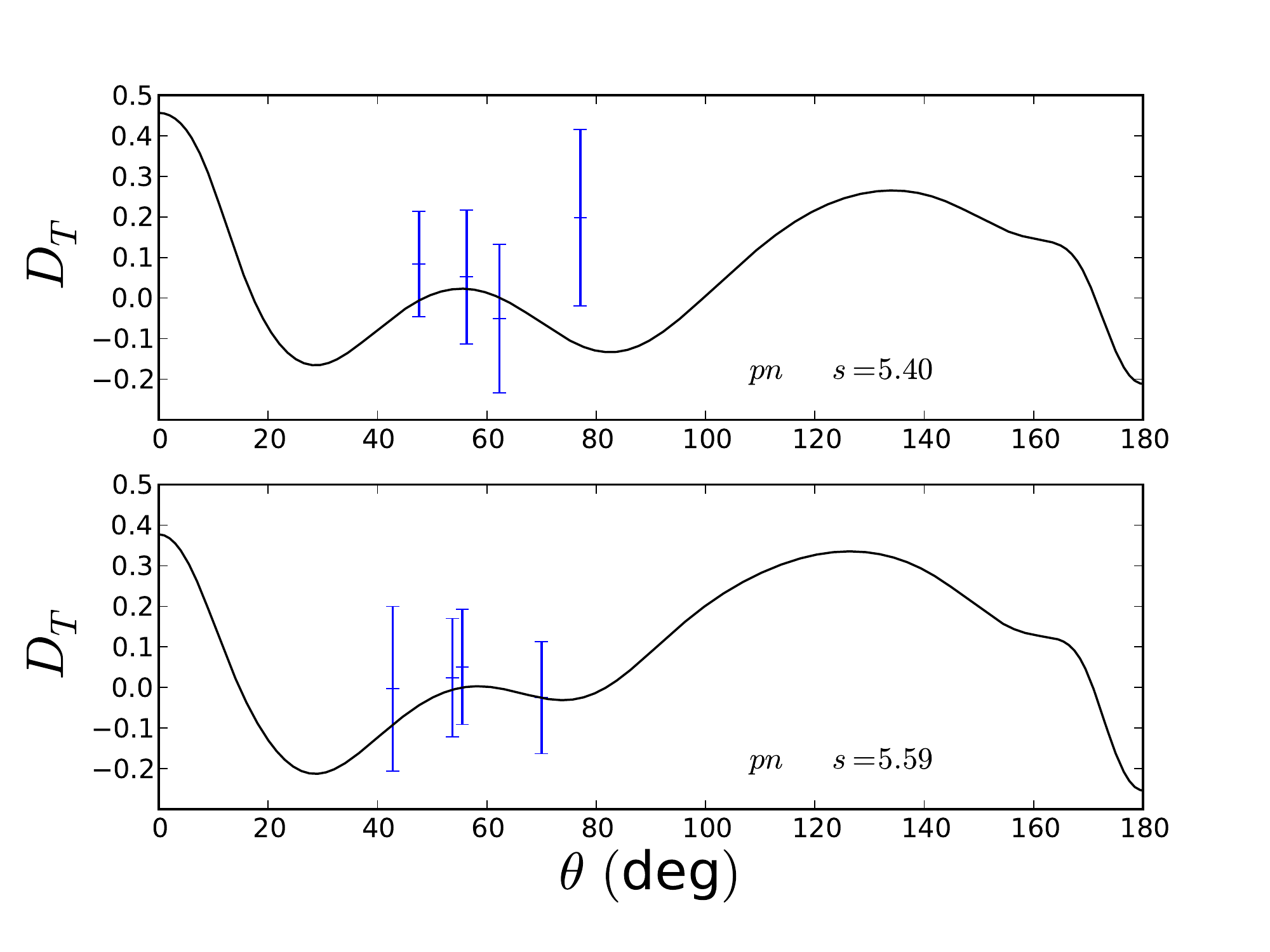} 
    \caption{Double polarization observable $D_{T}$ for proton-neutron scattering as a function of center of mass angle $\theta$.}
    \label{fig:DTpn}   
\end{figure}

$D_T$ is presented in Figs. \ref{fig:DTpp1} - \ref{fig:DTpp2} for proton proton and \ref{fig:DTpn} for proton-neutron. The model works where data are available, however, it should again be mentioned the very minimal amount of pn data with only 8 points.

\clearpage
\clearpage
\clearpage
%-----------------------------------------------------------------
%------------Double Polarization Observables----------------------
%-----------------------------------------------------------------
%
%-----------------------------------------------------------------
%-----------------------------AXX---------------------------------
%-----------------------------------------------------------------
%-----------------------------------------------------------------
%-----------------------------AZX---------------------------------
%-----------------------------------------------------------------
%-----------------------------------------------------------------
%-----------------------------AZZ---------------------------------
%-----------------------------------------------------------------
\begin{figure}
%\begin{center}
    \includegraphics[width=15cm]{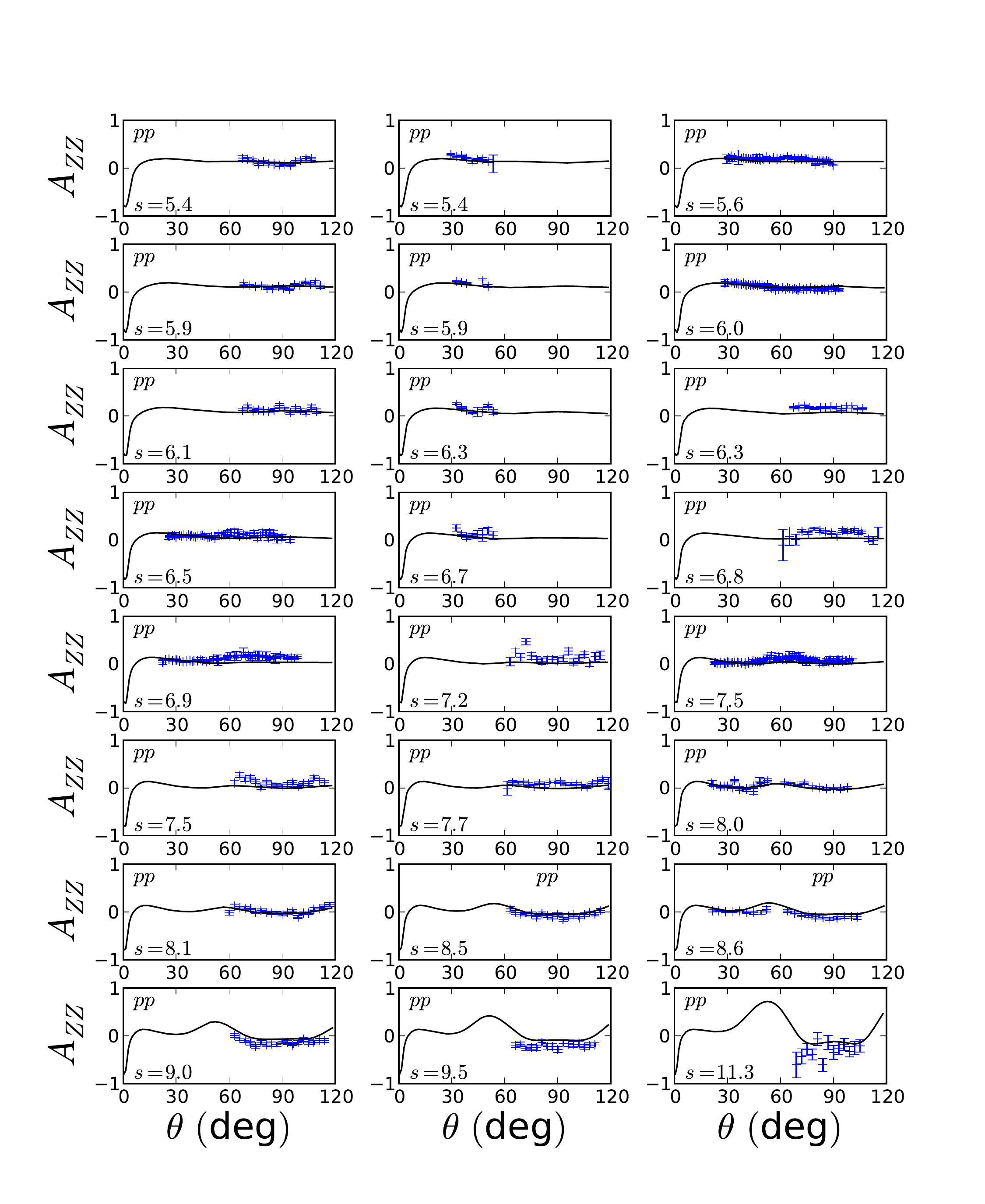} 
    \caption{Double polarization observable $A_{ZZ}$ for proton-proton scattering as a function of center of mass angle $\theta$.}
    \label{fig:AZZpp1}
%\end{center}    
\end{figure}
\begin{figure}
    \includegraphics[width=15cm]{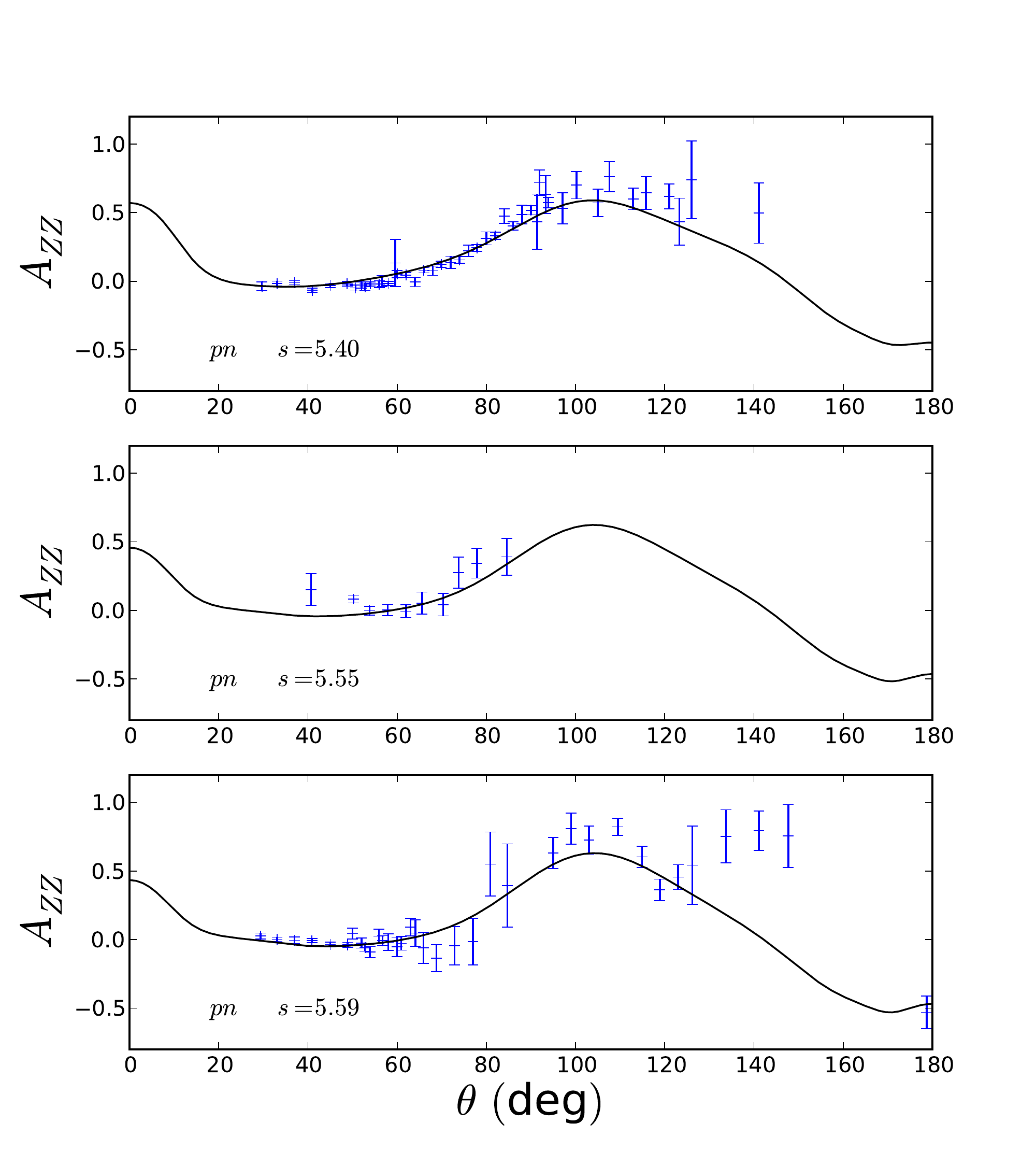} 
    \caption{Double polarization observable $A_{ZZ}$ for proton-neutron scattering as a function of center of mass angle $\theta$.}
    \label{fig:AZZpn}   
\end{figure}
%-----------------------------------------------------------------
%-----------------------------AYY---------------------------------
%-----------------------------------------------------------------
%-----------------------------------------------------------------
%-----------------------------D-----------------------------------
%-----------------------------------------------------------------
%-----------------------------------------------------------------
%-----------------------------DT----------------------------------
%-----------------------------------------------------------------
\newpage
%-----------------------------------------------------------------
%-------------------------Chapter---------------------------------
%-----------------------------------------------------------------
\chapter{Final state interactions in deuteron electrodisintegration}\label{sec:deepfsi}
The initial interest in modeling the $NN$ amplitudes was to describe final state interactions for the process of electrodisintegration of the deuteron.
In this section the application of the Regge model to this reaction is studied \cite{FJVO}.
%A calculation describing this reaction was performed and utilizes the scattering amplitudes as input. 
For this process there are many observables which can be investigated. Observables of interest, which were chosen to be calculated can be categorized as follows: 
\begin{itemize}
\item Unpolarized Hadrons \cite{JVO_2008_newcalc}
\begin{itemize}
	\item $\frac{d^5\sigma}{d\Omega_ed\Omega_p dE'}$, $A_{LT}$, ${A_{LT'}}$, $A_{TT}$
\end{itemize}
\item Target Polarization \cite{JVO_2009_tar_pol}
	\begin{itemize}
		\item ${A^V_d}$, $A^T_d$, $A^V_{ed}$, and ${A^T_{ed}}$
	\end{itemize}
\item Ejectile (proton) Polarization \cite{JVO_2009_ejec_pol}
	\begin{itemize}
		\item ${A^{n'}_{p}}$, ${A^{l'}_{p}}$, ${A^{s'}_{p}}$, $A^{n'}_{ep}$, $A^{l'}_{ep}$ and $A^{s'}_{ep}$ 
	\end{itemize}
\end{itemize}
Details of the calculation, and definitions of the asymmetries for these three cases can be found in the references provided in the above bulleted list.
%Four observables $\frac{d^5\sigma}{d\Omega_ed\Omega_p dE'}$, $A_{LT}$, ${A_{LT'}}$, and $A_{TT}$ , in which no hadrons are polarized \cite{JVO_2008_newcalc}. 
%Allowing for target polarization ${A^V_d}$, $A^T_d$, $A^V_{ed}$, and ${A^T_{ed}}$ \cite{JVO_2009_tar_pol} were calculated. Finally one can detect the polarization of the ejected proton, and six more observables are calculated ${A^{n'}_{p}}$, ${A^{l'}_{p}}$, ${A^{s'}_{p}}$, $A^{n'}_{ep}$, $A^{l'}_{ep}$ and $A^{s'}_{ep}$ \cite{JVO_2009_ejec_pol}. 

\begin{table} \centering
\caption{The kinematics for the $x=1$ and $x=1.3$ kinematics used in the electrodisintegration calculations presented here.}
\begin{tabular}{crr}
\hline\hline
kinematics & 1 &2\\\hline
$x$ & 1.0 & 1.3\\
$Q^2\ {\rm (GeV^2)}$& 2.4 & 4.5 \\
$E_{beam}\ {\rm (GeV)}$& 6.25& 8.6\\
$s\ {\rm (GeV^2)}$ & 5.91 & 5.93\\
$\theta_e\ {\rm (deg)}$&15.97& 16.0 \\
$\frac{Q^2}{q^2}$ & 0.60 & 0.57 \\
\hline\hline
\end{tabular}\label{tab:kinematics}
\end{table}

Figures of the various observables are presented here for deuteron electrodisintegration and comparisons are made between using the Regge model and SAID parametrizations of the final state interactions.
The necessary kinematics to describe the process are the electron beam energy $E_{beam}$, electron scattering angle $\theta_e$, the negative transferred four momentum squared of the electron $Q^2$, three momentum of the transfer $q$, and Bjorken $x = \frac{Q^2}{2m\nu}$, where $m$ is the nucleon mass and $\nu$ is the energy transfer of the electron. Mandelstam $s$ is the square of the sum of the four momentum of the nucleons as defined previously.  

Two different kinematics are used, one for Bjorken $x$ of $x=1.0$ and the other for $x=1.3$ corresponding to quasi-elastic scattering and non non quasi-elastic scattering respectively.  Various kinematical variables for the two kinematics are shown in Table \ref{tab:kinematics}. The kinematics are chosen such that $s$, the electron scattering angle $\theta_e$ and the ratio of the square of four-momentum transfer to three-momentum transfer $Q^2/q^2$ are approximately equal for the two cases. The values of $s$ are close to the upper range available from SAID and are at the lower end of the fitting range for the Regge parameterization. In all cases the onshell approximation for the final state interactions (FSI), as described in \cite{JVO_2008_newcalc}, is used. 
%
%An offshell prescription for the SAID FSI was proposed in \cite{JVO_2008_newcalc}, but a more complete approach is possible for the Regge parameterization and will be considered in a future paper. Since both the SAID and Regge amplitudes are fit to onshell data, using only the onshell approximation is the most reasonable way to compare the two methods.
%%

\begin{figure}
    \includegraphics[width=6in]{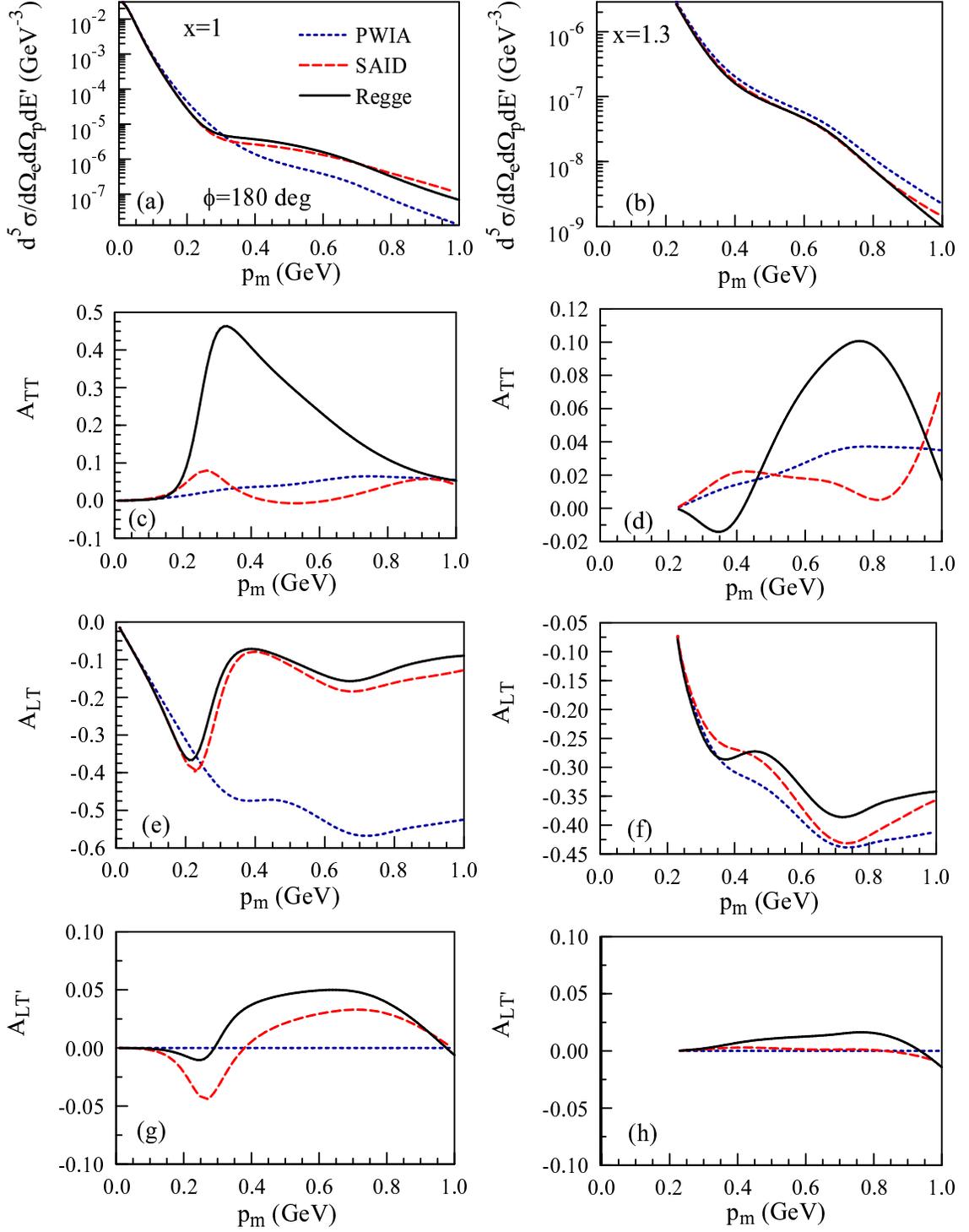}
    \caption{Spin observables for unpolarized hadrons. Short-dashed lines represent the PWIA contribution. Long-dashed lines include the SAID FSI and solid lines include the Regge FSI. Plots in the left-hand column are for the $x=1$ kinematics and plots in the right-hand column are for the $x=1.3$ kinematics.}
    \label{fig:stat_un}
\end{figure}

\begin{figure}
    \includegraphics[width=6in]{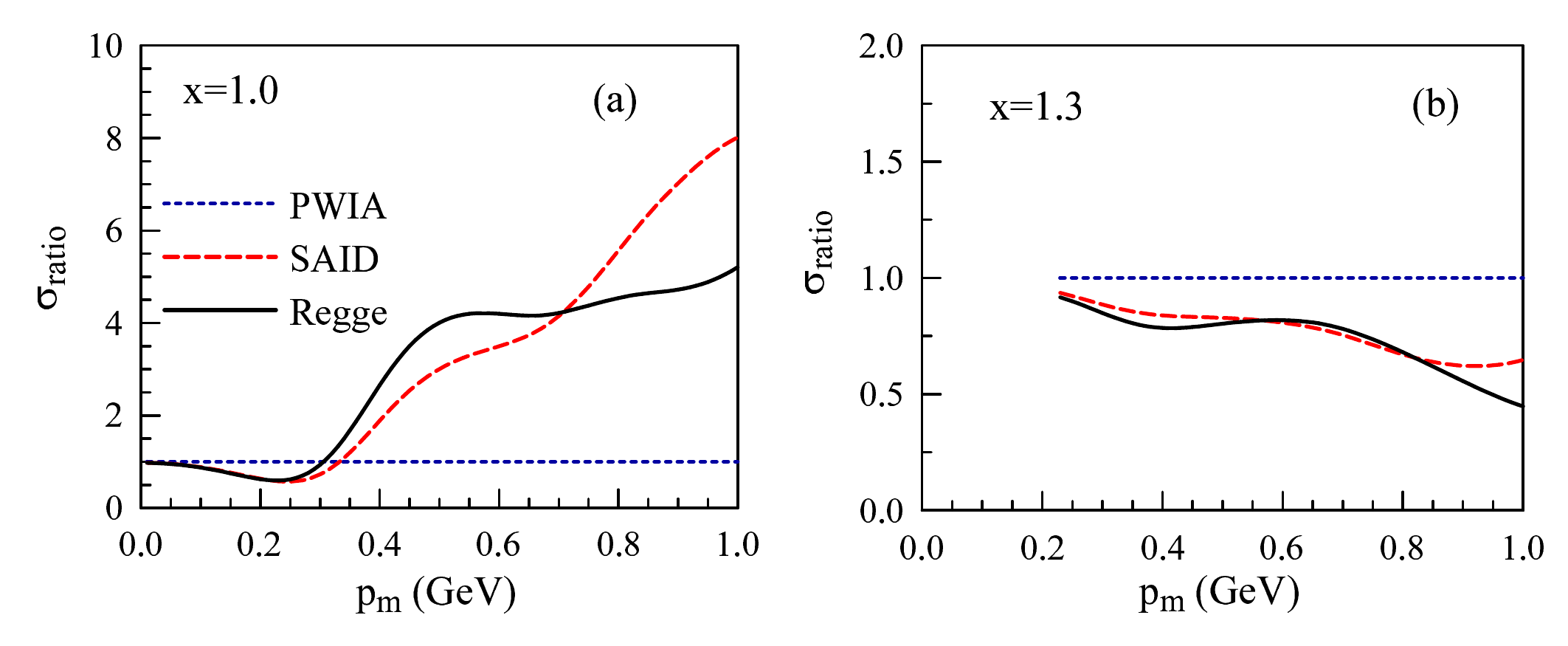}
    \caption{Ratios of the differential cross sections to the PWIA approximation. Lines are represented as in Fig. \ref{fig:stat_un}.}
    \label{fig:ratio}
\end{figure}

Figure \ref{fig:stat_un}  shows the observables for the case where neither the deuteron target nor the ejected proton are polarized, and where the azimuthal angle is chosen to be $\phi=180^\circ$. 
%A summary of the various calculated quantities is contained in the appendix. 
Figures \ref{fig:stat_un}(a)and (b) show the differential cross sections as a function of missing (neutron) momentum $p_m$ for the plane wave impulse approximation (PWIA) and for the SAID and Regge FSI for the $x=1$ kinematics and the $x=1.3$ kinematics respectively.
The size and shape of the two FSI calculations are similar in each case. Since these are semi-log plots, a more accurate evaluation of the differences is given by the ratio of distorted wave to PWIA cross section $\sigma_{ratio}$ as is shown in Fig. \ref{fig:ratio}.
In Fig. \ref{fig:ratio}(a) for the $x=1$ kinematics, the SAID and Regge results are very similar for $p_m<0.3\ {\rm GeV}$ but differ by up to 50 percent from the PWIA result. At higher missing momenta both the SAID and Regge results become increasingly large compared to the PWIA, and reach a value of approximately 8 times the PWIA at $p_m=1.0\ {\rm GeV}$ for the SAID FSI and approximately 5 times for the Regge FSI. For the $x=1.3$ kinematics, shown in Fig. \ref{fig:ratio}(b), the difference between the SAID and Regge FSI are much smaller and they are both much closer to the PWIA. Note that for $x=1$ both of the FSI lie above the PWIA but for $x=1.3$ they are below. This suggests that it may be possible to find kinematics at which the FSI effects are minimal and may allow for an approximate extraction of the deuteron ground-state momentum distribution, as has been suggested previously \cite{Boeglin_highmom}.

\begin{figure} \centering
    \includegraphics[width=5in]{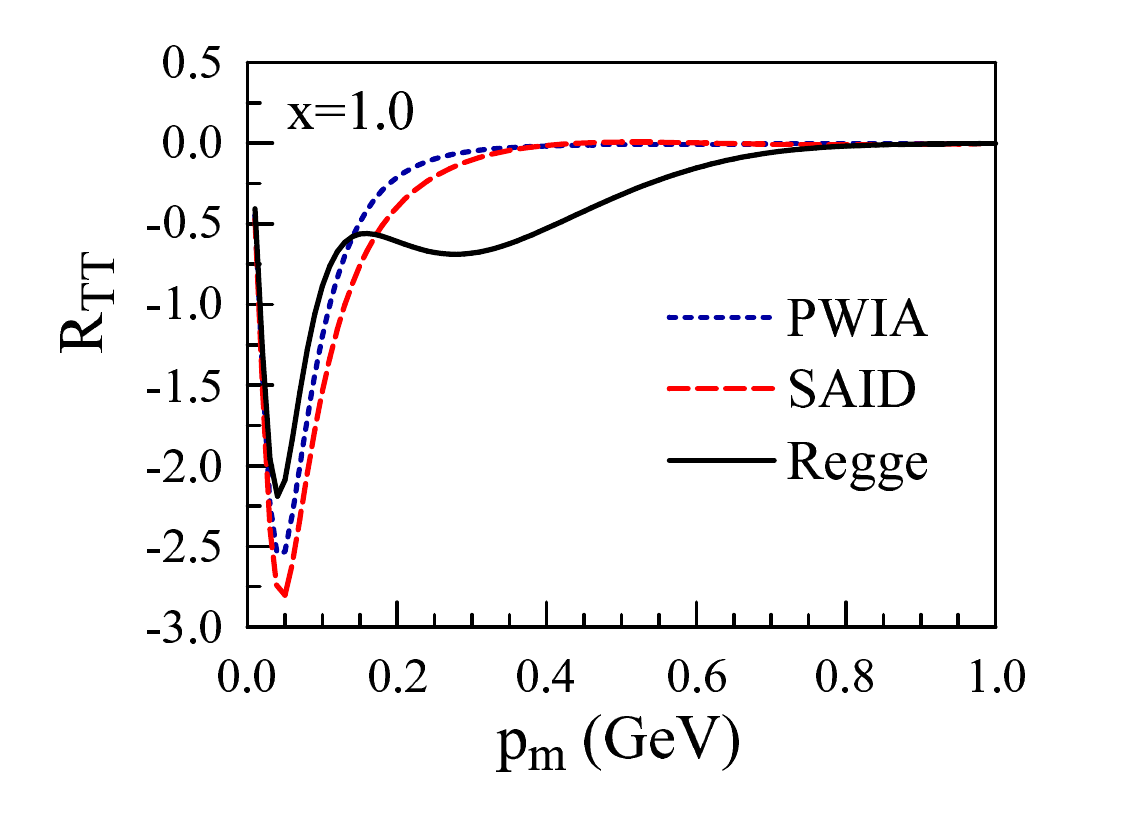}
    \caption{The transverse-transverse response function $R_{TT}$ for the $x=1$ kinematics. Lines are represented as in Fig. \ref{fig:stat_un}.}
    \label{fig:RTT}
\end{figure}

Figures \ref{fig:stat_un}(c) and (d) show the transverse-transverse asymmetry $A_{TT}$ for the $x=1$ and $x=1.3$ kinematics, respectively. This asymmetry, which is proportional to the $R_{TT}$ response function, is generally assumed to be small since $R_{TT}$ has generally been shown to be small. This is the case in \ref{fig:stat_un}(c) for the PWIA and SAID results, but the asymmetry for the Regge FSI is large for intermediate values of $p_m$. The reason for this can be seen from Fig. \ref{fig:RTT} which shows $R_{TT}$ for the $x=1$ kinematics.
Note that all three calculations have a minimum at around $p_m=0.05\ {\rm GeV}$ where the cross section is large. However, while the PWIA and SAID results fall smoothly to 0 with increasing $p_m$, the Regge results show a second minimum in a region where it is comparable in magnitude to the rapidly falling cross section. This results in the large values for $A_{TT}$ which involves a ratio of the transverse-transverse contribution to the cross section to the sum of the longitudinal and transverse contributions. It should be noted that the relationship between the Fermi invariants and the response functions is very complicated and can involve interferences between the various contributions. As a result is has not been possible to isolate a single source for the second peak in the Regge $R_{TT}$ response function. Interference response functions and their associated asymmetries may show unpredictable sensitivities to small differences in the Fermi invariants. Ascertaining the significance of these differences requires that the errors in fitting parameters for the scattering amplitudes be propagated to the electrodisintegration calculations. This can be done for the Regge case since the hessian matrix can be generated for the fit, and is planned as a future work. Unfortunately, there is not access to similar information about the SAID helicity amplitudes.

Figures \ref{fig:stat_un}(e) and (f) show the longitudinal-transverse asymmetry $A_{LT}$ for the $x=1$ and $x=1.3$ kinematics respectively. At $x=1$, this asymmetry is relatively large and the two FSI models give comparable results and differ substantially form the PWIA. At $x=1.3$, the two FSI models have similar form but tend to be in less agreement than in the $x=1$ case. Both, however, are much closer to the PWIA result.

Figures \ref{fig:stat_un}(g) and (h) show the longitudinal-transverse asymmetry $A_{LT'}$ for the $x=1$ and $x=1.3$ kinematics respectively. Measurement of the asymmetry requires a polarized electron beam. Since this response is odd under the combination of time reversal and parity, its value is 0 in PWIA. For both kinematics it is small for both FSI models. The significance of the differences between the SAID and Regge results is unclear.

\begin{figure}
    \includegraphics[width=6in]{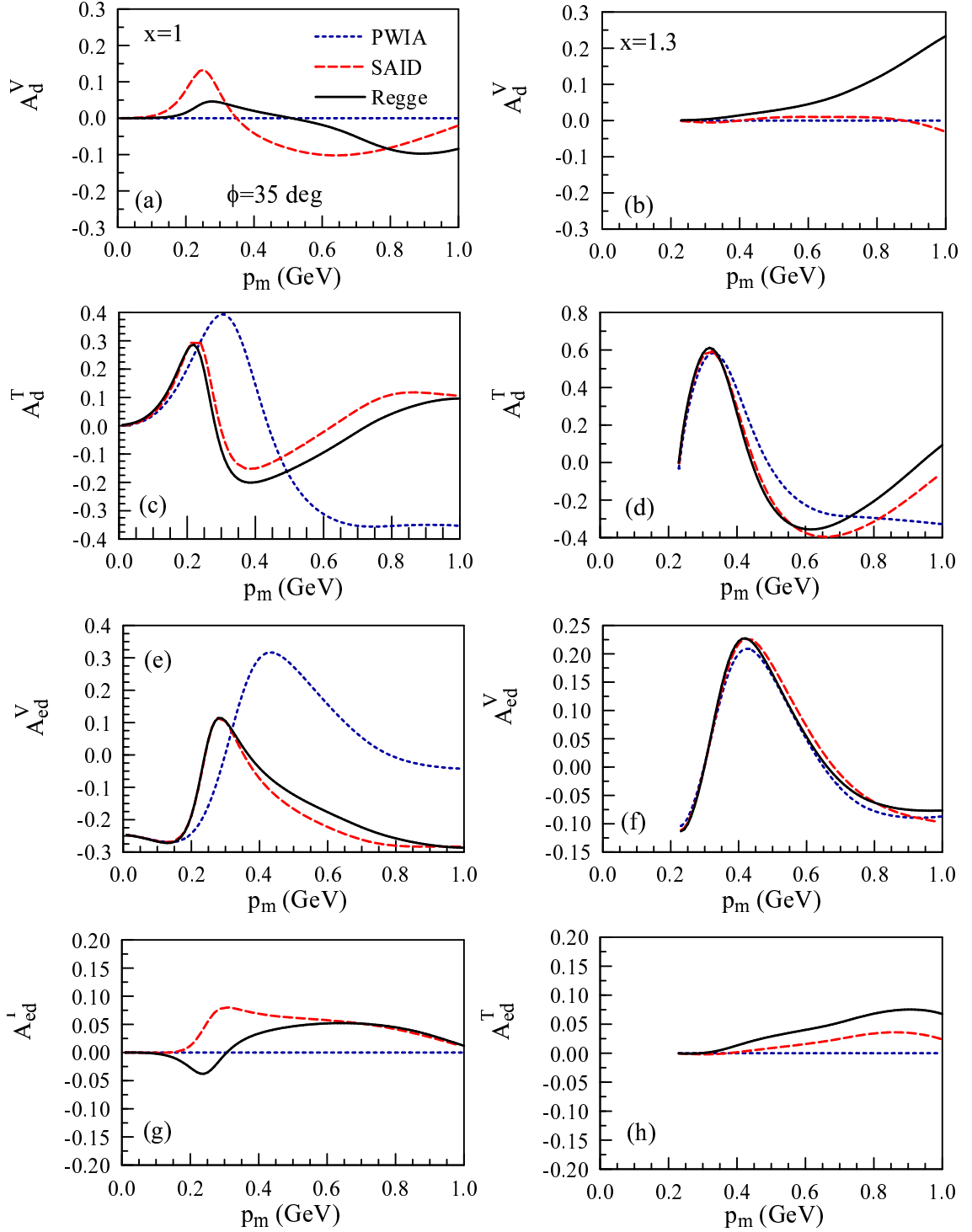}
     \caption{Single and double spin asymmetries for vector and tensor polarizations along the beam axis. Plots in the left-hand column are for the $x=1$ kinematics and plots in the right-hand column are for the $x=1.3$ kinematics. Lines are represented as in Fig. \ref{fig:stat_un}.}
    \label{fig:stat_targ}
\end{figure}

Figure \ref{fig:stat_targ} shows the single and double spin asymmetries for vector and tensor polarization of the target deuteron along the direction of the electron beam at an azimuthal angle of $\phi=35^\circ$. There is reasonable agreement in these observables between the Regge and SAID approach, as well as strong effects from the FSI. This suggests that target polarization asymmetries can provide insight to the effects of FSI while masking the model dependence of how these are calculated.

Figures \ref{fig:stat_targ}(a) and (b) shows the vector polarized target asymmetry $A^{V}_{d}$, for $x = 1$ and $x = 1.3$ kinematics respectively. Note that this asymmetry is zero in the absence of final state interactions. Qualitatively the Regge and SAID approaches are similar.

Figures \ref{fig:stat_targ}(c) and (d) show tensor polarized target asymmetry $A^{T}_{d}$, for $x = 1$ and $x = 1.3$ kinematics respectively. Here the two approaches are in excellent agreement, and there is a dramatic change in behavior for $x = 1$. The FSI contributions to this observable are minimal at the $x = 1.3$ kinematics.

Figures \ref{fig:stat_targ} (e) and (f) show the double spin asymmetry for vector polarized target and polarized beam $A^{V}_{ed}$, for $x = 1$ and $x = 1.3$ kinematics respectively. Note again that the two approaches are in excellent agreement and observe that the FSI contributions are minimal at the $x = 1.3$.

Figures \ref{fig:stat_targ}(g) and (h) show the double spin asymmetry for tensor polarized target and polarized beam $A^{T}_{ed}$, for $x = 1$ and $x = 1.3$ kinematics respectively. This asymmetry is zero in the PWIA. Qualitatively the approaches yield similar results, and while the FSI do cause a non zero contribution the value is relatively small.

\begin{figure}
    \includegraphics[width=6in]{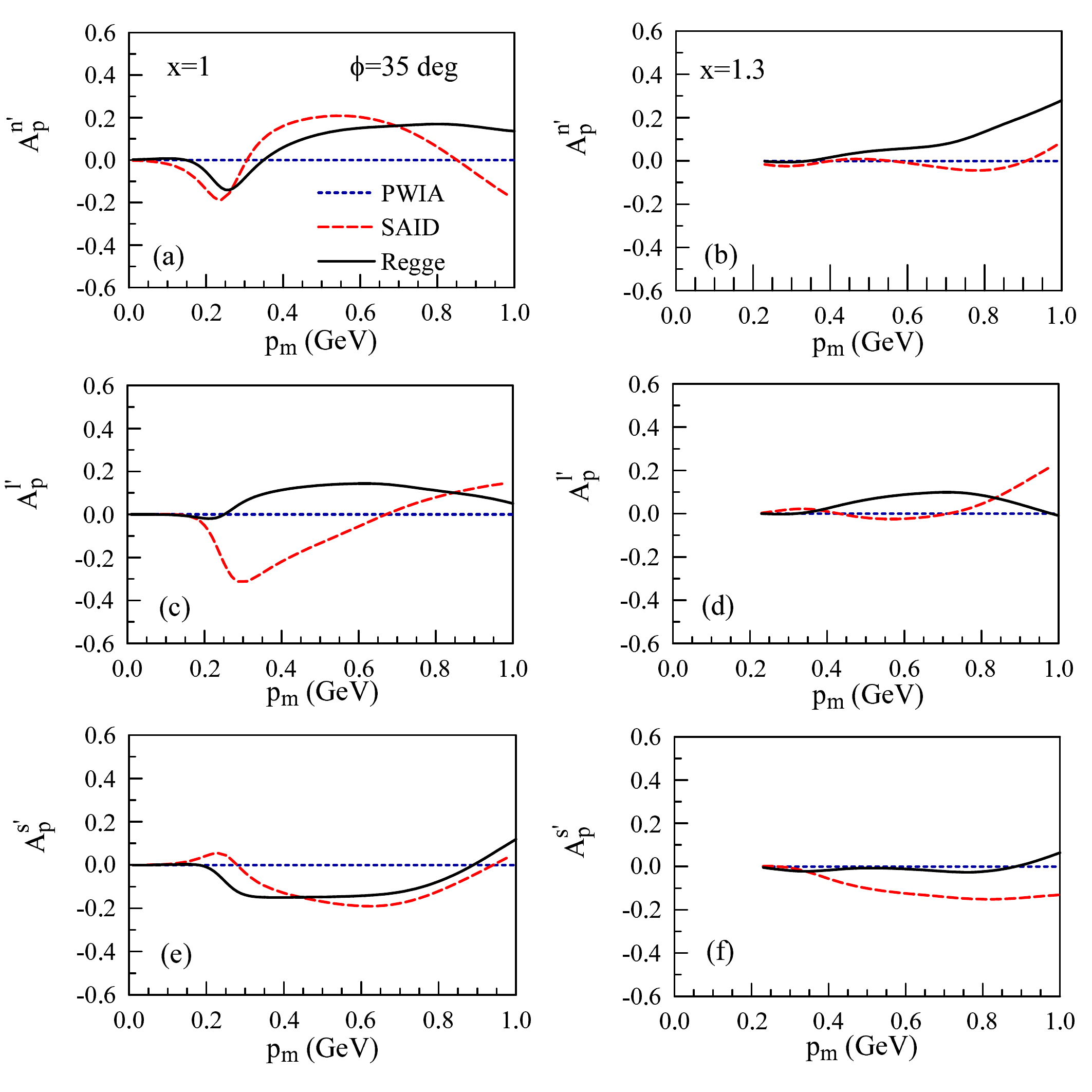}
    \caption{Single spin asymmetries for ejected protons polarized along the $\hat{n}'$, $\hat{l'}$ and $\hat{s}'$ directions. Plots in the left-hand column are for the $x=1$ kinematics and plots in the right-hand column are for the $x=1.3$ kinematics. Lines are represented as in Fig. \ref{fig:stat_un}.}
    \label{fig:stat_eject1}
\end{figure}

In Figure \ref{fig:stat_eject1} the results for polarized ejected proton are presented at an azimuthal angle of $\phi=35^\circ$. All asymmetries in Figure \ref{fig:stat_eject1} are zero for PWIA, thus presenting an ideal set of asymmetries for exploring the contribution of FSI.

Figures \ref{fig:stat_eject1}(a) and (b) show the asymmetry $A^{n'}_{p}$, for $x = 1$ and $x = 1.3$ kinematics respectively, and there is good agreement between the models. Figures \ref{fig:stat_eject1}(c) and (d) show the asymmetry $A^{l'}_{p}$. For the $x = 1$ kinematics it is observed that the model approaches are similar in magnitude but differ in sign. Because of the strong model dependence evident in this observable, and due to the relatively large value, this would provide an interesting measurement, which could shed light on the role of FSI as well as the various models used to calculate them. FSI effects at $x = 1.3$ are less pronounced. Figures \ref{fig:stat_eject1}(e) and (f) show the asymmetry $A^{s'}_{p}$. Here again there is good qualitative agreement between the two models.

\begin{figure}
    \includegraphics[width=6in]{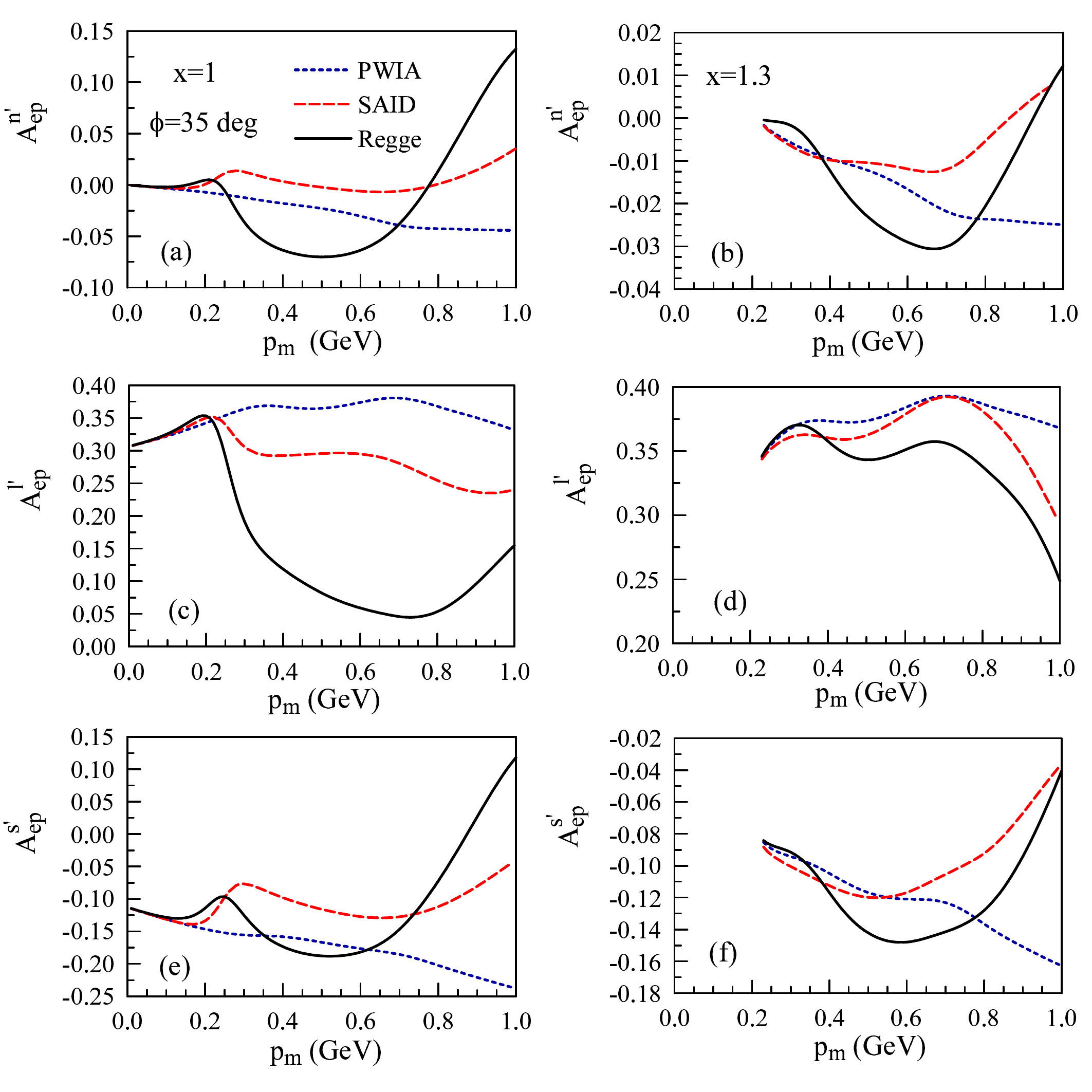}
    \caption{Double spin asymmetries for ejected protons polarized along the $\hat{n}'$, $\hat{l'}$ and $\hat{s}'$ directions. Plots in the left-hand column are for the $x=1$ kinematics and plots in the right-hand column are for the $x=1.3$ kinematics. Lines are represented as in Fig. \ref{fig:stat_un}.}
    \label{fig:stat_eject2}
\end{figure}

Figure \ref{fig:stat_eject2} shows the double spin asymmetries for polarized beam and polarized ejected proton at an azimuthal angle of $\phi=35^\circ$.
Figures \ref{fig:stat_eject2}(a) and (b) show the asymmetry $A^{n'}_{ep}$, for $x = 1$ and $x = 1.3$ kinematics respectively. Note that this asymmetry is highly sensitive to FSI model dependence at $x = 1$, causing deviation in opposite directions to the PWIA, although the magnitude of the deviation is relatively small. The same behavior is observed for $x = 1.3$, however less dramatic.

Figures \ref{fig:stat_eject2}(c) and (d) show the asymmetry $A^{l'}_{ep}$, for $x = 1$ and $x = 1.3$ kinematics respectively. In this case the same behavior between the SAID and Regge approaches is observed qualitatively, although the Regge model is much more drastic at $x = 1$. At $x = 1.3$ there are similar, albeit less pronounced effects. Due to the large differences between the approaches, it can again be noted that this observable is sensitive to the model dependence of the FSI, and it should be pointed out that measurements of this asymmetry would prove insightful.

Figures \ref{fig:stat_eject2}(e) and (f) show the asymmetry $A^{s'}_{ep}$, for $x = 1$ and $x = 1.3$ kinematics respectively. For both kinematics the two models are qualitatively similar, with relatively small deviations from the PWIA and each other.

%In this paper we have presented a new method of calculating final state interactions for the electrodisintegration of the deuteron. The FSI are calculated using a Regge model, which was fit to available NN scattering data. The model is fully relativistic and incorporates full spin dependence. With the addition of this new method we have significantly extended the kinematic range of the deuteron electrodisintegration calculation which was previously limited due to the absence of high energy NN amplitudes from SAID. We have presented results for kinematic regions where the SAID and Regge approaches overlap for comparison.

The comparisons suggest that for most of the observables there is good agreement between the two approaches. The expectation is that most discrepancies between the two models would be within error bands were they available. There is anticipation of being able to propagate the error for the Regge model once sufficient resources are available, however, an error analysis requires information that is unavailable from SAID. Propagation of the $NN$ fitting error to the electrodisintegration observables will require a substantial amount of computational resources.

The results are consistent with expectations that FSI play a vital role in understanding the reaction mechanism. It is observed that from the results there may be kinematic regions where FSI are minimized and PWIA is a valid approximation. This is most evident in the cross section and polarized target asymmetries. Asymmetries have been identified which have large FSI contributions and significant sensitivity to the model dependence of the two approaches, in particular $A_{TT}$, $A^{l'}_{p}$ and $A^{l'}_{ep}$. While almost all observables are sensitive to FSI and measurements would prove useful, these are particularly interesting because of the discrepancies between the two models.

\chapter{Conclusions}\label{sec:conclusions}
In this thesis a model has been presented to calculate mid to high energy nucleon-nucleon interactions.
No such models exist at these energies, which can be readily obtained from the community, therefore this work has sought to fill in this gap.
An adequate description at these energies is necessary in order to account for final state interactions for a variety of processes.
In order to describe this process effectively a relativistic, fully spin dependent model was developed.
The model was designed using Regge theory because of its ability to scale to higher energies. 
Especially for proton-neutron scattering there is a limited amount of data available at higher energies. 
It was therefore necessary that the model have the ability to allow confidence in extrapolating the results, and Regge theory facilitates this need well.
%Because of the ability of Regge theory to scale, we are able to extrapolate the results to $s \approx 20$ GeV$^2$ with reasonable confidence.

In chapter \ref{sec:theoretical framework} it was shown how the nucleon-nucleon scattering amplitudes can be parametrized in terms of Regge exchanges. 
Relating to the Fermi invariants allows us to calculate all spin dependence directly, 
while ensuring that a Regge exchange with definite quantum numbers contributes appropriately to the amplitude. 
The Regge analysis reduces to the spinless case, eliminating the need to worry about complicated crossing relations. 

In chapter \ref{sec:NNresults} the results were presented of the fit to available nucleon-nucleon scattering observables.
A discussion of the fitting procedure was included. This was developed in order to fit to such a large data set encompassing many observables over a large energy range.
Figures showing the solution of the fit to the data set were presented, and it should be noted that the model could be improved if more data were available. 
The fit results are acceptable, that is the model describes the data well with acceptable $\chi^2$ values, and the fit to observables was discussed. 
A comprehensive error analysis of the model would be extremely beneficial. 

In chapter \ref{sec:deepfsi} an application of the model to describe final state interactions in deuteron electrodisintegration was presented. 
The Regge model was compared to the previous method of calculating the final state interactions using the SAID parametrization.
Good agreement is noted between the methods for most observables. Observables which demonstrate large discrepancies between the approaches have been highlighted.
These observables are particularly interesting in that they may be able to shed light on the process due to the strong model dependence they exhibit in addition to the large role that final state interactions play.
It has been noted that for most of the observables of this system final state interactions cannot be ignored, 
although one may find particular kinematic regimes where they are minimized. 
Future work includes further study of this system at kinematics which will be accessible after the Jefferson Lab 12 GeV upgrade is complete. 
In addition off-shell effects should be taken into account, and the Regge model allows for a natural off-shell extrapolation \cite{FVO_offshell}. 

Work has begun to implement this model into a calculation for neutrino deuteron scattering. Specifically understanding the reaction $\nu +d \rightarrow \mu + p +p$, has been suggested as a way to more accurately determine neutrino flux, which is always an issue in neutrino experiments. As in the case of deuteron electrodisintegration, final state interactions are expected to play a significant role. 

The model is also planned to be utilized in a calculation of electrodisintegration of $^3\mathrm{He}$ and $^4\mathrm{He}$ \cite{Schiavilla:2005hz,Schiavilla:2004xa}. In these calculations the final state interactions are taken into account by a Glauber model, which requires the proton-proton and proton-neutron amplitudes as input. Implementing the Regge model in this calculation can significantly extend the applicable kinematic range.

It is anticipated that this model will be useful for future applications in the nuclear physics community. The model is intended to be available for others to utilize as needed. It is expected that providing this model, which is unique in the energy range that it describes, will prove to be a useful tool for many future applications.

\bibliography{PHD_thesis.bib} 
\bibliographystyle{apsrev4-1}

% \begin{thebibliography}{Bliggs}
% 
% \addtocontents{toc}{\vspace*{12pt}}
% 
% \addcontentsline{toc}{chapter}{BIBLIOGRAPHY}
% %
% \bibitem{KeisterandPolyzou} B. D. Keister and W. N. Polyzou, Adv. Nucl.
%                   Phys. {\bf 20}, 225 (1961).
% %
% \bibitem{Chung} P. L. Chung, F. Coester, B. D. Keister, and W. N.
%                  Polyzou, Phys. Rev. C {\bf 37}, 2000 (1988).
% %
% \bibitem{BetheSalpeter} E. E. Salpeter and H. A. Bethe, Phys. Rev. {\bf 84},
% 1232 (1951).
% 
% \end{thebibliography}

\appendix
\chapter{Parameters}
Parameter values are given in Table \ref{ta:params_pol}. A naming convention was utilized as follows. If the trajectory was obtained from the meson spectrum the trajectory name was choses to match the meson name. All ``effective'' trajectories, where the trajectory parameters were also fit parameters, are denoted $X_i$.  % and Table \ref{ta:params_unpol} respectively.
%\begin{turnpage}
\begin{landscape}
\begin{table}[htbp]\centering 
\caption{Parameter values for polarized solution. The fit parameters are indicated in bold.}
\label{ta:params_pol}
\scriptsize
\begin{tabular}{cccccccccccc} 
\hline \hline
$\beta_0$ & $\beta_1$ & $\delta$ & $\gamma$ & $\alpha_0$ & $\alpha_1$ & Isospin & Parity & G-Parity & Type & Residue & Name  \\  
\hline
\bm{$-2.3014\times10^2$} & \bm{$3.0982\times10^0$} & \multicolumn{1}{r}{$0.0000\times10^0$} &  & $1.0800\times10^0$ & $2.5000\times10^{-1}$ & 0 & $+$ & $+$ & 1 & $I$ & $\Pom$ \\ 
\bm{$3.3606\times10^1$} & \bm{$2.5208\times10^0$} & \bm{$-1.3505\times10^0$} & $3.0000\times10^0$ & \bm{$1.2915\times10^0$} & \bm{$3.0031\times10^{-1}$} & 0 & $+$ & $+$ & 1 & $II$ & $X_1$ \\ 
\bm{$-1.4315\times10^0$} & \bm{$4.2364\times10^{-1}$} & \bm{$-3.2163\times10^0$} & $3.3330\times10^{-1}$ & \bm{$1.2228\times10^0$} & \bm{$7.6208\times10^{-2}$} & 0 & $+$ & $+$ & 1 & $II$ & $X_2$ \\ 
\bm{$-4.2550\times10^2$} & \bm{$8.7558\times10^0$} & \bm{$-2.8650\times10^{-1}$} &  & $6.7000\times10^{-1}$ & $8.2000\times10^{-1}$ & 0 & $+$ & $+$ & 1 & $I$ & $f$ \\ 
\bm{$3.0579\times10^3$} & \bm{$3.5797\times10^0$} & \bm{$-5.0940\times10^0$} &  & \bm{$-7.1114\times10^{-1}$} & \bm{$1.1570\times10^0$} & 0 & $+$ & $+$ & 1 & $I$ & $X_3$ \\ 
\bm{$-5.1011\times10^1$} & \bm{$3.9362\times10^{-1}$} & \multicolumn{1}{r}{$0.0000\times10^0$} &  & $4.3000\times10^{-1}$ & $9.2000\times10^{-1}$ & 0 & $-$ & $-$ & 2 & $I$ & $\omega_a$ \\ 
\bm{$-8.3319\times10^2$} & \bm{$6.0000\times10^0$} & \bm{$-1.8189\times10^0$} &  & $1.3000\times10^{-1}$ & $8.3000\times10^{-1}$ & 0 & $-$ & $-$ & 2 & $I$ & $\phi_a$ \\ 
\bm{$3.3968\times10^5$} & \bm{$4.0113\times10^1$} & \bm{$2.0543\times10^0$} &  & \bm{$-8.3722\times10^0$} & \bm{$1.1658\times10^{-3}$} & 0 & $-$ & $-$ & 2 & $I$ & $X_4$ \\ 
\bm{$3.6954\times10^2$} & \bm{$1.0385\times10^1$} & \bm{$-2.8048\times10^0$} &  & $-2.3000\times10^{-1}$ & $8.6000\times10^{-1}$ & 0 & $+$ & $-$ & 3 & $I$ & $h$ \\ 
\bm{$-1.7985\times10^2$} & \bm{$1.4258\times10^0$} & \bm{$1.4192\times10^0$} &  & \bm{$5.5908\times10^{-1}$} & $8.2000\times10^{-1}$ & 0 & $+$ & $-$ & 3 & $I$ & $X_5$ \\ 
\bm{$-3.2225\times10^1$} & \bm{$3.5689\times10^{-2}$} & \bm{$-4.1374\times10^0$} &  & $4.3000\times10^{-1}$ & $9.2000\times10^{-1}$ & 0 & $-$ & $-$ & 4 & $I$ & $\omega_b$ \\ 
\bm{$-8.5937\times10^3$} & \bm{$6.0000\times10^0$} & \bm{$9.0817\times10^{-1}$} &  & $1.3000\times10^{-1}$ & $8.3000\times10^{-1}$ & 0 & $-$ & $-$ & 4 & $I$ & $\phi_b$ \\ 
\bm{$1.8226\times10^3$} & \bm{$9.5443\times10^{-1}$} & \bm{$3.2746\times10^0$} &  & \bm{$-6.5816\times10^0$} & \bm{$8.1649\times10^{-4}$} & 0 & $-$ & $-$ & 4 & $I$ & $X_6$ \\ 
\bm{$-5.0967\times10^2$} & \bm{$1.6424\times10^0$} & \bm{$-4.0676\times10^{-1}$} &  & $-2.3000\times10^{-1}$ & $8.6000\times10^{-1}$ & 0 & $-$ & $+$ & 5 & $I$ & $\eta$ \\ 
\bm{$3.4250\times10^2$} & \bm{$1.2447\times10^0$} & \bm{$-1.4211\times10^0$} &  & \bm{$5.9469\times10^{-3}$} & \bm{$2.4531\times10^{-1}$} & 0 & $-$ & $+$ & 5 & $I$ & $X_7$ \\ 
\bm{$7.7744\times10^1$} & \bm{$3.5462\times10^1$} & \bm{$2.5834\times10^{-1}$} &  & $-4.0000\times10^{-2}$ & $7.2000\times10^{-1}$ & 1 & $+$ & $+$ & 1 & $I$ & $b$ \\ 
\bm{$-5.9219\times10^2$} & \bm{$1.3809\times10^0$} & \bm{$-3.0997\times10^0$} &  & \bm{$-5.5996\times10^{-1}$} & \bm{$4.4269\times10^{-1}$} & 1 & $+$ & $+$ & 1 & $I$ & $X_8$ \\ 
\bm{$3.4534\times10^2$} & \bm{$1.2989\times10^0$} & \bm{$-5.2535\times10^0$} &  & $-4.0000\times10^{-2}$ & $7.2000\times10^{-1}$ & 1 & $-$ & $-$ & 2 & $I$ & $\pi_a$ \\ 
\hline \hline
\end{tabular}
\end{table}
\end{landscape}
%\end{turnpage}
\newpage
\begin{landscape}
\begin{table*}%[htbp]
\caption*{TABLE \ref{ta:params_pol}. (Continued)}
%\caption{Parameter values for polarized solution (continued)}
\scriptsize
\begin{tabular}{cccccccccccc}
\hline \hline
$\beta_0$ & $\beta_1$ & $\delta$ & $\gamma$ & $\alpha_0$ & $\alpha_1$ & Isospin & Parity & G-Parity & Type & Residue & Name  \\  
\hline
\bm{$9.1120\times10^2$} & \bm{$6.8368\times10^{-1}$} & \bm{$1.1157\times10^{-1}$} &  & \bm{$-1.1372\times10^1$} & \bm{$1.7908\times10^{-5}$} & 1 & $-$ & $-$ & 2 & $I$ & $X_9$ \\ 
\bm{$6.8240\times10^2$} & \bm{$4.3461\times10^{-1}$} & \bm{$-5.3158\times10^{-1}$} &  & $4.7000\times10^{-1}$ & $8.9000\times10^{-1}$ & 1 & $+$ & $-$ & 3 & $I$ & $a$ \\ 
\bm{$6.6961\times10^2$} & \bm{$5.4994\times10^{-1}$} & \bm{$2.3584\times10^0$} &  & \bm{$4.5559\times10^{-1}$} & $8.9000\times10^{-1}$ & 1 & $+$ & $-$ & 3 & $I$ & $X_{10}$ \\ 
\bm{$2.9363\times10^2$} & \bm{$4.4182\times10^{-1}$} & \bm{$-2.3057\times10^0$} &  & \bm{$2.8195\times10^{-1}$} & \bm{$8.9330\times10^{-1}$} & 1 & $+$ & $-$ & 3 & $I$ & $X_{11}$ \\ 
\bm{$-2.6347\times10^1$} & \bm{$1.7260\times10^{-3}$} & \bm{$3.1945\times10^0$} &  & $-4.0000\times10^{-2}$ & $7.2000\times10^{-1}$ & 1 & $-$ & $-$ & 4 & $I$ & $\pi_b$ \\ 
\bm{$-1.5941\times10^3$} & \bm{$1.6051\times10^0$} & \bm{$-3.8402\times10^0$} &  & \bm{$-1.3492\times10^0$} & \bm{$1.5166\times10^{-4}$} & 1 & $-$ & $-$ & 4 & $I$ & $X_{12}$ \\ 
\bm{$1.1653\times10^2$} & \bm{$6.0278\times10^1$} & \bm{$-5.5234\times10^0$} &  & \bm{$7.0679\times10^{-2}$} & \bm{$8.7412\times10^{-2}$} & 1 & $-$ & $+$ & 5 & $I$ & $X_{13}$ \\ 
\bm{$1.3195\times10^4$} & \bm{$4.1072\times10^0$} & \bm{$-1.5099\times10^0$} &  & $1.3000\times10^{-1}$ & $8.3000\times10^{-1}$ & 0 & $-$ & $-$ & 2 & $III$ & $X_{14}$  \\ 
\bm{$-1.4708\times10^3$} & \bm{$1.3829\times10^0$} & \bm{$1.9703\times10^0$} &  & $-4.0000\times10^{-2}$ & $7.2000\times10^{-1}$ & 1 & $+$ & $+$ & 1 & $III$ & $X_{15}$ \\ 
\bm{$-2.2575\times10^3$} & \bm{$1.7071\times10^0$} & \bm{$-2.5369\times10^0$} &  & $-4.0000\times10^{-2}$ & $7.2000\times10^{-1}$ & 1 & $-$ & $-$ & 2 & $III$ & $X_{16}$ \\ 
\bm{$4.2280\times10^3$} & \bm{$1.0389\times10^0$} & \bm{$-5.7058\times10^{-1}$} &  & \bm{$-1.8954\times10^0$} & \bm{$2.0775\times10^{-1}$} & 0 & $+$ & $+$ & 1 & $III$ & $X_{17}$ \\ 
\bm{$-6.3292\times10^2$} & \bm{$7.2339\times10^{-1}$} & \bm{$5.0303\times10^0$} &  & \bm{$-6.0089\times10^0$} & \bm{$4.8252\times10^{-1}$} & 0 & $-$ & $-$ & 2 & $III$ & $X_{18}$ \\ 
\bm{$-2.5071\times10^4$} & \bm{$1.9199\times10^0$} & \bm{$2.3593\times10^{-1}$} &  & \bm{$-1.2202\times10^1$} & \bm{$1.0805\times10^{-5}$} & 1 & $+$ & $+$ & 1 & $III$ & $X_{19}$ \\ 
\bm{$9.6450\times10^2$} & \bm{$7.5662\times10^{-1}$} & \bm{$1.1350\times10^{-1}$} &  & \bm{$-4.4005\times10^0$} & \bm{$4.4305\times10^{-3}$} & 0 & $+$ & $-$ & 3 & $III$ & $X_{20}$ \\ 
\bm{$4.6080\times10^1$} & \bm{$1.5845\times10^{-1}$} & \multicolumn{1}{r}{$0.0000\times10^0$} &  & \bm{$-2.1192\times10^{-1}$} & \bm{$3.7327\times10^{-1}$} & 0 & $-$ & $-$ & 4 & $III$ & $X_{21}$ \\ 
\bm{$1.3204\times10^0$}    &  & \bm{$3.0270\times10^0$} &  &  &  &  &  &  &  &  & $EM_a$ \\ 
\bm{$5.2979\times10^{-1}$} &  & \bm{$3.1599\times10^0$} &  &  &  &  &  &  &  &  & $EM_b$ \\ 
\bm{$4.6382\times10^{-1}$} &  & \bm{$3.7668\times10^0$} &  &  &  &  &  &  &  &  & $EM_c$ \\
\hline \hline
\end{tabular}
\end{table*}
\end{landscape}

\chapter{Amplitudes and Observables}
All observables can be written in terms of the five independent helicity amplitudes \cite{Bystricky} given in (\ref{eq:amplitudes(abcde)}). 
The $NN$ observables relevant to this paper are,
\begin{align}
 \sigma &= \frac{-2m^2}{\sqrt{s(s-4m^2)}}\Im\left[ a + c \right]_{t = 0}   \\
 \frac{d\sigma}{dt} &= \frac{m^4}{2 \pi s(s-4m^2)}\left( |a|^2 +4|b|^2 +|c|^2 +|d|^2 + |e|^2 \right) \\  
 %\frac{d\sigma}{d\Omega} &= \frac{m^4}{2 \pi s(s-4m^2)}\left( |a|^2 +4|b|^2 +|c|^2 +|d|^2 + |e|^2 \right) \\ 
 \tilde{\sigma} &=  \frac{1}{2}\left(|a|^2 +4|b|^2 +|c|^2 +|d|^2 + |e|^2 \right) \\
 \frac{d\sigma}{dt} &= \frac{m^4}{ \pi s(s-4m^2)}\tilde{\sigma}   \\ 
 \tilde{\sigma}P    &= \tilde{\sigma}A_N = -\Im[b^{*}(a + c + d - e)] \\
 \tilde{\sigma}A_{XX} &=  \Re(a^{*}d + c^{*}e) \\
 \tilde{\sigma}A_{ZX} &=  -\Re[b^{*}(a+d-c+e)] \\
 \tilde{\sigma}A_{ZZ} &= -\frac{1}{2}\left(|a|^2  +|d|^2 - |c|^2 - |e|^2 \right) \\
 \tilde{\sigma}A_{YY} &=  \Re(a^{*}d - c^*e) +2|b|^2 \\
 \tilde{\sigma}D      &=  \Re(a^{*}c - d^*e) +2|b|^2 \\
 \tilde{\sigma}D_{T}  &=  \Re(a^{*}e - d^*c) +2|b|^2
\end{align}
\chapter{Helicity Spinors}
The metric, spinor normalization, form of the gamma matrices, and other conventions are those used by Bjorken and Drell \cite{bjorken2013relativistic}.
In the center of momentum frame the helicity spinors are,
\begin{align} \label{eq:spinors}
 u( \pm \mathbf{p}, \lambda) &= N \left( \begin{array}{c} 1  \\  2 \lambda \tilde{p} \end{array} \right) \chi_{\pm \lambda}(\mathbf{\hat{p}}) , \\
 v( \pm \mathbf{p}, \lambda) &= N \left( \begin{array}{c} -2 \lambda \tilde{p}  \\ 1 \end{array} \right) \chi_{\mp \lambda}(\mathbf{\hat{p}}) ,
\end{align}
where $N =  \sqrt{\frac{E + m}{2m}}$, $\tilde{p} = \frac{|\mathbf{p}|}{E + m}$, $\mathbf{\hat{p}}$ is a unit vector in the direction of $\mathbf{p}$, and $\chi_{\pm \lambda}(\mathbf{\hat{p}})$ are given in Table \ref{ta:spinors}.
\begin{table}[h] \centering
\caption{Two component spinors of (\ref{eq:spinors})} 
\begin{tabular}{|c|c c} 
                &  $\chi_{\frac{1}{2}}(\mathbf{\hat{p}})$ &  $\chi_{-\frac{1}{2}}(\mathbf{\hat{p}})$  \\ \hline \hline \\
initial state   &   $ \left( \begin{array}{c} 1 \\ 0 \end{array} \right)$    & $ \left( \begin{array}{c} 0 \\ 1 \end{array} \right) $   \\ \\ \hline   \\  
final state	&   $ \left( \begin{array}{c} \cos{\frac{\theta}{2}} \\  \sin{\frac{\theta}{2}} \end{array} \right)   $ 
     & $ \left( \begin{array}{c}    -\sin{\frac{\theta}{2}}   \\ \cos{\frac{\theta}{2}}    \end{array} \right)  $
\end{tabular}
\label{ta:spinors}
\end{table}
\chapter{Amplitudes to Fermi Invariants}
The helicity dependent matrices which relate the Fermi invariants to the helicity amplitudes are,
\begin{equation} \label{eq:C_t} \small
 C_{ij}^{t} = 
\left( \begin{array}{ccccc}
1+\frac{t}{s-4m^2}      &      C^{t}_{12}  & -2 + \frac{2t}{s-4m^2} & 0  & C^{t}_{15} \\ \\
C^{t}_{21}              & C^{t}_{21}                                           & 2C^{t}_{21}            & 0  &-C^{t}_{21}                             \\ \\
1+\frac{t}{s-4m^2}      &  C^{t}_{32} & 2+\frac{2t}{s-4m^2}   & 0  & -C^{t}_{32}  \\ \\
\frac{st}{4m^2(s-4m^2)} & \frac{t}{s-4m^2}   & \frac{s-2m^2}{m^2}\left( 2+\frac{t}{s-4m^2} \right)      & \frac{t}{4m^2} &  -2-\frac{t}{s-4m^2}         \\ \\
\frac{-st}{4m^2(s-4m^2)}& \frac{-t}{s-4m^2}  &  \frac{-2t}{s-4m^2}                         & \frac{t}{4m^2}           & \frac{t}{s-4m^2} \\ \\
\end{array} \right)
\end{equation}

\begin{equation}
	C^{t}_{12}= -1 + \frac{s}{2m^2} +\frac{t}{s-4m^2}   
\end{equation}
\begin{equation}
	C^{t}_{15} = -1 + \frac{s}{2m^2} - \frac{t}{s-4m^2} 
\end{equation}

\begin{equation}
C^{t}_{21}=-\frac{\sqrt{s}}{4m}\sin(\theta) =  -\frac{\sqrt{s}}{2m}\sqrt{\frac{-t}{s-4m^2}} +\frac{\sqrt{s}}{4m}\left(\frac{-t}{s-4m^2} \right)^\frac{3}{2}
\end{equation}
\begin{equation}
C^{t}_{32}= \frac{1}{2m^2}(s-2m^2)\left(1+\frac{t}{s-4m^2}\right)
\end{equation}
\begin{align} \label{eq:C_u} \footnotesize
 C_{ij}^{u} = 
\left( \begin{array}{ccccc}
-1-\frac{u}{s-4m^2}     &  C^u_{12} & 2 - \frac{2u}{s-4m^2} & 0  & C^u_{15} \\ \\
C^u_{21} & C^u_{21} & 2C^u_{21} & 0 & -C^u_{21} \\ \\
\frac{-su}{4m^2(s-4m^2)} & \frac{-u}{s-4m^2}  & \frac{-2u}{s-4m^2}  & \frac{u}{4m^2}  & \frac{u}{s-4m^2}  \\ \\
\frac{-su}{4m^2(s-4m^2)} & \frac{-u}{s-4m^2}   & \frac{s-2m^2}{m^2}\left(-2-\frac{u}{s-4m^2} \right)   & -\frac{u}{4m^2} & 2+\frac{u}{s-4m^2} \\ \\
1+\frac{u}{s-4m^2}       & C^u_{52}  &  2+\frac{2u}{s-4m^2} & 0 & C^u_{52} \\ \\
\end{array} \right)
\end{align}
\begin{equation}
	C^u_{12} = 1 - \frac{s}{2m^2} -\frac{u}{s-4m^2}  
\end{equation}
\begin{equation}
	C^u_{15} = 1 - \frac{s}{2m^2} + \frac{u}{s-4m^2}
\end{equation}
\begin{equation}
	C^u_{21} = -\sqrt{\frac{(4m^2-s-u)su}{4m^2(s-4m^2)^2}}
\end{equation}
\begin{equation}
C^u_{52} = \frac{s-2m^2}{m^2}\left(-1-\frac{u}{s-4m^2} \right)
\end{equation}

\newpage
\vita{
\section*{Education}
\begin{itemize}
	\item \textbf{Ph.D.} Physics, Old Dominion University, Norfolk VA
	\item \textbf{M.S.} Physics, Wright State University, Dayton OH
	\item \textbf{B.S.} Engineering Physics, Wright State University, Dayton OH
\end{itemize}
\section*{Honors and Awards}
\begin{itemize}
	\item Old Dominion University, College of Sciences, Graduate Student Fellowship Spring 2013. 
	\item JSA/JLab Graduate Fellowship for academic years 2010-2011 and 2011-2012. 
	\item Old Dominion University Graduate Fellowship for academic year 2007-2008. 
	\item Old Dominion University Physics Department Award for outstanding achievement in the 2007 Ph.D. candidacy exam.
\end{itemize}
%\section*{Presentations}
%{\noindent
%``Regge Model for NN Scattering and Applications to Final State Interactions'', \\
%Nuclear Physics Seminar, the Ohio State University, Columbus, OH, March 2013 \\ \\
%%
%``NN Relativistic Scattering Amplitudes and Their Application to Final State Interactions'', \\
%Nuclear Physics Seminar, Old Dominion University, Norfolk, VA, February 2013 \\ \\
%%
%``A Regge Model for Nucleon-Nucleon Scattering Amplitudes'', \\
%Talk at the 2012 Elba XII workshop on Electron-Nucleus Scattering, \\ Elba, Italy, June 2012 \\ \\
%%
% ``A Regge Model for Nucleon-Nucleon Scattering Amplitudes'', \\
%Theory Seminar, University of Tuebingen, Tuebingen, Germany, June 2011 \\ \\
%%
%``A Regge Model for Nucleon-Nucleon Scattering Amplitudes'', \\
%Theory Seminar, Jefferson Laboratory, Newport News, VA, December 2010 \\ \\
%%
%``A Regge Model for Nucleon-Nucleon Scattering Amplitudes'', \\
%Talk at the 2010 Annual Fall Meeting of the APS Division of Nuclear Physics, \\ Santa Fe, NM, November 2010 \\ \\
%%
%``A Regge Model for Nucleon-Nucleon Scattering Amplitudes'', \\
%Poster at the 2010 Gordon Conference on Photonuclear Reactions, \\ Tilton, NH, August 2010
%}
\section*{Publications}
%\bibverse{Ford:2012dg}
{\noindent
William P. Ford, Sabine Jeschonnek, J. W. Van Orden \\
``$^2$H$(e,e'p)$ observables using a Regge model parametrization of final state interactions'' 
\href{http://prc.aps.org/abstract/PRC/v87/i5/e054006}{Phys. Rev. C 87, 054006 (2013)}
\\ \\
William P. Ford and J. W. Van Orden \\
``Regge model for nucleon-nucleon spin dependent amplitudes'' \\
\href{http://prc.aps.org/abstract/PRC/v87/i1/e014004}{Phys. Rev. C 87, 014004 (2013)}
%\\ \\
%William P.~Ford \\
%``Development of a Fourier transform far infrared (FTFIR) spectrometer to characterize broadband transmission properties of common materials in the terahertz region'' \\
%\href{http://etd.ohiolink.edu/view.cgi?acc_num=wright1158508606}{Masters Thesis}, Wright State University, (2006). 
}
}
 \vitapage

\end{document}